\documentclass{article}
\pdfoutput=1
\usepackage{amsmath, amssymb, bm}
\usepackage{authblk}
\usepackage[labelfont=bf]{caption}
\usepackage{float} 
\usepackage[flushmargin]{footmisc}
\usepackage{textcomp} 
\usepackage{gensymb}
\usepackage[a4paper]{geometry}
\usepackage{graphbox} 
\usepackage{graphicx}
\usepackage{hyperref}
\usepackage[utf8]{inputenc}
\usepackage{makecell}
\usepackage{multirow}
\usepackage{natbib} 
\usepackage{rotating}
\usepackage{subcaption} 
\usepackage{tabularx}
\usepackage{units} 
\usepackage{url}
\usepackage{verbatimbox} 
\usepackage[dvipsnames]{xcolor}
\usepackage{nicefrac}

\newcolumntype{C}{>{\raggedright\arraybackslash}X}

\newcommand\Includegraphics[2][]{\addvbuffer[3pt 0pt]{\includegraphics[#1]{#2}}} 

\providecommand{\keywords}[1]
{
  \small	
  \textbf{\textit{Keywords---}} #1
}

\def\bfA{\mathbf A}

\def\bfe{\mathbf e}

\def\bfw{\mathbf w}

\def\bfbeta   {\bm \beta}

\def\bftheta  {\bm \theta}

\def\bfPsi     {\mathbf \Psi}

\def\bfSigma  {\mathbf \Sigma}

\def\boldfacefake#1{\kern-4pt
	\hbox{ \mathsurround=0pt
		\hbox to 0.4pt{$#1$\hss}\hbox to 0.4pt{$#1$\hss}\hbox {$#1$}}}

\newcommand{\be}{\begin{eqnarray}}
	\newcommand{\ee}{\end{eqnarray}}
\newcommand{\ba}{\begin{eqnarray*}}
	\newcommand{\ea}{\end{eqnarray*}}
\newcommand{\bc}{\begin{center}}
	\newcommand{\ec}{\end{center}}
\newcommand{\btab}[1]{\begin{tabular}{#1}}
	\newcommand{\etab}{\end{tabular}}

\title{Spatial Bayesian GLM on the cortical surface produces reliable task activations in individuals and groups}
\author[1]{Daniel Spencer\thanks{Corresponding author: Daniel Spencer, danieladamspencer@gmail.com}}
\affil[1]{Department of Statistics, Indiana University, Myles Brand Hall E104 901 E. 10th Street Bloomington, IN, 47408, USA}
\author[2]{Yu (Ryan) Yue}
\affil[2]{Paul H. Chook  Department of Information Systems and Statistics, Baruch College, The City University of New York, New York, NY, 10010, USA}
\author[3]{David Bolin}
\affil[3]{CEMSE Division, King Abdullah University of Science and Technology, Thuwal, Makkah Province, 23955-6900, Saudi Arabia}
\author[4]{Sarah Ryan}
\affil[4]{Department of Biostatistics, Epidemiology, and Informatics; University of Pennsylvania Perelman School of Medicine, Philadelphia, PA, 19104, USA}
\author[1]{Amanda F. Mejia}
\date{\today}

\begin{document}
\maketitle
\vspace{-5mm}
\begin{abstract}
    The general linear model (GLM) is a widely popular and convenient tool for estimating the functional brain response and identifying areas of significant activation during a task or stimulus. However, the classical GLM is based on a massive univariate approach that does not explicitly leverage the similarity of activation patterns among neighboring brain locations. As a result, it tends to produce noisy estimates and be underpowered to detect significant activations, particularly in individual subjects and small groups. A recently proposed alternative, a cortical surface-based spatial Bayesian GLM, leverages spatial dependencies among neighboring cortical vertices to produce more accurate estimates and areas of functional activation. The spatial Bayesian GLM can be applied to individual and group-level analysis. In this study, we assess the reliability and power of individual and group-average measures of task activation produced via the surface-based spatial Bayesian GLM. We analyze motor task data from 45 subjects in the Human Connectome Project (HCP) and HCP Retest datasets.  We also extend the model to multi-run analysis and employ subject-specific cortical surfaces rather than surfaces inflated to a sphere for more accurate distance-based modeling. Results show that the surface-based spatial Bayesian GLM produces highly reliable activations in individual subjects and is powerful enough to detect trait-like functional topologies. Additionally, spatial Bayesian modeling enhances reliability of group-level analysis even in moderately sized samples ($n=45$). Notably, the power of the spatial Bayesian GLM to detect activations above a scientifically meaningful effect size is nearly invariant to sample size, exhibiting high power even in small samples ($n=10$). The spatial Bayesian GLM is computationally efficient in individuals and groups and is convenient to implement with the open-source \texttt{BayesfMRI} R package.
\end{abstract}


\newpage

\section{Introduction}

The functional topology of the human brain has been shown to be highly individualized, thanks to recent studies collecting large amounts of functional magnetic resonance imaging (fMRI) data on individual subjects \citep{gordon2020msc, laumann2015functional, choe2015reproducibility, braga2017parallel, kong2019spatial, barch2013function}. Functional boundaries have been shown to be consistent under task and rest conditions \citep{laumann2015functional} and to be predictive of behavior \citep{kong2019spatial}. Unfortunately, uncovering individualized functional topology has often relied on collecting vast amounts of data on individual subjects, which is infeasible in many settings due to practical constraints and participant considerations.  Some populations of vital research and clinical interest are generally unable to undergo long or frequent scans, including young children, the elderly, or those with developmental disorders or suffering from neurodegenerative disease. Recently, practical Bayesian techniques have been proposed as a way to extract reliable and predictive measures of brain organization in individuals based on much shorter scan duration by leveraging information shared across multiple observations to improve estimation \citep{bzdok2017inference, kong2019spatial, mejia2020template}.  

However, most analyses of task fMRI continue to focus on estimating group-level effects using conventional analytical methods, particularly in the absence of large amounts of data on individuals. Group-level analyses are favored in part because individual-level measures of task activation produced using the classical ``massive univarite'' general linear model (GLM) have been found to be unreliable \citep{elliott2020test}. In the classical GLM, which is popular due to its simplicity and computational efficiency, a separate linear model is fit at every location of the brain relating observed blood oxygenation level dependent (BOLD) activity to the expected hemodynamic response to a series of tasks or stimuli \citep{friston1995analysis}.  Task activation amplitude shows strong local spatial dependence, but information shared across locations is not leveraged at this stage except through ad-hoc smoothing \citep{mikl2008effects}. To identify areas of activation due to each task, a \textit{t}-test is then performed at each location, which requires correcting for the massive number of multiple comparisons this involves. This correction, combined with failure to fully leverage information shared across brain locations, often results in a lack of power to detect many true activations at the group level for small sample sizes (e.g., $n=20$ to $30$), and even more so in individual subjects \citep{lindquist2015zen, cremers2017relation}.

While group-level discoveries using task fMRI have greatly advanced general understanding of brain function and organization, as well as systematic differences related to disease, condition, and normal development and aging, individual-level measures are vital for advancing fMRI-based research to new frontiers. Longitudinal modeling, biomarker discovery, therapeutic clinical trials, translation of research findings into clinical practice, and pre-surgical planning all depend on extracting accurate measures of brain function and organization in individual subjects, often without the luxury of long or multiple sessions of data. Therefore, it is vital to develop more powerful and accurate methods. A promising direction is to incorporate expected patterns of spatial dependence and sparsity in activation amplitude \citep{zhang2015bayesian} through spatial Bayesian models. Several such models have been proposed for volumetric (typically slice-wise) analysis \citep{spencer2020joint,zhang2014spatio,zhang2015bayesian,zhang2016spatiotemporal}. Yet there is growing evidence in favor of surface-based analyses to improve sensitivity, power and reproducibility \citep{fischl1999cortical, anticevic2008comparing, tucholka2012empirical, glasser2013minimal, brodoehl2020surface}. Importantly, surface-based analysis avoids spurious activations induced by mixing signals across distinct cortical areas \citep{glasser2013minimal, brodoehl2020surface}, which can occur in standard volumetric smoothing as well as spatial Bayesian models applied to volumetric data, since they implicitly smooth activations. 

Recently, \citet{mejia2020bayesian} proposed a novel surface-based spatial Bayesian GLM, which combines the benefits of spatial modeling and the advantages of cortical surface analysis. In this framework, activation amplitudes are based on the mean of the posterior distribution for each task, which incorporates spatial dependence and sparsity from the prior, yielding smoother and more focal regions of peak activation.  Areas of activation are identified using the joint posterior distribution through an excursions set approach \citep{bolin2015excursion}, avoiding the need for multiple comparisons correction and greatly increasing power. Group effects can be estimated through a computationally efficient approach based on combining the results of each subject-level model in a principled way. The Bayesian computation is performed using integrated nested Laplace approximations (INLA) \citep{rue2009approximate}, which is computationally efficient and does not suffer from the inaccuracies common to variational Bayesian approaches \citep{siden2017fast}, which have been commonly used in volumetric spatial Bayesian analyses. 

This approach was validated by \citet{mejia2020bayesian} through simulation studies and a study of twenty individuals from the Human Connectome Project (HCP) \citep{barch2013function}. These analyses showed a major improvement to estimation efficiency and power, relative to the classical massive univariate GLM.  However, this approach has not yet been fully validated with real data, and how accurately it estimates brain function and organization reflecting individualized functional topology remains to be determined.  In this paper, we extensively validate the surface-based spatial Bayesian GLM framework in terms of reliability and power. We do not perform a simulation studies here, as extensive simulations were performed within \cite{mejia2020bayesian}, showing improvements in the accuracy and power of the spatial Bayesian GLM for both subject-level and group analysis. We illustrate the gain in power and statistical efficiency of the Bayesian approach for both subject-level and group-level analyses, particularly for shorter scan durations.  To assess the ability of the Bayesian approach to extract unique individual-level insights, we examine the reliability of the the Bayesian estimates and areas of activation. 

We also extend the original model proposed by \cite{mejia2020bayesian} in two important ways.  First, in the original model, the spherical surfaces from each subject were used as the spatial domain.  Inflation to the sphere, while useful for inter-subject registration, distorts the distances between neighboring vertices up to 3-fold (\textbf{Appendix Figure \ref{fig:sphere_dist}}).  This has implications for the smoothing of task activations performed implicitly in the Bayesian model, because the degree of dependence between neighbors is a function of the distance between them.  In this work, we use the midthickness surface of each individual subject having been registered to the fsaverage32k template, which was created using the FreeSurfer software platform \citep{fischl2012freesurfer} and is freely available in the HCP data release \citep{barch2013function}. This surface geometry respects the individual anatomical features of each individual subject and preserves the geodesic distances between locations along the cortical surface, while aligning vertices across subjects for possible group-level analysis. 

Second, the original model was proposed for single-subject, single-run analysis (with group-level analysis possible through a principled post-hoc approach). Here, we generalize the model to multi-run analysis. In this framework, separate run-specific estimates and areas of activation are produced, along with cross-run averages.  A major advantage of the multi-run model is that hyperparameters controlling the spatial properties of each task activation field are shared across runs, improving estimation efficiency. 

The remainder of this paper is organized as follows. The \ref{sec:Methods} section will outline the surface-based spatial Bayesian GLM and the classical GLM, and will also include a description of the data, the model estimation procedure, and the reliability metrics.
Section \ref{sec:Results} will outline the application and results of analyses of the motor task data from the Human Connectome Project \citep{barch2013function} using the classical and Bayesian GLMs. Section \ref{sec:Discussion} will summarize the findings.

\section{Methods}
\label{sec:Methods}

\subsection{Surface-Based Spatial Bayesian GLM}
\label{sec:SBSB_GLM}

\begin{figure}[t]
    \centering
    \includegraphics[width = 1\textwidth]{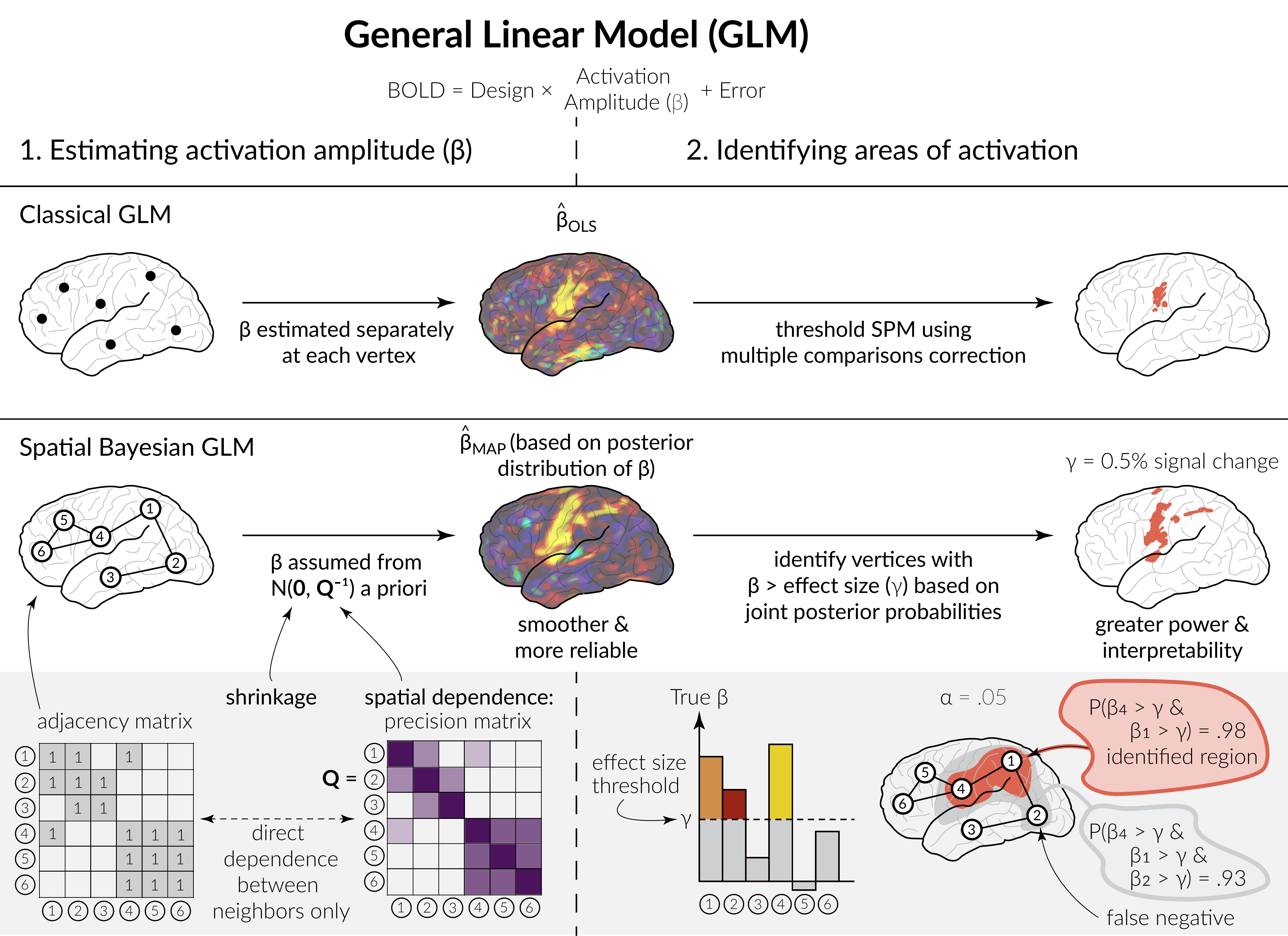}
    \caption{The surface-based spatial Bayesian GLM compared with the classical GLM. Both GLMs consist of two stages: (1) estimating activation amplitude and (2) identifying areas of activation.  At stage 1, the Bayesian GLM incorporates spatial dependence and performs shrinkage of background locations through a prior on $\bfbeta$, resulting in smoother and more reliable estimates of activation, given by the maximum-a-posteriori (MAP) value from the posterior distribution of $\bfbeta$.  At stage 2, the Bayesian GLM identifies the collection of vertices with activation amplitude above a specified effect size, based on joint posterior probabilities.  This results in greater power to detect true activations.}
    \label{fig:graphical_abstract}
\end{figure}

The subject-level surface-based spatial Bayesian (SBSB) GLM proposed by \cite{mejia2020bayesian} consists of two stages: model estimation and identifying areas of activation.  \textbf{Figure \ref{fig:graphical_abstract}} illustrates both stages in contrast with the classical GLM.  Below, we describe each stage briefly, including our novel multi-run extension.  The SBSB GLM also allows for computationally efficient group-level estimation, described below. For more details on the mathematical construction and Bayesian computation of the SBSB model, see \cite{mejia2020bayesian}. The Bernstein-von Mises theorem, which gives asymptotic guarantees concerning convergence for Bayesian models, holds in the Bayesian implementation of the model as long as regularity conditions about the mean are met and the model itself is not misspecified \cite{van2000asymptotic}. INLA centers the posterior around the maximum likelihood estimator with covariance equal to the inverse Fisher information matrix using a normal prior distribution as the sample size approaches infinity. This satisfies the regularity conditions imposed by the Bernstein-von Mises Theorem because the density function is continuous and twice differentiable everywhere. Please see \cite{van2000asymptotic} for further details about the conditions of the Bernstein-von Mises theorem.

\subsubsection{Single-subject modeling}
\label{sec:single_subject_model}

\textbf{Single-run model.} Let $N$ be the number of vertices on the cortical surface where BOLD signal is measured, and let $T$ be the duration of the fMRI timeseries. The classical GLM \citep{friston1995analysis} adapted to the cortical surface is based on fitting a separate regression model at each vertex.  In each model, the response is the observed BOLD activity, and the predictors are the expected BOLD response due to each of $K$ tasks or stimuli, which is constructed by convolving the timeseries of stimulus presentation with a haemodynamic response function (HRF).  For simplicity, assume that nuisance signals (e.g. head motion parameters, drift) have been regressed from both the response and task predictors, and assume that the data has been prewhitened to remove temporal autocorrelation in the model residuals and to eliminate spatial heterogeneity in the residual variance.  Then, the classical GLM at vertex $v$ can be represented as 
\begin{align}\label{eq:basicGLMlocv}
    \mathbf{y}_v = \mathbf{X}_v \boldsymbol{\beta}_v + \boldsymbol{\epsilon}_v, 
\end{align}
where $\mathbf{y}_v \in \mathbb{R}^T$ is the observed BOLD timeseries, $\mathbf{X}_v \in \mathbb{R}^{T \times K}$ contains the expected response to each of the $K$ stimuli, $\boldsymbol{\beta}_v \in \mathbb{R}^K$ are the coefficient values representing the activation amplitude for each stimulus at a single vertex $v$, and $\boldsymbol{\epsilon}_v \overset{ind}{\sim} \text{Normal}(\mathbf{0},\sigma^2\mathbf{I}_T)$ are white-noise residuals. Note that $\mathbf{X}_v$ may vary across vertices due to prewhitening, which involves pre-multiplying the original design matrix by a vertex-specific whitening matrix. If no prewhitening is performed, then $\mathbf{X}_1 = \cdots = \mathbf{X}_N$. While computationally convenient and simple, fitting thousands of separate models is clearly suboptimal, since neighboring vertices are known to exhibit similar patterns of task activation. If these similarities are not explicitly modeled, the estimates of activation will contain high levels of noise due to reduced statistical efficiency. 

A spatial Bayesian GLM addresses this by treating the \textit{image} of activation amplitudes in response to task $k$, $\boldsymbol{\beta}_k = (\beta_{1k},\ldots,\beta_{Nk})' \in \mathbb{R}^{N}$, as a latent field across the $N$ data locations, and assuming a spatial prior to incorporate prior knowledge of local spatial dependence and sparsity \citep{zhang2015bayesian}. The surface-based spatial Bayesian GLM proposed by \cite{mejia2020bayesian} makes use of a particular class of Gaussian Markov Random Field (GMRF) priors called stochastic partial differential equation (SPDE) priors, which approximate a continuous Mat\'ern random field by a GMRF \citep{lindgren2011explicit,bolin2013comparison}. SPDE priors are particularly well-suited to model cs-fMRI data for several reasons: they have sparse precision (required for high dimensional contexts), they are built on a triangular mesh (the format of cs-fMRI data, see \textbf{Appendix Figure \ref{fig:cifti_mesh}}), they have two separate parameters to control the scale and the smoothness, they are invariant to finite resamplings, and the parameters are interpretable given the relationship with the Mat\'ern covariance function. The SPDE prior uses the Mat\'ern kernel to inform the spatial covariance for the latent field of task coefficients. The Mat\'ern covariance for a pair of vertices $\mathbf{u}$ and $\mathbf{v}$ is a function of their distance $||\mathbf{u} - \mathbf{v}||$, and is explicitly defined as:
    \begin{align*}
        \text{cov}(\mathbf{u}, \mathbf{v}) & = \sigma^2(\kappa ||\mathbf{u} - \mathbf{v}||)K_1(\kappa ||\mathbf{u} - \mathbf{v}||),
    \end{align*}
where $\sigma^2 > 0$ is the variance and $K_1(\cdot)$ is the modified Bessel function of the second kind of order 1, which decreases rapidly as the distance between two vertices increases (see \cite{mejia2020bayesian} for further details). It is important to note that $||\mathbf{u} - \mathbf{v}||$ represents the \textit{geodesic} distance between vertices $\mathbf{u}$ and $\mathbf{v}$, and not the Euclidean distance. The geodesic distances are calculated using the subject-specific cortical surfaces and takes the folded geometry of the cortex into account to find the distance along the surface.

An SPDE prior for a given Gaussian process $\boldsymbol{\beta}$ takes the form
\begin{align}
    \boldsymbol{\beta} & = \bfPsi \bfw,\quad
    \mathbf{w} \sim \text{Normal}(\mathbf{0},\mathbf{Q}_{\kappa,\tau}^{-1}), \label{eq:spde_prior1}\\
    \mathbf{Q}_{\kappa,\tau} & = \tau^2(\kappa^4\mathbf{C} + 2\kappa^2\mathbf{G} + \mathbf{GC}^{-1}\mathbf{G}), \label{eq:spde_prior2} 
\end{align}
where $\bfPsi$ is an $N\times n$ indicator matrix in which element $\psi_{i,j} = 1$ when data location $i$ corresponds to vertex $j$, and 0 for all $\psi_{i,j'}$, where $j \neq j'$. Often, $\boldsymbol{\Psi}$ is an identity matrix because cortical surface data locations are already on a mesh. However, in some cases, the mesh may contain additional locations to satisfy shape and size constraints and boundary locations to improve estimation along the data boundary. In our study, the cortical surface geometry is stored in the form of a triangular mesh, and so a mesh does not to be constructed. However, in the case when no cortical surface geometry is available, an automated procedure to construct a mesh is implemented in the \texttt{R-INLA} package \cite{martins2013bayesian}, which maximizes the minimum interior angle of the mesh triangles to make transitions between small and large triangles as smooth as possible. Please see \cite{mejia2020bayesian} for further details on the construction of the mesh and $\bfPsi$. In the SBSB GLM, additional mesh locations consist of the medial wall, which serves as a supplemental layer to improve estimation along the data boundary. The matrix $\mathbf{Q}_{\kappa,\tau}$ is a sparse precision (inverse covariance) matrix with a fixed set of non-zero elements whose value are determined by $\kappa$ and $\tau$. The smoothness of the latent field is controlled by the $\kappa$ parameter, which controls how much distance dependence there should be in the field. A larger value for $\kappa$ corresponds to a ``smoother" latent field with a larger range of spatial dependence. The precision parameter $\tau$ determines the level of precision in the latent field such that lower values correspond to lower precision values (higher variance) in the prior. The matrix $\mathbf{G}$ is a sparse, symmetric adjacency matrix in which non-zero entries exist only on the diagonal and in cells corresponding to neighboring locations, and $\mathbf{C}$ is a diagonal matrix \citep{bolin2013comparison}. 

The single-run SBSB GLM assumes independent SPDE priors on each of the $K$ latent fields. Let $\mathbf{y} =$  $(\mathbf{y}_1',\ldots,\mathbf{y}_N')'$ $\in \mathbb{R}^{NT \times 1}$, let $\mathbf{X} = \text{block-diagonal}(\mathbf{X}_1,\ldots,\mathbf{X}_N) \in \mathbb{R}^{NT \times NK}$, and let $\boldsymbol{\beta} = (\boldsymbol{\beta}_1',\ldots,\boldsymbol{\beta}_N')' \in \mathbb{R}^{N \times K}$. The surface-based spatial Bayesian GLM is given by
\begin{align}
\begin{split}
    \mathbf{y}|\boldsymbol{\beta} & = \mathbf{X}\boldsymbol{\beta} + \mathbf{e},\quad \mathbf{e} \sim \text{Normal}(\mathbf{0},\sigma^2\mathbf{I}_{NT}) \\
    \boldsymbol{\beta}_k &= \bfPsi \bfw_k, \quad \bfw_k|\kappa_k,\tau_k \sim \text{Normal}(\mathbf{0},\mathbf{Q}_{\kappa_k,\tau_k}^{-1}),\quad k=1,\dots,K \\
    \boldsymbol{\theta} &= (\kappa_1,\tau_1, \ldots, \kappa_K,\tau_K,\sigma^2)\sim \pi(\bftheta),
\end{split}
    \label{eq:spatialBayesmodel} 
\end{align}
where $\bftheta$ contains all the model hyperparameters, and $\pi(\bftheta)$ is their joint prior. We adopt the default priors set by the R-INLA framework: Independent log-Normal distributions are used for all $\kappa_k$ and $\tau_k$, and a Gamma distribution is used for the precision ($\frac{1}{\sigma^2}$). 

\textbf{Multi-run model.} If multiple runs of task data are available from a given subject, it is beneficial to leverage those repeated measures to more accurately estimate the model hyperparameters (e.g. the parameters $\kappa_k$ and $\tau_k$ controlling the spatial properties of the activation amplitude for each task $k$, and the residual variance $\sigma^2$). We therefore propose a multi-run spatial Bayesian GLM to jointly model runs $j = 1,\ldots,J$. Let $\mathbf{y}_j$, $\mathbf{X}_j$, $\boldsymbol{\beta}_j$, $\boldsymbol{\beta}_{j,k}$, and $\bfe_j$ be the run $j$-specific quantities in equation (\ref{eq:spatialBayesmodel}).  The multi-run SBSB GLM can be represented as
\begin{align}
\begin{split}
    \mathbf{y}_j | \boldsymbol{\beta}_j & = \mathbf{X}_j\boldsymbol{\beta}_j + \mathbf{e}_j,,\quad \bfe_j \sim \text{Normal}(\mathbf{0},\sigma^2\mathbf{I}_{NT}),\quad j=1,\dots, J \\
    \boldsymbol{\beta}_{j,k} &= \bfPsi \bfw_{j,k}, \quad \bfw_{j,k}|\kappa_k,\tau_k \sim \text{Normal}(\mathbf{0},\mathbf{Q}_{\kappa_k,\tau_k}^{-1}),\quad k=1,\dots,K \\
    \boldsymbol{\theta} &= (\kappa_1,\tau_1, \ldots, \kappa_K,\tau_K,\sigma^2)\sim \pi(\bftheta),
\end{split}
\end{align}
where we again assume for simplicity nuisance signals have been regressed from both the response and task predictors, and assume that the data has been prewhitened to remove temporal autocorrelation in the model residuals and to eliminate spatial heterogeneity in the residual variance. Note that the run-specific activation amplitudes, $\bfbeta_{j,k}$, are estimated individually, while sharing a common prior determined by the parameters $\kappa_k$ and $\tau_k$. Additionally, the between-run average amplitude can be estimated. These provide more statistically efficient estimates of activation amplitudes if differences across runs are not of interest. These averages are constructed as linear combinations of the run-specific latent fields, so their posterior distribution is available and can be used to identify areas of activation as described in Section \ref{sec:id_activations} below.

\subsubsection{Group-level modeling}
\label{sec:group_model}

Previously proposed spatial Bayesian GLMs for volumetric fMRI data were limited to single-subject analysis, in large part due to the computational burden associated with analyzing data from many subjects concurrently.  \cite{mejia2020bayesian} proposed a computationally efficient ``joint'' group-level modeling approach based on first estimating each subject-level model separately, and then combining the results in a principled way. Here, we generalize this approach to any number of runs, $J\geq 1$. For simplicity of notation, assume that all subjects have the same number of runs $J$.  Let $\boldsymbol{\beta}_{m} \in \mathbb{R}^{KJM}$  represent the activation amplitudes for subject $m$ for all tasks and all runs. The joint group-level modeling approach is based on specifying a group-level contrast  matrix $\bfA$, so that the quantity of interest can be expressed as a linear combination of the subject-level parameter estimates. $\mathbf{A}$ is based on a group contrast vector $\mathbf{a} \in \mathbb{R}^{MJK}$ which specifies the desired contrast across subjects, runs, and tasks. For example, $\mathbf{a}$ can be constructed to represent the group average activation amplitude in response to a particular task, the difference in average amplitude across two groups of subjects or different conditions, or a contrast across tasks. Next, the contrast matrix is created as $\bfA = \mathbf{a}' \otimes \mathbf{I}_N$, where $\otimes$ represents the Kronecker product. The group-level effect is defined as $\boldsymbol{\beta}_{G} = \mathbf{A}\boldsymbol{\beta} \in \mathbb{R}^{N}$, where $\boldsymbol{\beta} = (\boldsymbol{\beta}_{1}',\ldots,\boldsymbol{\beta}_{M}')' \in \mathbb{R}^{NKJM}$ is the concatenated activation amplitudes across all runs, subjects, and tasks. A specific example is given in Section \ref{sec:Estimation} below. See \cite{mejia2020bayesian} for details on the posterior computation of $\boldsymbol{\beta}_G$. Having obtained its posterior distribution, the estimate of $\boldsymbol{\beta}_G$ is given by its posterior mean or other summary metric, and we can identify group-level areas of activation as described in the following section.

Group modeling under the classical GLM was carried out by averaging the estimates of the task coefficients from each subject, i.e. 
\begin{align*}
    \boldsymbol{\beta}_{G,k} = \frac{1}{M} \sum_{i = 1}^M \boldsymbol{\beta}_{i,k},
\end{align*}
where $\boldsymbol{\beta}_{m,k}$ is the length $N$ vector of coefficient estimates for subject $m$ and task $k$. Since all runs across subjects and sessions were of the same length with the same repetition time, this is equivalent to concatenating the data across all runs and performing the group classical GLM following the same steps as in the single-subject classical GLM. Activation was determined using t-tests, as in the single-subject classical GLM model.

\subsubsection{Identifying areas of activation}
\label{sec:id_activations}

In the SBSB GLM, areas of activation are identified based on the joint posterior distribution of activation across all locations using an excursions set approach \citep{bolin2015excursion}, implemented in the \texttt{excursions} R package \citep{bolin2018Rexcursions}. In brief, the excursions method works by determining the probability that a set of vertices all have activation amplitude greater than a set threshold $\gamma$. This probability is based on the joint posterior distribution across vertices, which takes into account spatial dependencies. The largest set with probability at least $1 - \alpha$ is said to be the ``excursions set''.  As the joint posterior distribution is used, this approach avoids massive multiple comparisons and the consequent need for multiplicity correction. As a result, power is increased compared with previously proposed spatial Bayesian models, which used the marginal posterior distribution at each location to identify areas of activation and hence required multiplicity correction \citep{marchini2004comparing}. While this approach does rely on setting a probability level via $\alpha$ and a threshold $\gamma$, so too do other methods (note that $\gamma=0$ is typically implicit in the classical GLM), and researchers need to take proper care choosing and reporting these parameters. 

In the classical GLM, by contrast, areas of activation in response to each task or stimulus are typically identified by performing a $t$-test at every vertex, followed by correction for multiple comparisons.  Traditionally the null hypothesis for a one-sided test is that $\beta_{v,k}\leq 0$, corresponding to $\gamma = 0\%$.  For a general value of $\gamma$, the null hypothesis is simply modified as $\beta_{v,k}\leq \gamma$, with the alternative hypothesis that $\beta_{v,k} > \gamma$.  The test statistic is then computed as $(\hat{\beta} - \gamma)/{SE_{\hat\beta}}$; based on this value, the $p$-value is computed as usual, namely as the upper tail area of the $t$ distribution with $T-K-1$ degrees of freedom. (Note that since the data has been prewhitened, we assume temporally-independent residuals.) Multiplicity correction in the classical GLM typically aims to control the family-wise error rate (FWER) or the false discovery rate (FDR) \citep{benjamini1995controlling}.  Many techniques have been proposed to account for spatial dependence and encourage spatial contiguity at the correction stage, including permutation tests, random field theory \citep{worsley1992three,worsley1996unified}, and threshold-free cluster enhancement \citep{smith2009threshold}.  However, these techniques are all limited by the shortcomings of massive univariate modeling at the model estimation stage, and as a result will have reduced power to detect activations. The correction itself has the effect of further diminishing power to detect activations. Recently, \cite{bowring2019spatial} proposed a confidence set method for group analyses based on Cohen's $d$ statistic, which was expanded upon in \cite{bowring2021confidence}. This method is very promising for large datasets, but is only shown to be accurate in sample sizes as low as 60. As our validation is focused on single-subject and group analyses for a sample size of up to 45 subjects, this method is not used for comparison.

Since in the SBSB GLM spatial dependence is accounted for and leveraged at both estimation and inference, and because multiplicity correction is not required, its power to detect activations tends to be quite high. This can result in large areas of activation with small but non-zero effect size being identified. This is consistent with previous work that has found that when power is high, such as large group studies, the traditional choice of $\gamma = 0\%$ can lead to identification of ``significant'' activations in areas with small effect size that are not of scientific interest \citep{bowring2021confidence}. To avoid this, it is common to specify a scientifically relevant activation threshold, $\gamma$, above which activations are of interest.  For example, if the data are scaled to represent percent signal change, an activation threshold of $\gamma=0.1\%$ to $2\%$ may be reasonable, depending on the magnitude of signal change evoked by a particular task. This does not mean that using a threshold of $\gamma = 0\%$ is never appropriate. However, it is important that researchers are aware of the increase in power using the SBSB GLM and the higher likelihood of subtle activations being labelled as significant. Higher thresholds will generally result in more localized areas of activation.  

For a given value of $\gamma$ and significance level $\alpha$, the areas of activation identified can be said to have activation greater than or equal to $\gamma$ with probability at least $1-\alpha$, based on the joint posterior distribution across all vertices.  That is, there is probability of $\alpha$ or less that at least one vertex in the activated region is a false positive. The excursions set approach therefore controls the FWER at level $\alpha$, but with typically much greater power to detect true activations than in the classical GLM. Both the excursions set approach in the Bayesian GLM and Bonferroni correction in the classical GLM, applied within a single hemisphere, control the probability of a single false positive at $\alpha$. Therefore, by controlling the FWER within each hemisphere, we are effectively controlling the probability of a false positive across both hemispheres at $1 - (1-\alpha)^2$.  If $\alpha=0.01$ as in our analysis, this equals $0.0199\approx 0.02$. Therefore, the whole-brain FWER is controlled at approximately $2\alpha$.

\subsection{Data and Model Estimation}
\label{sec:Data}

\subsubsection{Data Collection}

We perform an extensive reliability study using cortical surface task fMRI data from the Human Connectome Project (HCP) \citep{van2013wu, barch2013function}. To compare the reliability of estimates and areas of task activation produced using the SBSB GLM with the classical GLM, we analyze 180 task fMRI runs from 45 subjects who participated in the Human Connectome Project (HCP) and the HCP Retest Dataset.  The sample of 45 subjects included 31 females, with 4 subjects between the ages of 22 and 25, 14 subjects between the ages of 26 and 30, and 27 subjects between the ages of 31 and 35. Each subject was scanned while performing a motor task \citep{barch2013function}. The study used a 3-second visual cue to alert the subject of the type of motor task that they were expected to complete. Subjects were instructed to tap their fingers (left or right hand), squeeze their toes (left or right foot), or move their tongue for 12 seconds after being prompted by the cue. Each of the five motor tasks was repeated twice during each run. Two runs (acquired with opposing LR and RL phase-encoding directions) were collected at each of two visits, resulting in four runs per subject. 

\subsubsection{Preprocessing and Prewhitening}
\label{sec:preproc_prewhite}

The task fMRI data was preprocessed according to the HCP minimal surface preprocessing pipelines, including projection to the cortical surface and registration to a common surface template  \citep{glasser2013minimal}. These pipelines also include generation of a subject-specific high-resolution 164k native surface mesh based on the high-resolution T1-weighted and T2-weighted structural images for each subject, registration to the \textit{fsaverage} mesh, and resampling to a lower-resolution 32k mesh to approximately match the original fMRI voxel resolution. For spatial modeling we utilize the \textit{midthickness} surface, which represents the midpoint of the cortical ribbon between the white matter and pial surfaces. As part of the HCP minimal preprocessing pipelines, the fMRI timeseries were slightly smoothed along the midthickness surface to regularize the mapping process using a $2$mm full-width half-maximum (FWHM) Gaussian kernel with the GEO\_GAUSS\_AREA smoothing method implemented within the Connectome Workbench software \footnote{https://www.humanconnectome.org/software/connectome-workbench} \footnote{https://www.humanconnectome.org/software/workbench-command/-cifti-smoothing}. 

Prior to model estimation, several additional processing steps are performed: smoothing (for the classical GLM only), resampling, centering and scaling, nuisance regression, and prewhitening.  Each surface is then resampled from approximately 32,000 vertices to approximately 5,000 vertices per hemisphere using barycentric interpolation, which minimizes blurring \citep{glasser2013minimal}. This greatly improves computational efficiency for the SBSB GLM, as well as for the vertex-wise prewhitening employed in both the SBSB and classical GLMs, which we describe below.   After resampling, vertex size remains small relative to the size of expected activations.  See \textbf{Appendix \ref{app:resampling_smoothing}} for details on the effects of resampling and smoothing.

In order to fairly compare methods, the cs-fMRI data used in the classical GLM analyses are first smoothed via the Connectome Workbench \citep{glasser2013minimal} using a Gaussian kernel with a full-width half-maximum (FWHM) of 6mm, as this is commonly done in practice to increase statistical power. We adopt 6mm FWHM for two reasons.  First, this is a commonly used smoothing level in practice. Second, we compare a range of smoothing levels and find 6mm FWHM to be near-optimal for all tasks in terms of test-retest reliability, as illustrated in \textbf{Appendix Figure \ref{fig:smooth_cor_plot}}. However, it is important to note that traditional data smoothing imposes the same degree of smoothing across all tasks, unlike the spatial Bayesian GLM which estimates the inherent smoothness of each task activation field separately. In contexts where some task activation fields are much smoother than others (e.g., visual cue versus hand movement), this represents an advantage of spatial Bayesian modeling.

The BOLD timeseries at each vertex and each column of the design matrix is centered prior to model fitting. This eliminates the need for a baseline field, since the intercept of a linear model is zero when both each predictor and the response have mean zero. The BOLD timeseries at each vertex is also simultaneously scaled relative to the local average BOLD signal, which introduces units of percent signal change. The task design matrix is created by convolving the stimulus boxcar function with a canonical double-Gamma HRF \citep{friston1998event, glover1999deconvolution}. The design is then scaled by dividing the task covariates by their maximum over time, and then centered again. Twelve motion covariates (six rigid body realignment parameters and their first derivatives), along with linear and quadratic drift terms, are regressed from the BOLD data and task design matrix. Simultaneously with nuisance regression, the temporal derivative of each task design column is also regressed from the data and design matrix, in order to account for differences in the onset of the hemodynamic response across subjects, runs, tasks and areas of the brain.

Prewhitening is performed to satisfy the GLM assumption of residual independence. Prewhitening at vertex $v$ consists of estimating the residual covariance matrix $\bfSigma_v$, then pre-multiplying the BOLD data and design matrix by $\bfSigma_v^{-\frac{1}{2}}$. This induces a residual vector that is temporally independent and residual variance that is constant across all vertices. To estimate $\bfSigma_v$, we first estimate the residuals at vertex $v$ using the classical GLM after performing the processing steps described above.  We elect to use a high-order, spatially-varying autoregressive (AR) process to model the residual autocorrelation, since both are observed to be necessary based on exploratory analysis of the residuals (see \textbf{Appendix Figure \ref{fig:prewhitening}}). Specifically, an AR(6) model is fit to the each residual timeseries using the Yule-Walker equations \citep{brockwell2016introduction}. To regularize the estimates, the estimated AR coefficients and white noise variance are averaged over runs in the multi-run model for each subject and visit, and surface-smoothed using a Gaussian kernel with a FWHM of $6$mm (see \textbf{Appendix Figure \ref{fig:prewhitening}}). The resulting AR coefficient and white noise variance estimates at each vertex $v$ are used to compute $\bfSigma_v^{-\frac{1}{2}}$. This vertex-wise procedure results in a unique design matrix at each vertex, as in equation (\ref{eq:basicGLMlocv}).

\subsubsection{Model Estimation}
\label{sec:Estimation}

For each subject and visit, estimates and areas of activation are produced using the multi-run SBSB GLM described in Section \ref{sec:single_subject_model}. For comparison, the single-run SBSB GLM results are also obtained based on the LR runs. The two visits are analyzed independently to assess the test-retest reliability of the estimates and areas of activation. Since the surface meshes representing the left and right hemispheres do not intersect, each hemisphere is estimated separately. The preprocessing performed includes centering, scaling, and prewhitening at each vertex, which eliminates spatial dependence in the noise, both within and across hemispheres. For both the classical and Bayesian GLMs, lateralized tasks (e.g. left foot, right hand) are excluded in the model for the ipsilateral hemisphere, since minimal ipsilateral activation is expected during lateralized motor tasks. \cite{mejia2020bayesian} found that this approach is computationally advantageous for Bayesian modeling, while having negligible impact on model results. Thus, multiple testing corrections for the classical GLM are done within hemisphere, as analyses for each hemisphere were carried out separately to exclude the ipsilateral tasks, and this provides for the closest comparison in activation detection between the two methods. 

For each subject- and visit-specific model, we identify areas of activation using a significance level of $\alpha = 0.01$ at three different activation thresholds, $\gamma = (0\%,0.5\%, 1\%)$ using the excursions set approach described in Section \ref{sec:id_activations}. Using a range of activation thresholds allows identification of areas that exhibit even subtle activation separately from those that exhibit high levels of activation in response to each task.  Subject-level activations are identified for each run and for the average across runs for each visit.  

For the classical GLM, we identify areas of activation by performing a $t$-test at each vertex, followed by Bonferroni correction to control the FWER at $\alpha=0.01$. Correction is performed within each hemisphere to provide analogous false positive control to the Bayesian GLM. While Bonferroni correction is often considered overly conservative in a volumetric or full-resolution surface analysis, note that here the correction is only performed across approximately 5,000 resampled vertices within each hemisphere, and will therefore be much less so.  As illustrated in \textbf{Appendix Figure \ref{fig:determining_activations}} , we find that in our context Bonferroni correction produces similar results to nonparametric permutation testing \citep{nichols2002nonparametric}, with Bonferroni correction being slightly less conservative. \cite{lindquist2015zen} pointed out that it is generally accepted to adopt the less conservative of the two approaches, as they both control the FWER. Further, while more activations can be identified in the classical GLM by controlling the FDR instead of the FWER, FWER correction provides similar false positive rate guarantees as the excursions set approach adopted in the SBSB GLM, as described in Section \ref{sec:id_activations}.  For these two reasons, we adopt Bonferroni correction for comparison with the classical GLM. See \textbf{Appendix \ref{appendix:multiplecomparisons}} for further details on multiple comparisons for the classical GLM. Though traditionally the classical GLM implicitly assumes an activation threshold of $0\%$, corresponding to the traditional hypothesis testing approach, here we test all three activation thresholds ($\gamma = 0\%,0.5\%, 1\%$) to provide an apples-to-apples comparison with the areas of activation produced via the Bayesian GLM.

We apply the joint group-level modeling approach described in Section \ref{sec:group_model} to obtain estimates of group-average activation amplitude across all $45$ subjects. Each visit is analyzed independently to facilitate reliability analysis. We also assess the impact of sample size on reliability of group-level estimates and areas of activation, since smaller sample sizes are relatively common in fMRI studies.  To this effect, the group-level modeling is repeated on random subsets of 10, 20, and 30 subjects; for each sample size, ten different random samples of subjects are generated and analyzed.  Areas of activation are identified as in the single-subject case for both the classical and Bayesian GLMs. The average estimates across two sessions in the group model are found using a contrast matrix. Consider, for example, that the tongue task is the fourth task ($k=4$) in four tasks ($K=4$), and we study the average across $J=2$ sessions across all $M=45$ subjects. Following the construction of the group contrasts as specified in section \ref{sec:group_model}, the contrast matrix used is written as $\bfA = \mathbf{a}' \otimes \mathbf{I}_n$, where $\mathbf{a} = (\mathbf{a}_{1}',\mathbf{a}_{2}',\ldots,\mathbf{a}_{45}')'$, and $\mathbf{a}_{m} = (0,0,0,\frac{1}{2\times 45},0,0,0,\frac{1}{2 \times 45})'$. Contrasts for the average of other tasks is created following the same pattern, i.e. the contrast matrix for the average effect of the visual cue, the first task in the design matrix, is created as $\bfA = \mathbf{a}' \otimes \mathbf{I}_n$, where $\mathbf{a} = (\mathbf{a}_{1}',\mathbf{a}_{2}',\ldots,\mathbf{a}_{45}')'$, and $\mathbf{a}_{m} = (\frac{1}{2 \times 45},0,0,0,\frac{1}{2 \times 45},0,0,0)'$. The contrasts to find the averages across all subjects and runs for each task are created for by default for group modeling in the \texttt{BayesfMRI} software package.

We fit the subject- and group-level Bayesian and classical GLMs using the R package \texttt{BayesfMRI} (version 1.8.1) running on 6 parallel threads on a Mac Pro with a 2.7 GHz 24-Core Intel Xeon W processor and 512 GB of memory.  The \texttt{BayesfMRI} package is openly available via Github\footnote{\url{https://github.com/mandymejia/BayesfMRI/tree/1.8.1}} and performs model fitting using the \texttt{R-INLA} package \citep{lindgren2015RINLA} with the \texttt{PARDISO} sparse matrix library \citep{pardiso-7.2a,pardiso-7.2b,pardiso-7.2c}.  In sum, we fit 180 subject-level models (45 subjects, 2 visits, 2 hemispheres) and 124 group-level models ($N = 10, 20, 30, 45$, on 10 different subsamples for $N = 10, 20, 30$, for each of 2 visits, on 2 hemispheres). Note that the group analysis requires the output from the single-subject analyses, so the time shown to complete a group analysis is in addition to the requisite single-subject analyses.  Table \ref{tab:comptime} shows the mean and standard deviations of the amount of time in minutes taken to perform model analyses on both hemispheres for a given visit. 

\begin{table}[H]
    \centering
    \begin{tabular}{|c|c|c|c|c|c|c|}
        \hline
        \multicolumn{3}{|c|}{Single-subject Analyses} & \multicolumn{4}{c|}{Bayesian Group Analyses} \\
         \multicolumn{1}{|c}{Preprocessing} & \multicolumn{1}{c}{Bayesian}  & \multicolumn{1}{c|}{Classical}  & \multicolumn{1}{c}{$n = 10$} & \multicolumn{1}{c}{$n = 20$} & \multicolumn{1}{c}{$n = 30$} & \multicolumn{1}{c|}{$n = 45$} \\
         \hline
         3.93  & 12.50  & 0.14  & 84.79 & 175.8 & 230.2 & 335.87  \\
         (0.03) & (3.94) & (0.003) & (8.9) & (76.82) & (63.19) & (1.18) \\
         \hline
    \end{tabular}
    \caption{The mean (standard deviation) of the computing times in minutes for single-subject and group-level analyses. Each time reported corresponds to the time taken to analyze both hemispheres of the brain. Note that the time shown to complete a group analysis is in addition to the time required for the requisite single-subject analyses. The small standard deviation in the times to analyze the 45-subject groups is a result of only performing three separate full-sample group analyses.}
    \label{tab:comptime}
\end{table}

\subsection{Reliability analysis}
\label{sec:Reliability}

The SBSB GLM leverages spatial dependence and sparsity to produce estimates and areas of activation that should, in theory, more accurately reflect the true underlying patterns of activation. Here, we assess the ability of the SBSB GLM to deliver on that promise. We assess the extent to which the SBSB GLM produces estimates and areas of activation that reflect the unique activation features of individual subjects. To this end, we utilize the repeated visits available for each subject, which are analyzed independently as described in Section \ref{sec:Estimation}.  We use three types of metric to quantify reliability of subject-level measures of task activation: (1) intraclass correlation coefficient of estimates of activation amplitude, which quantifies the proportion of variability on the estimates attributable to unique and reliable subject-level features, (2) similarity of estimates to an unbiased ground truth proxy, and (3) test-retest overlap of areas of activation.  At the group-level, we also use the similarity of estimates to an unbiased ground truth proxy to assess reliability.  Finally, we assess power based on the size of areas of activation.

\subsubsection{Reliability of Amplitude Estimates}
\label{subsec:ICC}

To quantify reliability of subject-level amplitude estimates, we compute the intraclass correlation coefficient (ICC) \citep{bartko1966intraclass} at each vertex $v$ for each task, based on the estimates from all $45$ subjects. For each subject, the two separate visits serve as repeated measures. The ICC is equal to $\text{ICC} = {\sigma_b^2}/{\sigma_t^2}$, where $\sigma_b^2$ is the between-subject (signal) variance and $\sigma_t^2 \geq \sigma_b^2$ is the total variance, equal to the sum of the between-subject variance and the within-subject (noise) variance $\sigma_w^2$.  If a set of estimates are structured as $\mathbf{B} \in \mathbb{R}^{M \times 2}$, where $M$ is the number of subjects and the two columns correspond to repeated measurements, the variance components can be computed as 
\begin{align}
    \sigma_t^2 = \frac{1}{2} \left( \text{var}(\mathbf{B}_{\bullet,1}) + \text{var}(\mathbf{B}_{\bullet,2}) \right), &&
    \sigma_w^2 = \frac{1}{2} \text{var}(\mathbf{B}_{\bullet, 1} - \mathbf{B}_{\bullet,2}), &&
    \sigma_b^2 = \sigma_t^2 - \sigma_w^2, \label{eq:ICC}
\end{align}
where $\mathbf{B}_{\bullet,j}$ indicates the set of estimates from all subjects for measurement $j$ and $\text{var}(\mathbf{x}) = \frac{1}{n} \sum_{i=1}^n (x_i - \bar{x})^2$.

The interpretation of ICC is straightforward: a value of 1 happens when $\sigma_t^2=\sigma_b^2$, which indicates that there is no noise present in the amplitude estimates for a given subject; a value of 0 happens when $\sigma_b^2=0$, which indicates that there are no true differences between subjects, and all observed differences in a set of estimates are due to random noise. Note that negative ICC values are computationally possible given the estimation of $\sigma^2_b$ as a difference, especially when the true ICC is close to zero.  In activation amplitudes, this occurs most commonly outside of the areas of activation for a given task, where all subjects have essentially zero activation.  Since ICC truly ranges from $0$ to $1$, we truncate any negative values to zero. 

To assess the accuracy of group-level amplitude estimates, note that we cannot use the ICC to quantify the reliability of the group activation estimates because we only observe a single group. Instead, we use the visit 2 classical GLM estimates from models fit on unsmoothed cs-fMRI data as an unbiased proxy for the ground truth. Note that the classical GLM is used as the ground truth proxy for both the Bayesian and classical GLMs, providing two benefits: it is unbiased (though noisy), and it is common to both GLMs, avoiding any bias in favor of the Bayesian GLM or due to smoothing.  To quantify the similarity of estimates to this ground truth, we compute the mean squared error (MSE) and Pearson correlation for the visit 1 estimates, relative to this reference.  Lower MSE and higher Pearson correlation indicate better accuracy.

\subsubsection{Reliability of Areas of Activation}
\label{subsec:area_reliability}

To quantify the reliability of areas of activation produced from the Bayesian and classical GLMs, we utilize the Dice overlap coefficient \citep{dice1945measures}. Dice of two binary maps $A$ and $B$ is given by the number of overlapping locations across the maps, divided by the average number of locations in each map. 
$$
Dice(A,B)=\frac{2|A\cap B|}{|A|+|B|}
$$
For both the Bayesian and classical GLMs, we compute the Dice coefficient of test-retest overlap across visits for each subject at each activation threshold. We compute the test-retest overlap of both the run-specific areas of activation, as well as for the cross-run average areas of activation. 

\section{Results}
\label{sec:Results}

In this section, we examine the reliability of subject-level and group average estimates and areas of activation. Using the classical GLM as a benchmark, we provide visual illustrations and summary measures of the gain in reliability attained using the Bayesian GLM.  We also examine the power of both the classical and Bayesian GLM to identify areas of activation in individual subjects and at the group level. Results shown are based on the multi-run analyses of the visit 1 data, combining across the LR and RL runs. For brevity, only images of the tongue task are displayed. Corresponding figures for the remaining tasks and for single-run analysis using only the LR run are shown in \textbf{Appendix \ref{app:additional_results_figures}} and show similar patterns.

\subsection{Subject-level estimates of activation amplitude}
\label{sec:subj_estimate_results}

\begin{figure}
\centering
\begin{tabularx}{\textwidth}{c|X|X|X|}
    \multicolumn{1}{c}{} & 
    \multicolumn{1}{c}{\textbf{Classical GLM}} &
    \multicolumn{1}{c}{\textbf{Classical GLM}} &
    \multicolumn{1}{c}{\textbf{Bayesian GLM}}\\
	\multicolumn{1}{c}{} & 
	\multicolumn{1}{c}{\textbf{(unsmoothed)}} &
	\multicolumn{1}{c}{\textbf{(FWHM = 6mm)}} &
	\multicolumn{1}{c}{\textbf{(unsmoothed)}}  \\ \cline{2-4}
	\rotatebox[origin=c]{90}{\textbf{Subject A}\,} &
	\Includegraphics[width=0.3\textwidth, trim=0 125mm 170mm 0, clip]{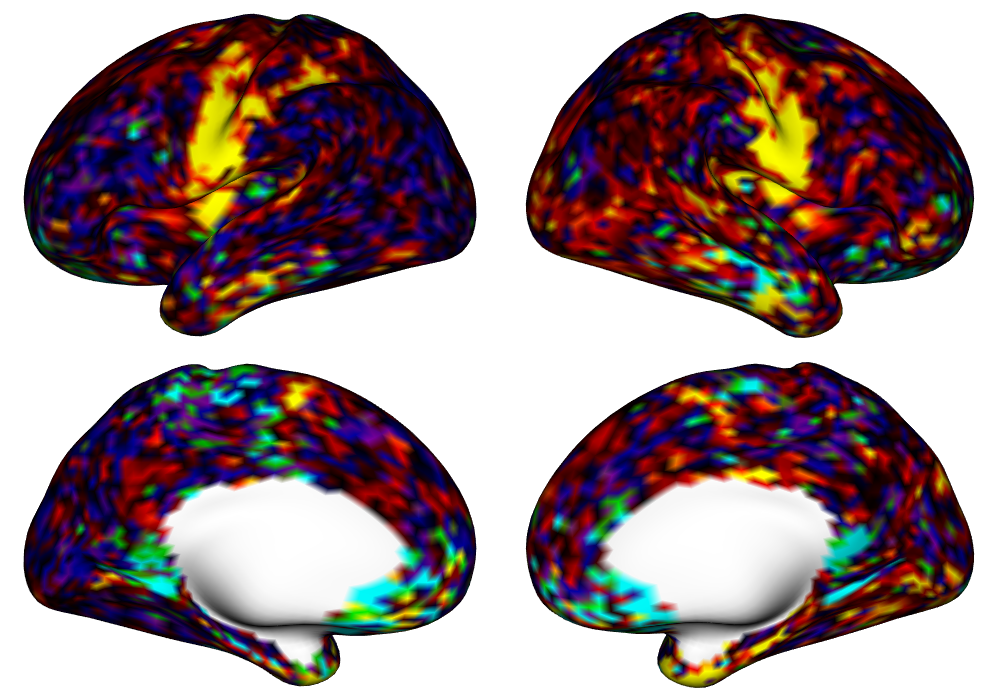} &
	\Includegraphics[width=0.3\textwidth, trim=0 125mm 170mm 0, clip]{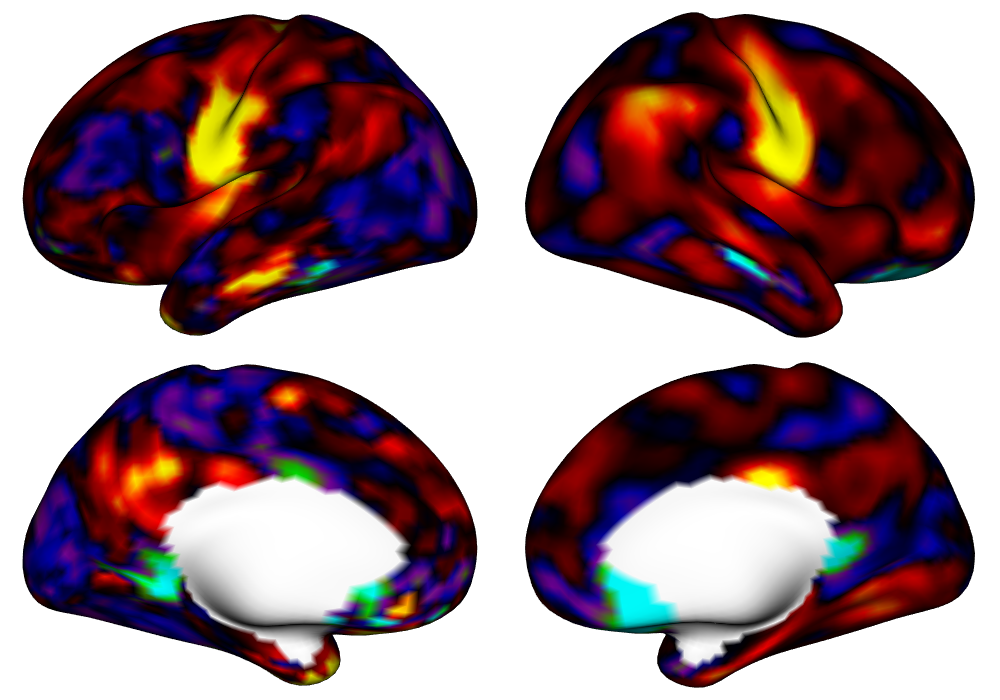} &
	\Includegraphics[width=0.3\textwidth, trim=0 125mm 170mm 0, clip]{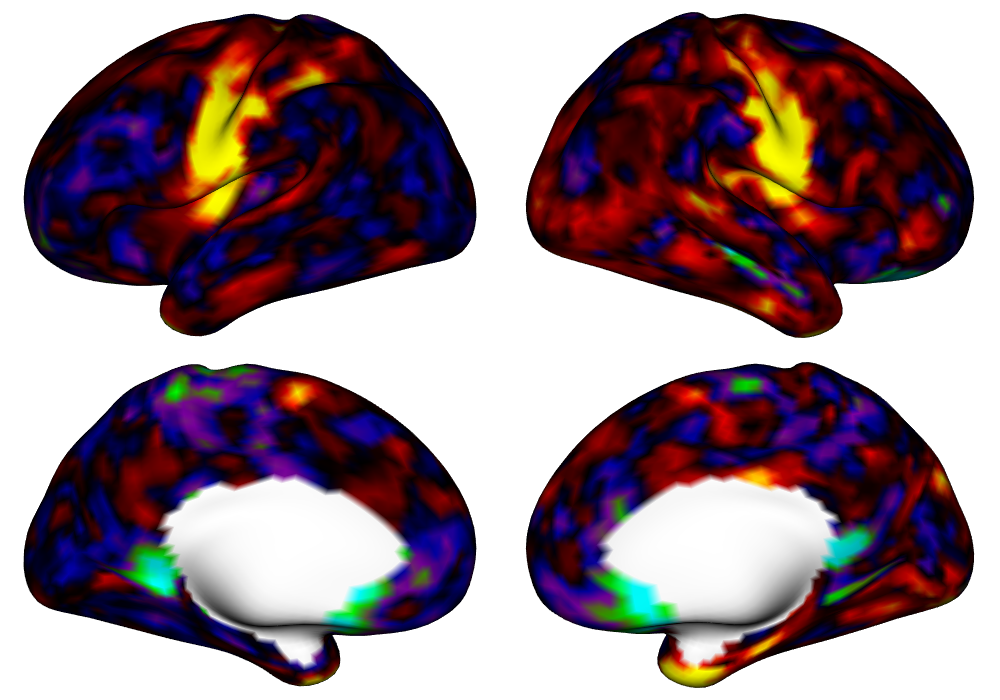} \\ \cline{2-4}
	\rotatebox[origin=c]{90}{\textbf{Subject B}} &
	\Includegraphics[width=0.3\textwidth, trim=0 125mm 170mm 0, clip]{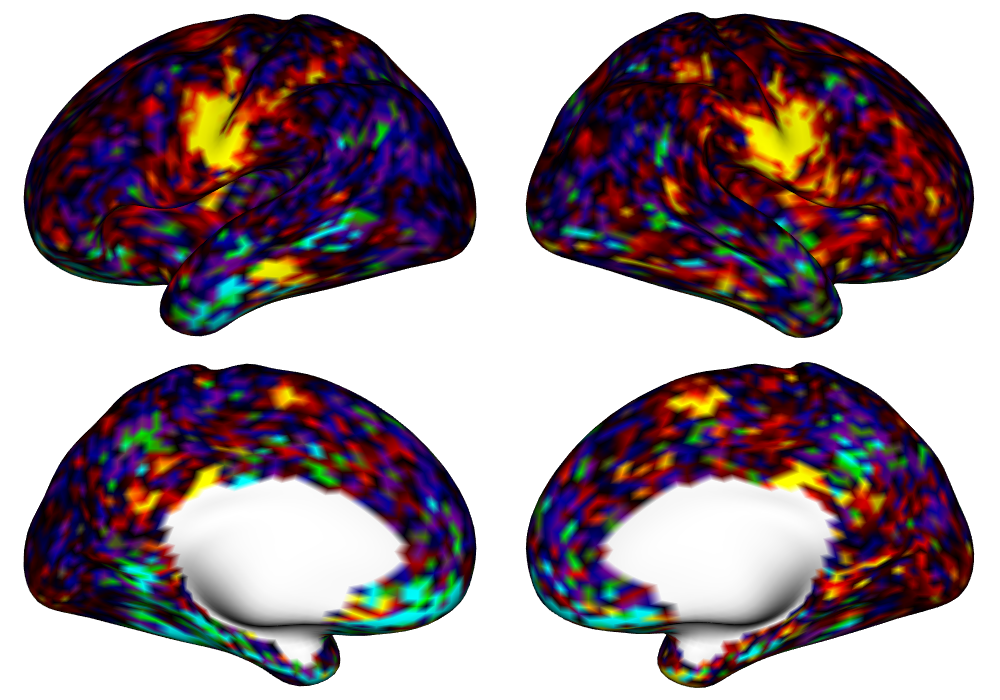} &
	\Includegraphics[width=0.3\textwidth, trim=0 125mm 170mm 0, clip]{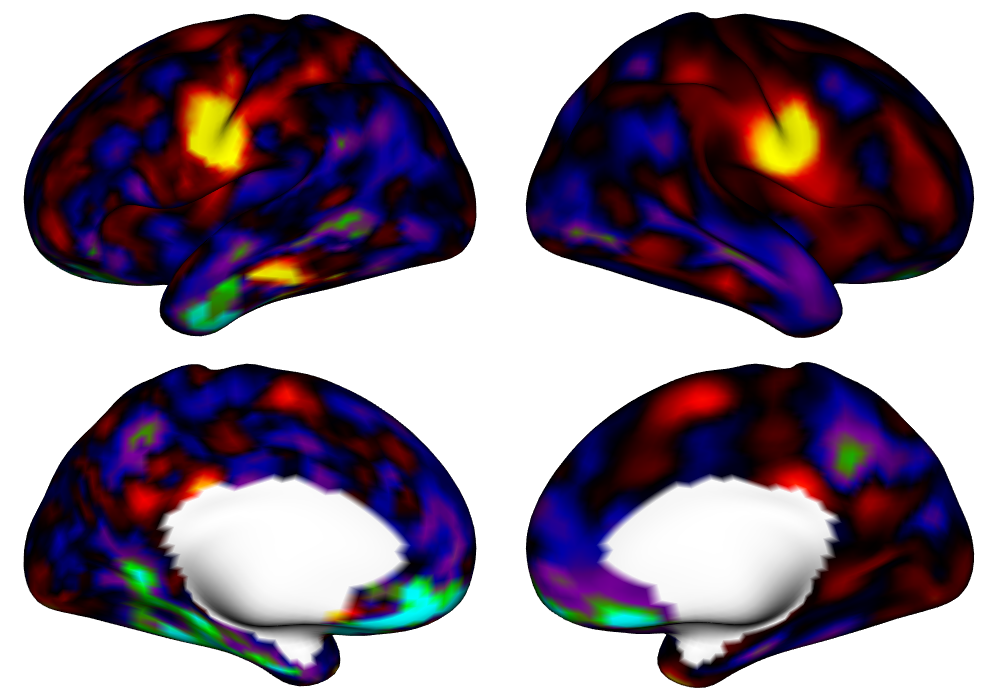} &
	\Includegraphics[width=0.3\textwidth, trim=0 125mm 170mm 0, clip]{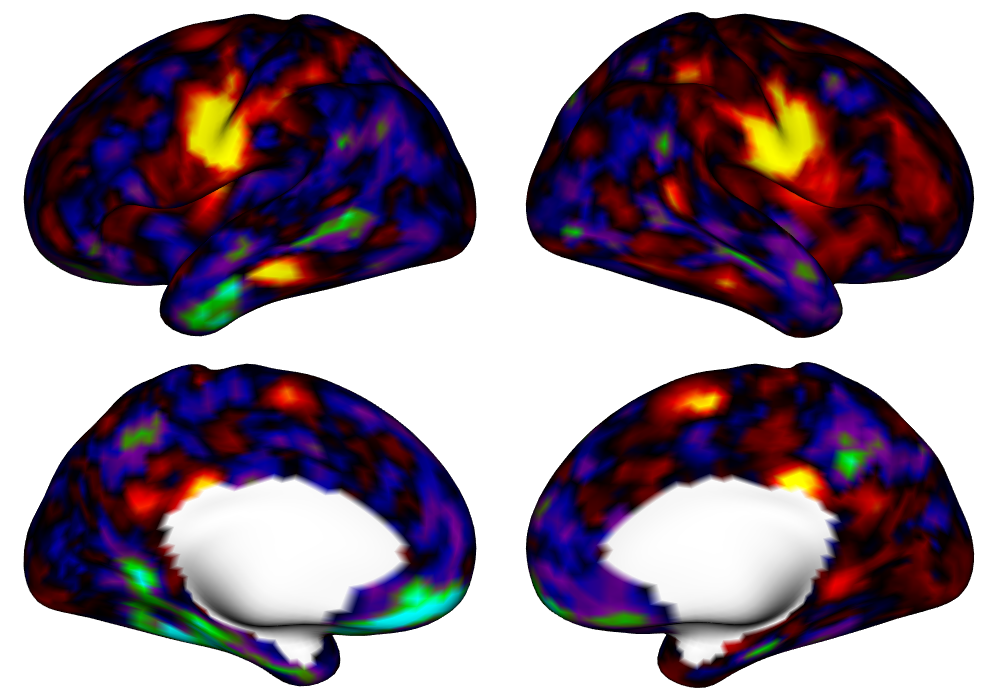} \\ \cline{2-4}
	\rotatebox[origin=c]{90}{\textbf{Subject C}} &
	\Includegraphics[width=0.3\textwidth, trim=0 125mm 170mm 0, clip]{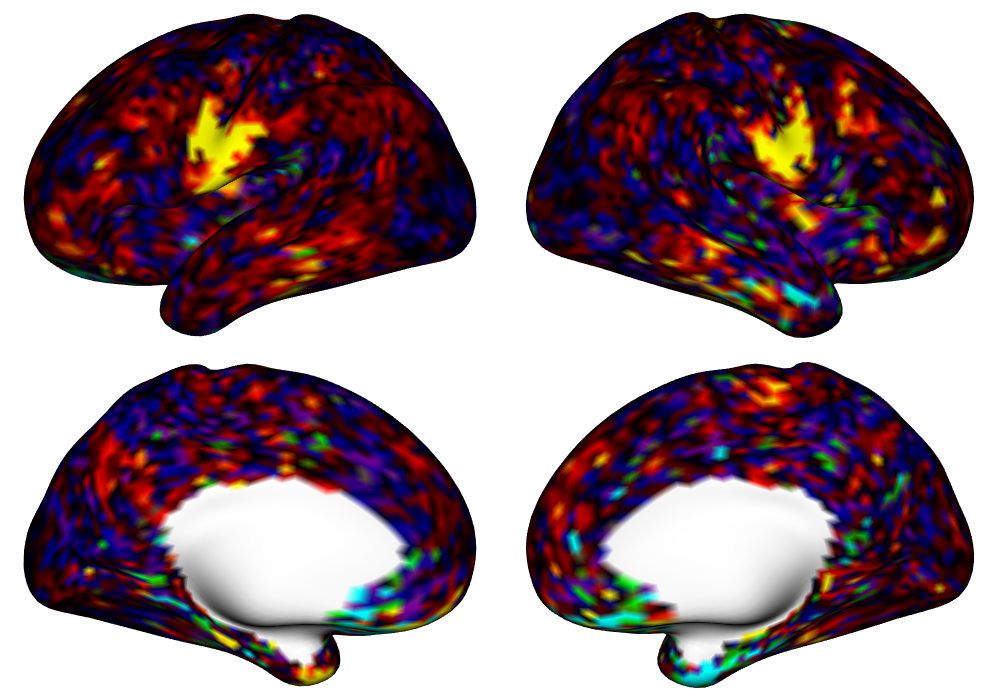} &
	\Includegraphics[width=0.3\textwidth, trim=0 125mm 170mm 0, clip]{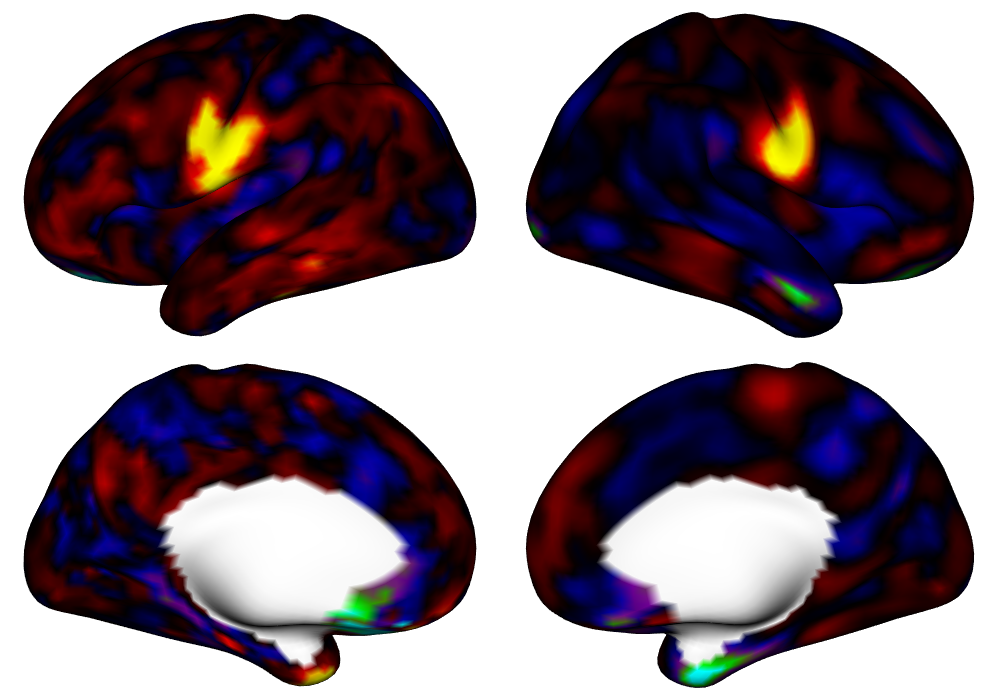} &
	\Includegraphics[width=0.3\textwidth, trim=0 125mm 170mm 0, clip]{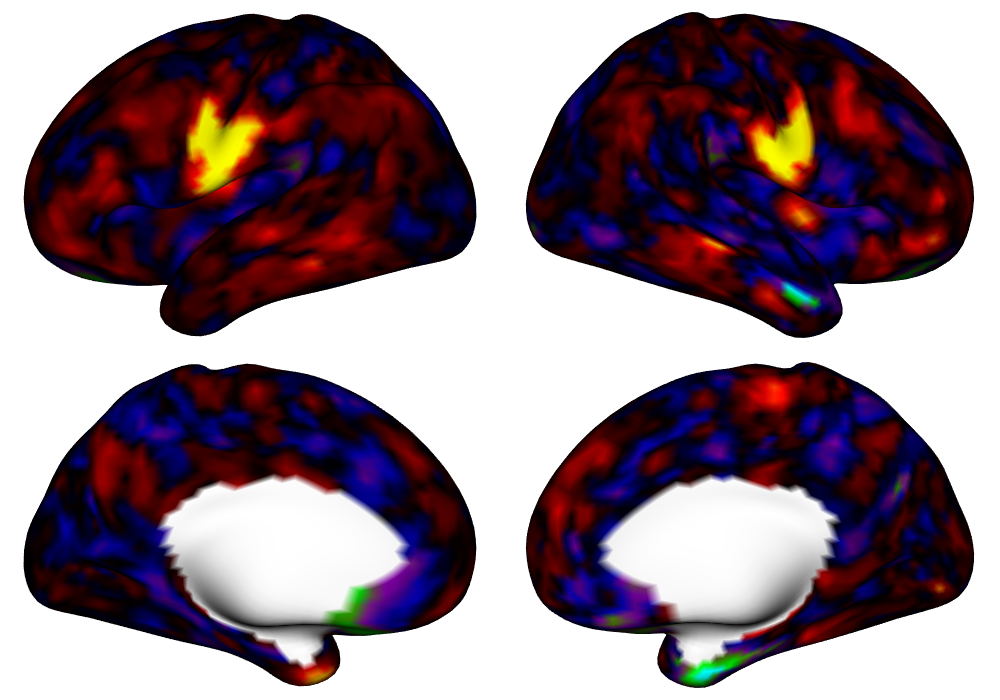} \\ \cline{2-4}
\end{tabularx}
\Includegraphics[width = 0.3\textwidth]{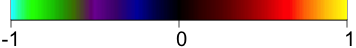}\hspace{-7mm}
\caption{\textbf{Subject-level estimates of activation amplitude in response to tongue movement, in units of percent signal change.} For the sake of space, only the lateral view of the left hemisphere is displayed. The Bayesian estimates are visually similar to those of the classical GLM when smoothing is applied in the latter case.  However, the Bayesian GLM appears to better preserve some subject-specific features, particularly in Subject A.}
\label{fig:tongue_est_subject}
\end{figure}

For three example subjects, \textbf{Figure \ref{fig:tongue_est_subject}} displays Bayesian and classical GLM estimates of activation amplitude for the tongue movement task. The Bayesian GLM and the classical GLM based on smoothed data produces estimates that are visually similar, due to the smoothing effect of the spatial prior in the Bayesian model. (Recall that the data analyzed in the Bayesian GLM are not smoothed.) However, some subject-specific activation features appear to be better preserved in the Bayesian GLM. These are particularly noticeable for Subject A, who exhibits a larger area of intense activation compared with the other two subjects.  

\begin{figure}[H]
    \centering
    \includegraphics[width=\textwidth]{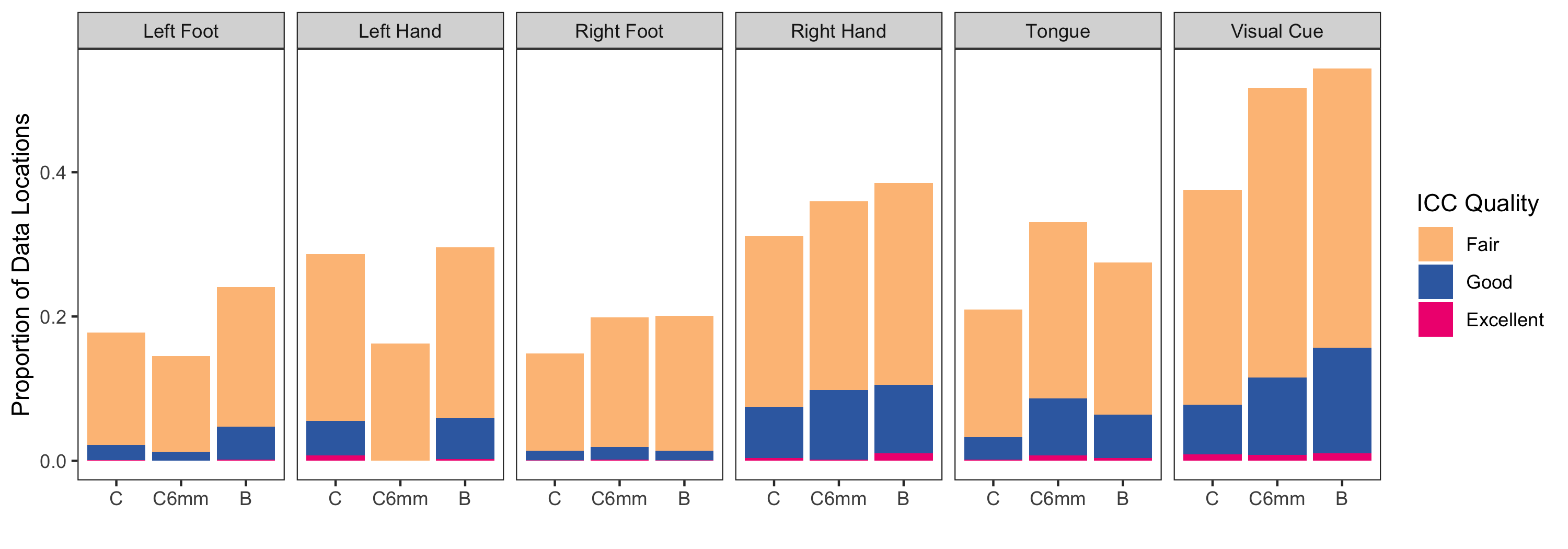}
    \caption{\textbf{Test-retest reliability of subject-level estimates of activation amplitude, in terms of ICC.} Each bar shows the proportion of vertices with estimates significantly greater than zero in the group analysis of the Bayesian result with ``fair'' (0.4 to 0.6), ``good'' (0.6 to 0.75) and ``excellent'' (over 0.75) ICC values, based on the independent estimates of activation from each visit.  Three models are compared: the classical GLM based on unsmoothed data (C), the classical GLM based on smoothed data (C6mm), and the Bayesian GLM (B). Recall that the Bayesian GLM is applied to unsmoothed data but implicitly smoothes each task activation field. Interestingly, data smoothing prior to applying the classical GLM has an inconsistent effect on reliability: smoothing improves reliability for the tongue and visual cue, worsens reliability for the left lateral tasks, and results in little change for the right lateral tasks. By contrast, the Bayesian GLM uniformly improves reliability compared with the classical GLM without smoothing. Comparing the classical GLM using smoothed data (C6mm) to the Bayesian GLM (B), reliability is higher based on the Bayesian GLM (B) for all tasks except the tongue task. These results suggest that the Bayesian GLM improves reliability of subject-level activation patterns while avoiding oversmoothing.\\[10pt]}
    \label{fig:icc_quality}
\end{figure}

We quantify the test-retest reliability of the estimates of activation amplitude via the ICC, as described in section \ref{subsec:ICC}. The repeated measurements are the estimates from the two different visits for each subject ($j = \{1,2\}$ in equation (\ref{eq:ICC})). Commonly-used ICC quality thresholds were established by \cite{cicchetti1994guidelines}: ICC below 0.4 is considered ``poor'', ICC between 0.4 and 0.6 is considered ``fair'', ICC between 0.6 and 0.75 is considered ``good'', and ICC over 0.75 is considered ``excellent''. In \textbf{Figure \ref{fig:icc_quality}}, we summarize the ICC of each image based on the proportion of vertices where fair, good and excellent ICC is achieved. Interestingly, smoothing the data prior to applying the classical GLM has a mixed effect on reliability: smoothing clearly improves reliability for the tongue and visual cue, clearly worsens reliability for the left lateral tasks, and results in little change for the right lateral tasks. By contrast, the Bayesian GLM uniformly improves reliability for all tasks compared with the classical GLM without smoothing. This illustrates that data smoothing may result in oversmoothing for some tasks and undersmoothing for other tasks, while the Bayesian GLM estimates the underlying smoothness of each task activation field and implicitly smoothes the estimates to the appropriate degree.  Comparing the Bayesian GLM to the classical GLM based on smoothed data, reliability is higher based on the Bayesian GLM for all tasks except for the tongue task.  These results suggest that the Bayesian GLM generally produces more reliable estimates of activation in individual subjects and better preserves unique features of individual subjects, compared with the classical GLM based on smoothed data.

\begin{figure}[H]
\centering
\begin{tabularx}{\textwidth}{c|X|X|X|}
    \multicolumn{1}{c}{} &
    \multicolumn{1}{c}{\textbf{Classical GLM}} &
    \multicolumn{1}{c}{\textbf{Classical GLM}} & 
    \multicolumn{1}{c}{\textbf{Bayesian GLM}} \\
	\multicolumn{1}{c}{} & 
	\multicolumn{1}{c}{\textbf{(unsmoothed)}} &
	\multicolumn{1}{c}{\textbf{(FWHM = 6mm)}} &
	\multicolumn{1}{c}{\textbf{(unsmoothed)}}  \\ 
	\cline{2-4}
	\rotatebox[origin=c]{90}{\textbf{Subject A}} & 
	\Includegraphics[width=0.3\textwidth]{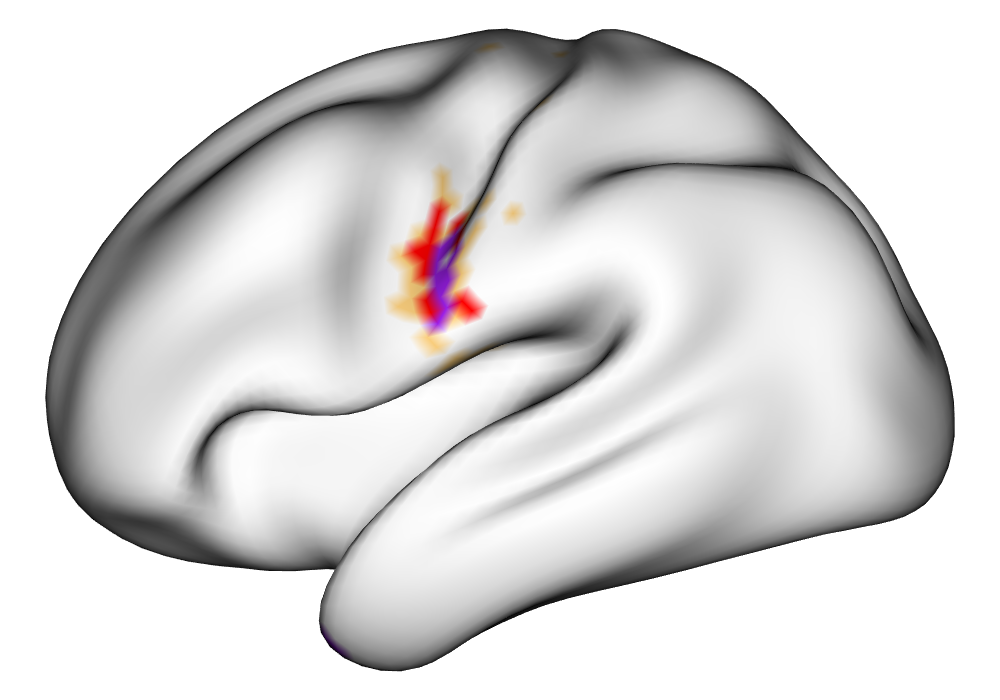} &
	\Includegraphics[width=0.3\textwidth, trim=0 125mm 170mm 0, clip]{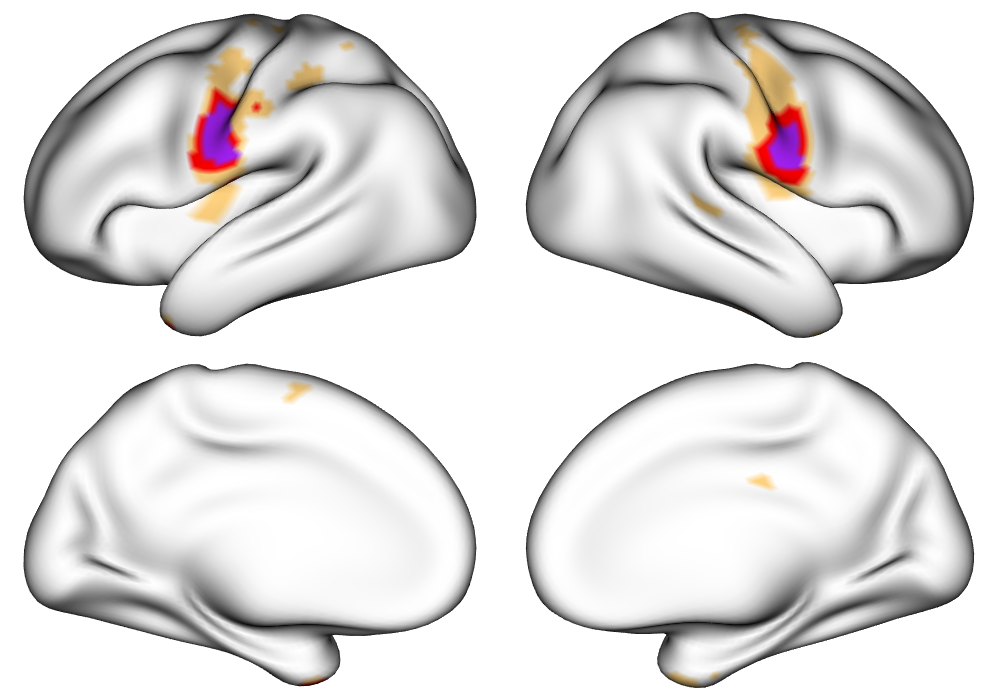} &
	\Includegraphics[width=0.3\textwidth, trim=0 125mm 170mm 0, clip]{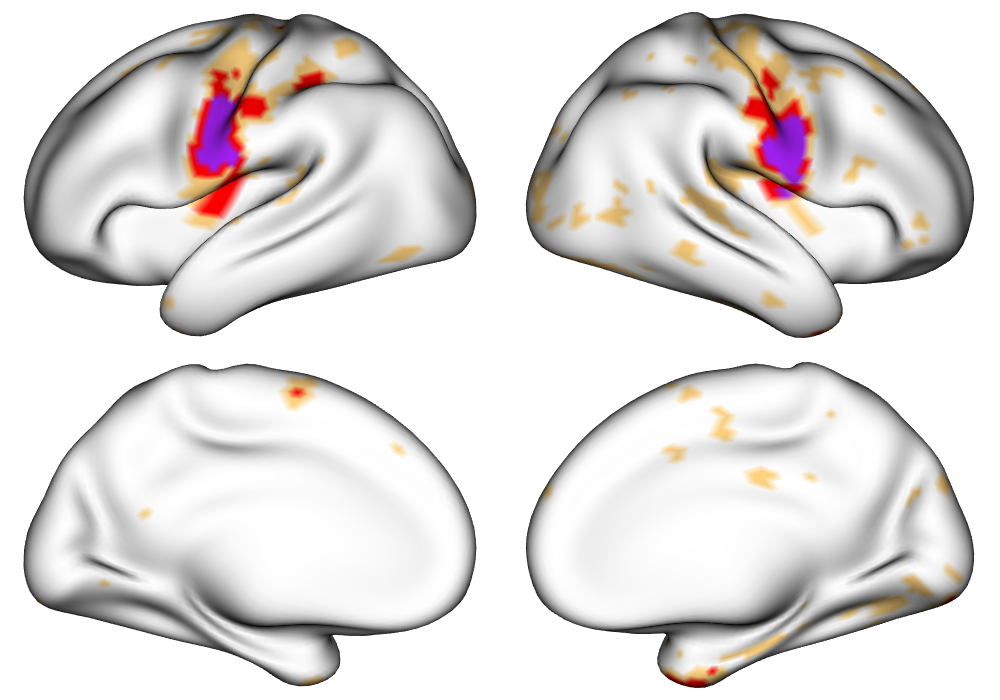} \\ 
	\cline{2-4}
	\rotatebox[origin=c]{90}{\textbf{Subject B}} & 
	\Includegraphics[width=0.3\textwidth]{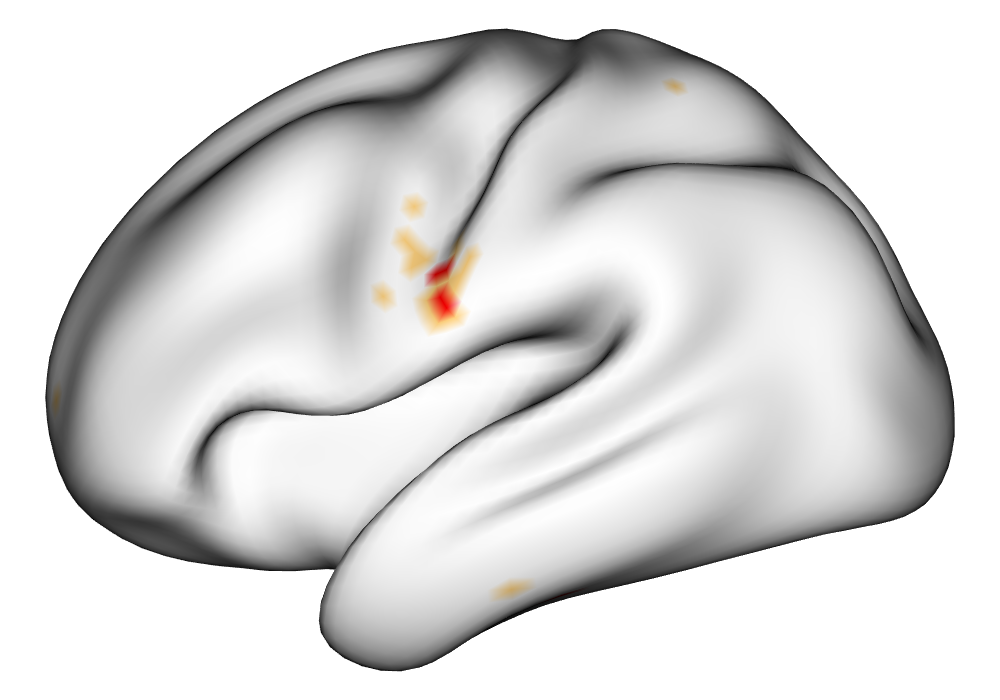} &
	\Includegraphics[width=0.3\textwidth, trim=0 125mm 170mm 0, clip]{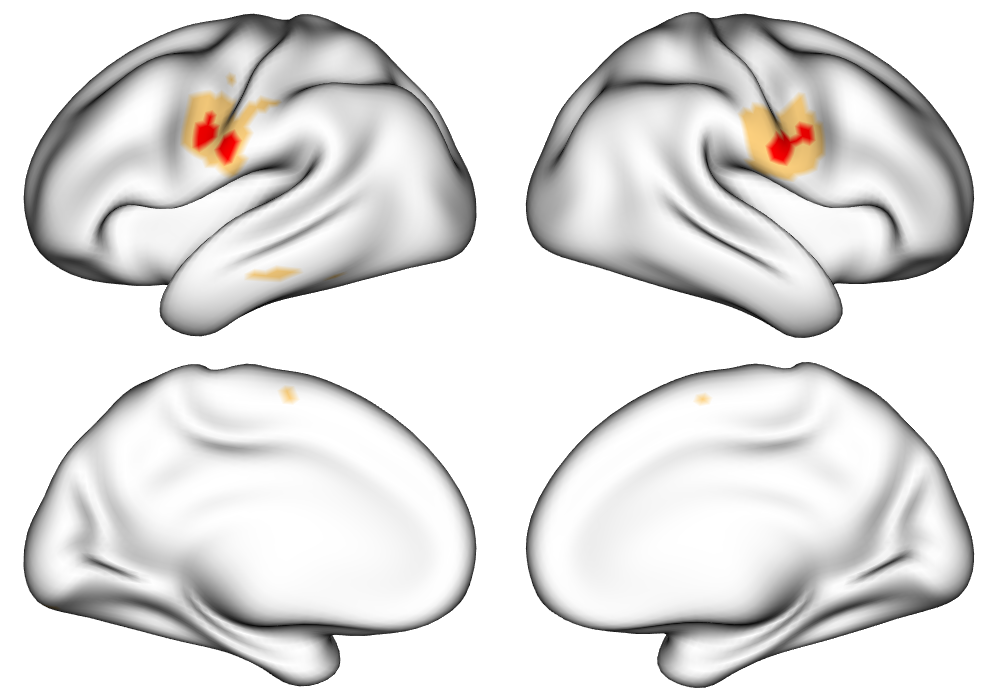} &
	\Includegraphics[width=0.3\textwidth, trim=0 125mm 170mm 0, clip]{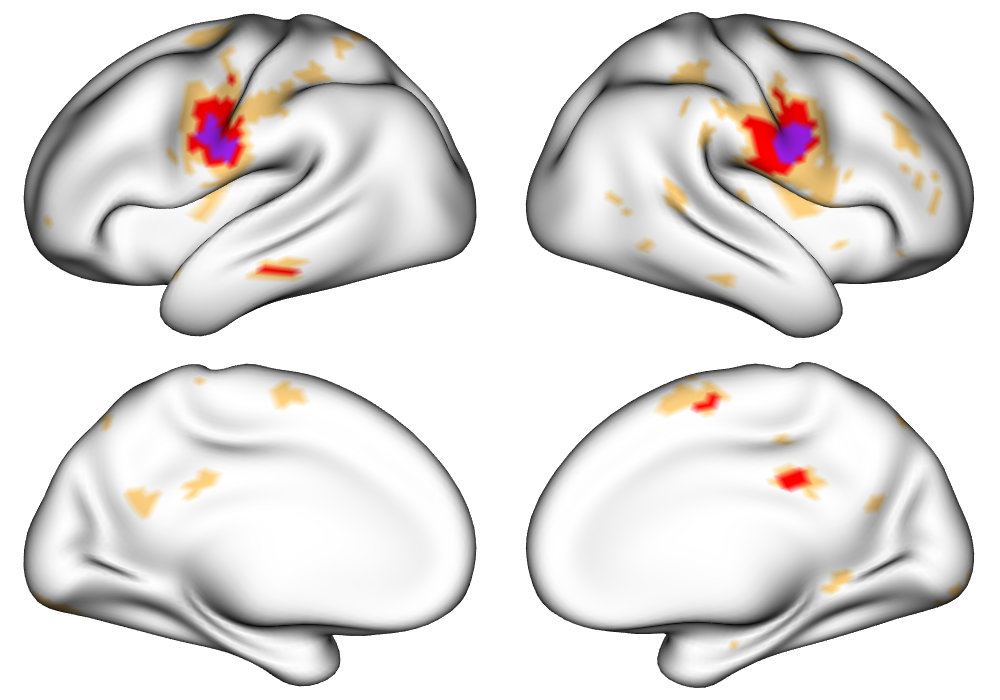} \\ \cline{2-4}
	\rotatebox[origin=c]{90}{\textbf{Subject C}} & 
	\Includegraphics[width=0.3\textwidth]{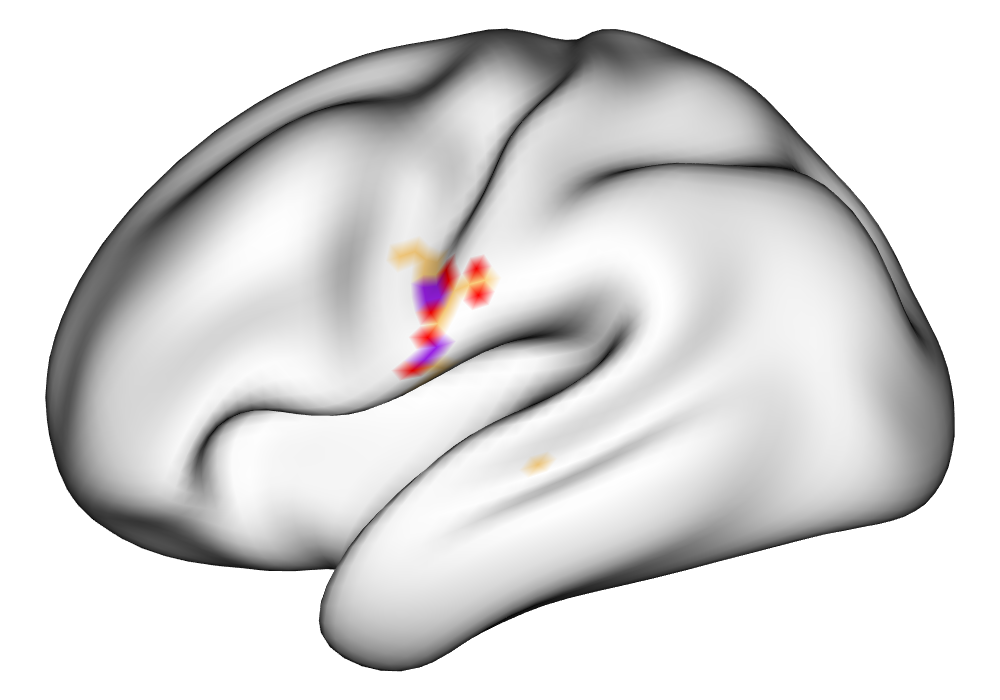} &
	\Includegraphics[width=0.3\textwidth, trim=0 125mm 170mm 0, clip]{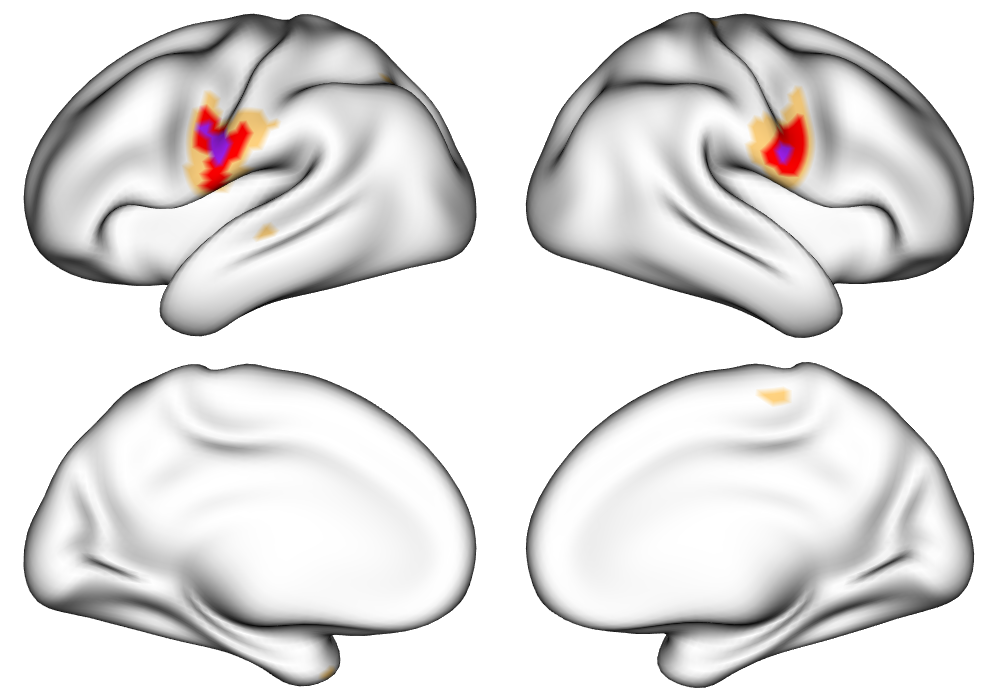} &
	\Includegraphics[width=0.3\textwidth, trim=0 125mm 170mm 0, clip]{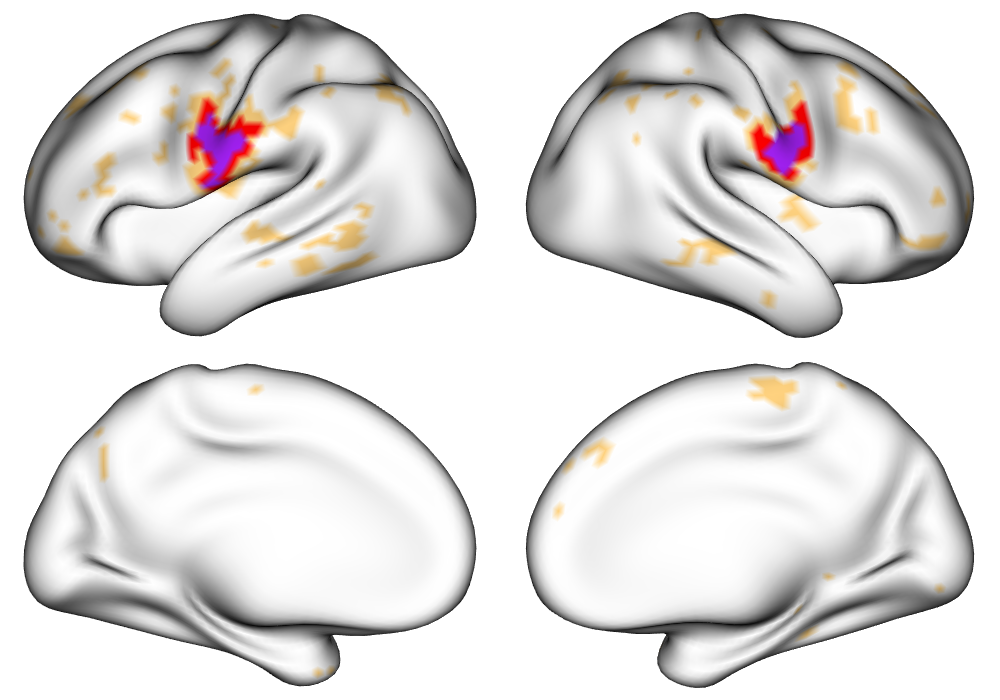} \\ \cline{2-4}
\end{tabularx}
$\gamma =$ \textcolor[HTML]{FFD27F}{$\blacksquare$} 0\% \textcolor[HTML]{FF0000}{$\blacksquare$} 0.5\% \textcolor[HTML]{A020F0}{$\blacksquare$} 1\% 
\caption{\textbf{Subject-level areas of activation during tongue movement.} For the sake of space, only the lateral view of the left hemisphere is shown. For the classical GLM, activations are based on controlling the FWER via Bonferroni correction; for the Bayesian GLM, areas of activation are based on the joint posterior distribution.  For all activations, the significance level is $\alpha = 0.01$.  In both the classical and Bayesian GLM, this is the probability of a single false positive among the activated vertices.  Three activation thresholds are considered: $\gamma = (0\%,0.5\%, 1\%)$ signal change.  Areas of activation at an activation threshold of $0\%$ represent areas exhibiting greater-than-zero amplitude, which corresponds to the traditional hypothesis testing approach in the classical GLM. The Bayesian GLM areas of activation are substantially larger than classical GLM ones (even when the data are smoothed), suggesting increased power to detect activations while maintaining strict false positive control.\\[20pt]}
\label{fig:tongue_act_subject}
\end{figure}

\begin{figure}[H]
	\begin{tabularx}{\textwidth}{c|C|C|}
		\multicolumn{1}{c}{} & \multicolumn{1}{c}{\boldsymbol{$\gamma = 0.5\%$}} & \multicolumn{1}{c}{\boldsymbol{$\gamma = 1\%$}}  \\ 
		\cline{2-3}
		\rotatebox[origin=c]{90}{\textbf{Subject A}} &
	    \Includegraphics[width=0.42\textwidth, trim=0 125mm 0 0, clip]{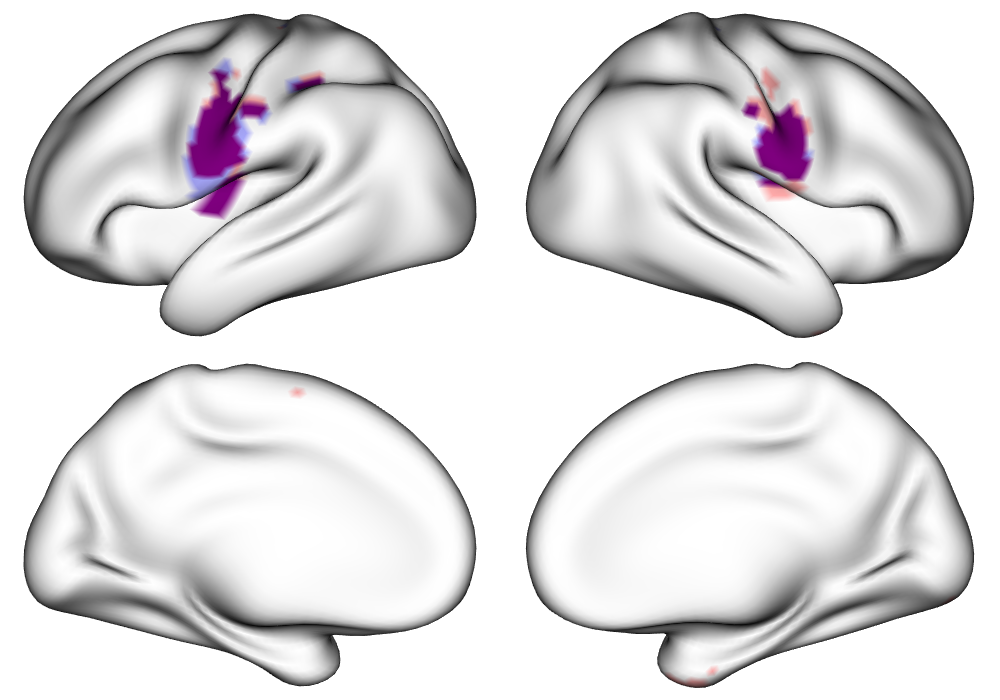} &
		\Includegraphics[width=0.42\textwidth, trim=0 125mm 0 0, clip]{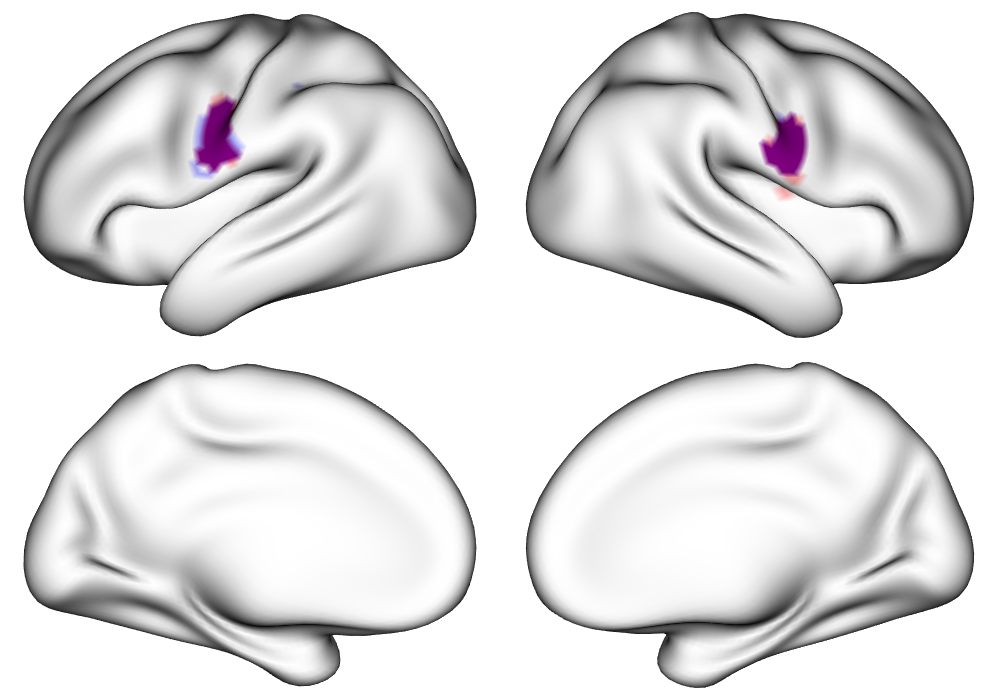} \\ 
		\cline{2-3}
		\rotatebox[origin=c]{90}{\textbf{Subject B}} &
		\Includegraphics[width=0.42\textwidth, trim=0 125mm 0 0, clip]{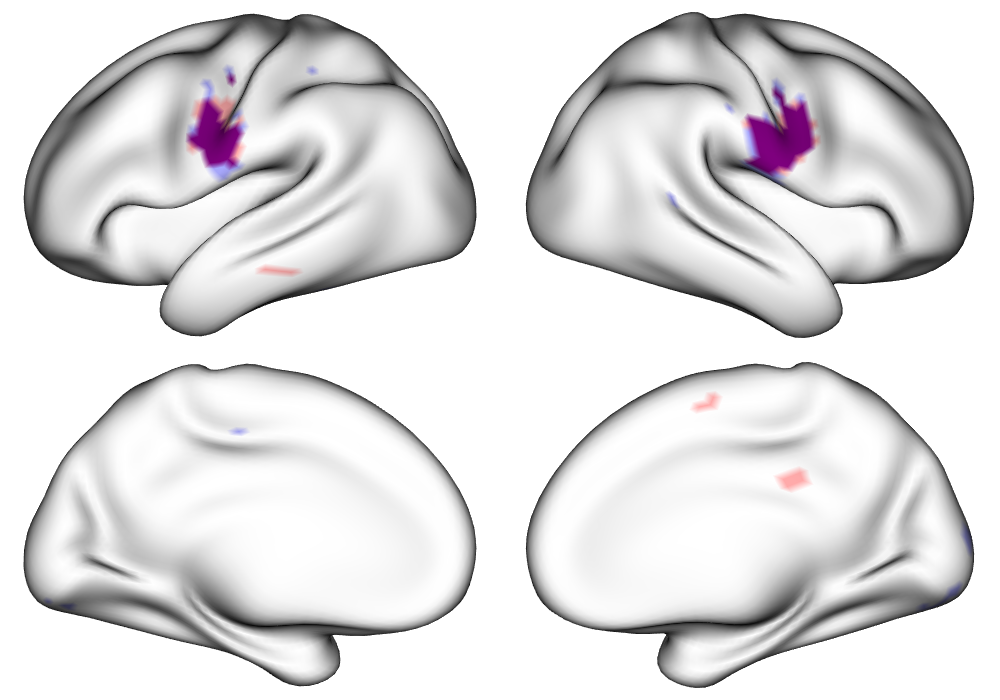} &
		\Includegraphics[width=0.42\textwidth, trim=0 125mm 0 0, clip]{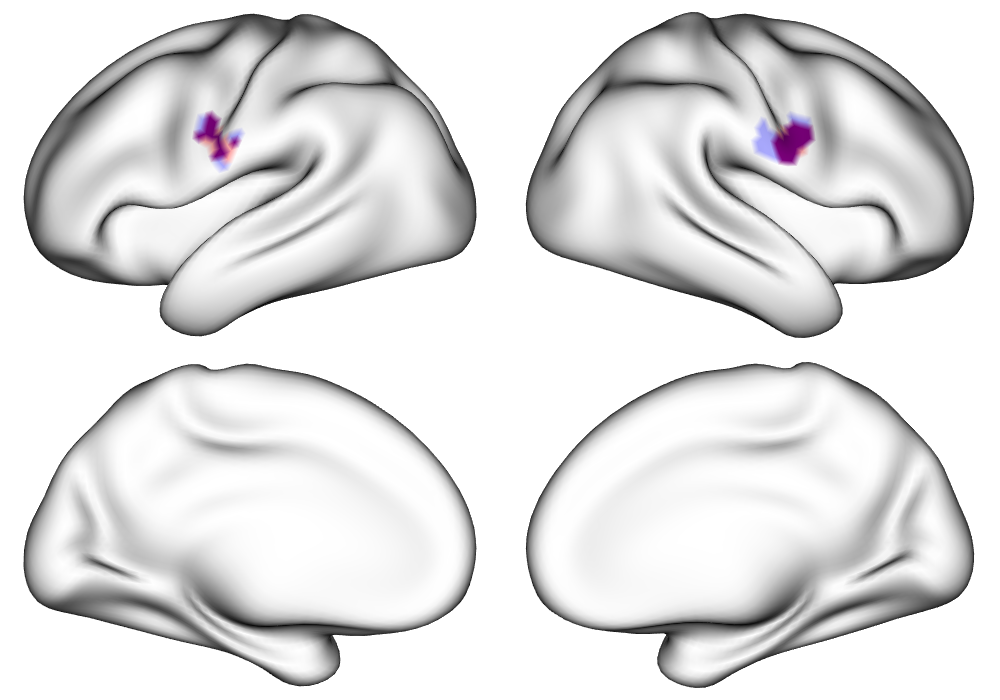} \\ 
		\cline{2-3} 
		\rotatebox[origin=c]{90}{\textbf{Subject C}} &
		\Includegraphics[width=0.42\textwidth, trim=0 125mm 0 0, clip]{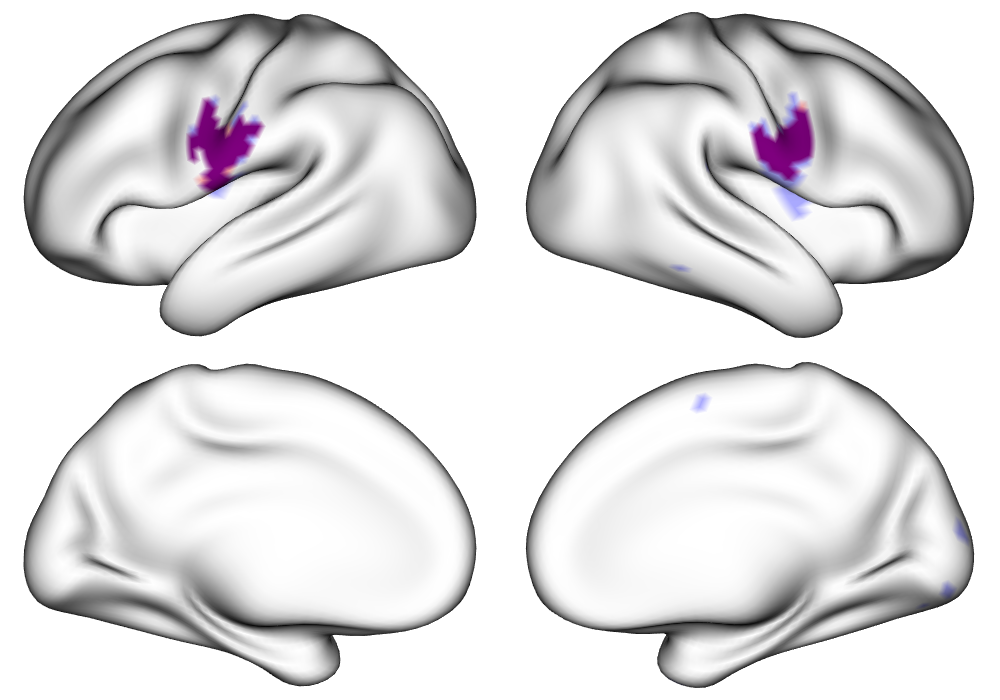} &
		\Includegraphics[width=0.42\textwidth, trim=0 125mm 0 0, clip]{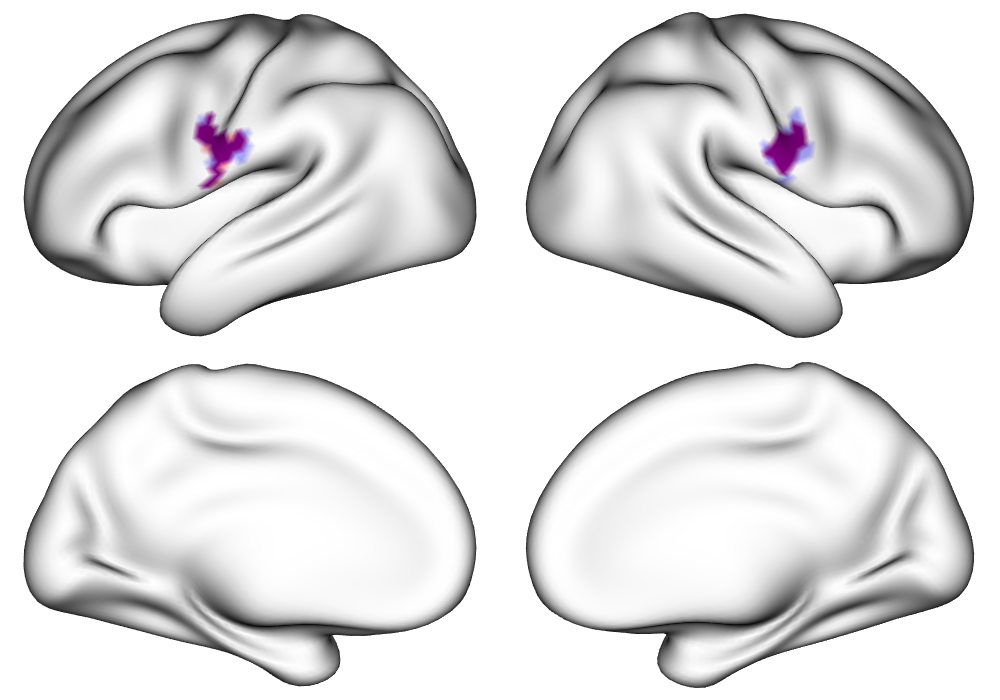} \\
		\cline{2-3}
        \multicolumn{1}{c}{} & \multicolumn{2}{c}{Activation detected in \quad \textcolor[HTML]{FFAAAA}{$\blacksquare$} visit 1 \quad
        \textcolor[HTML]{AAAAFF}{$\blacksquare$} visit 2 \quad
        \textcolor[HTML]{7F007F}{$\blacksquare$}  both visits }
	\end{tabularx}
	\caption{\textbf{Subject-level activations consistently detected across visits. } Areas of activation are found using the Bayesian GLM with activation thresholds $\gamma=(0.5\%,1\%)$ and significance level $\alpha = 0.01$. Areas of activation are highly consistent within each subject across visits. At the $0.5\%$ threshold, areas of activation closely mimic the regions of peak activation amplitude observed in \textbf{Figure \ref{fig:tongue_est_subject}}.}
	\label{fig:tongue_act_single_subject_two_visits}
\end{figure}

\subsection{Subject-level areas of activation}

For the same three example subjects as above, \textbf{Figure \ref{fig:tongue_act_subject}} displays classical and Bayesian areas of activation for the tongue movement task. \textbf{Appendix Figures \ref{fig:alltasks_subj_act_avg}} and \textbf{\ref{fig:alltasks_subj_act_LR}} show similar plots for all tasks in the multi-run and single-run cases. Areas that show statistically significant activation \textit{above} three activation thresholds ($\gamma=0\%$, $0.5\%$ and $1\%$) are displayed. The activation threshold $\gamma=0\%$ is analogous to a traditional hypothesis testing framework in the classical GLM, but is based on the joint posterior distribution of activation amplitude across all vertices. For both the classical and Bayesian GLMs, the significance level is set to $\alpha=0.01$, which represents an upper bound on the probability of observing a single false positive vertex, e.g. the FWER.

The most notable difference between the Bayesian and classical GLMs in \textbf{Figure \ref{fig:tongue_act_subject}} is that the Bayesian areas of activation are substantially larger at each activation threshold. Comparing with the estimates of activation amplitude for the same subjects shown in \textbf{Figure \ref{fig:tongue_est_subject}}, the Bayesian GLM areas of activation above $\gamma=0\%$ correspond well to both areas of intense and more subtle activation (red to yellow areas in \textbf{Figure \ref{fig:tongue_est_subject}}), while those exceeding $\gamma=0.5\%$ or $1\%$ signal change correspond well to areas of peak activation (yellow areas in \textbf{Figure \ref{fig:tongue_est_subject}}).  This suggests that the Bayesian GLM has good power to detect activations above a given effect size. In general, the classical GLM appears to be comparatively underpowered to detect activations at the subject level.

For our three example subjects, \textbf{Figure \ref{fig:tongue_act_single_subject_two_visits}} shows test-retest overlap of Bayesian areas of activation for the tongue task at thresholds of $\gamma=0.5\%$ and $1\%$. Other tasks are shown in \textbf{Appendix Figure \ref{fig:alltasks_act_single_subject_two_visits}}. Areas displayed in dark purple correspond to overlap across both visits, while areas displayed in semi-transparent blue and red correspond to areas detected in only a single visit.  These overlaps show remarkably strong within-subject, across-visit consistency of areas of activation with the Bayesian GLM.  We also observe unique patterns of individual functional topology, particularly at the $\gamma=0.5\%$ threshold. This suggests that while individuals react to stimuli in broadly similar regions, the extent and shape of their activations vary considerably. The Bayesian GLM appears able to discover these individualized patterns of functional activation, due to both its high power and ability to apply an appropriate level of smoothing to specific tasks.

\textbf{Appendix Figure \ref{fig:dice_improvements}} quantifies the test-retest reliability of areas of activation in terms of the Dice coefficient of overlap, described in Section \ref{subsec:area_reliability}.  Panel (a) displays the test-retest overlap of the subject-level Bayesian and classical areas of activation. The average over subjects is shown, along with error bars indicating 95\% bootstrap confidence intervals. The most reliable activations are produced with the Bayesian GLM using an activation threshold of $0.5\%$, achieving a Dice overlap of near or above 0.6 for all tasks. For the classical GLM, the most reliable activations tend to be produced at the standard activation threshold of $0\%$, which corresponds to the traditional hypothesis testing approach. A reference line indicates this scenario, which serves as a baseline. Panel (b) shows the size of activations versus test-retest overlap. Note that while the Bayesian and classical GLMs sometimes produce activations with similar test-retest overlap (e.g. with the $0\%$ activation threshold for the hand, foot and tongue tasks), the Bayesian areas of activation are substantially larger.  Overall, the Bayesian GLM produces subject-level areas of activation that are generally both larger and more reliable compared with the classical GLM. 

\begin{figure}[H]
    \centering
    \includegraphics[width=\textwidth]{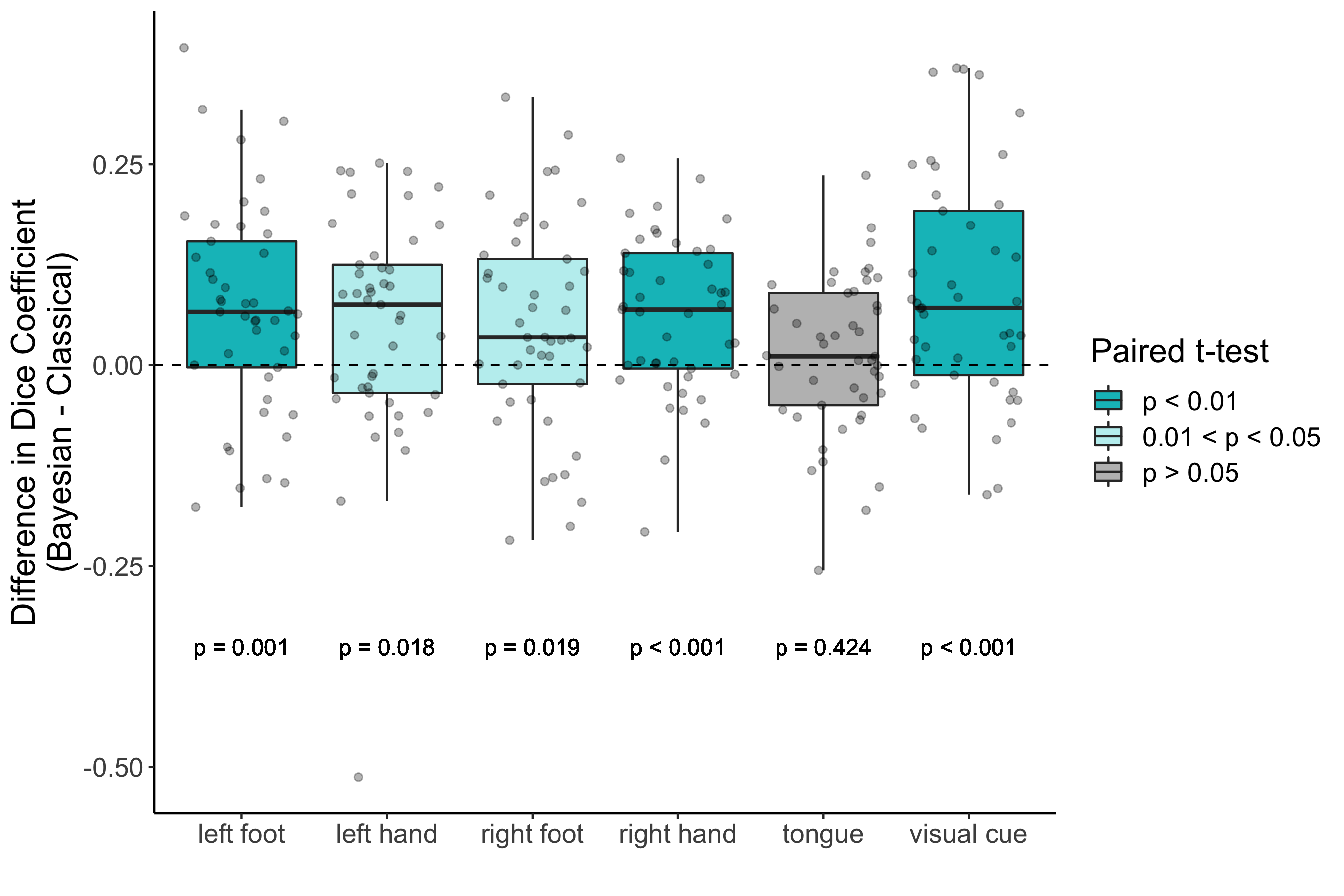}
    \caption{\textbf{Improvement in reliability of areas of activation with the Bayesian GLM.} The paired difference of the Dice coefficients found for each subject. Dice coefficients are found using the threshold $\gamma = 0.5\%$ for the Bayesian GLM, and $\gamma = 0\%$ for the classical GLM. Classical GLM results are based on smoothed data. These two thresholds were chosen for each GLM because they produce activations of roughly similar size and are the most reliable threshold for each GLM. The Bayesian GLM produces more reliable subject-level areas of activation on average for all tasks, and this difference is statistically significant for all but the tongue task.}
    \label{fig:dice_paired_difference}
\end{figure}

\textbf{Figure \ref{fig:dice_paired_difference}} directly compares the reliability of activations produced with the Bayesian and classical GLMs, in terms of the difference in the Dice coefficient of test-retest overlap between the Bayesian GLM using activation threshold of $\gamma = 0.5\%$ and the classical GLM using activation threshold of $\gamma = 0\%$. These two thresholds were chosen for each GLM because they produce activations of roughly similar size and are the most reliable threshold for each GLM (see \textbf{Appendix Figure \ref{fig:dice_improvements}}). The Bayesian GLM produces more consistent areas of activation on average for all tasks. Paired $t$-tests indicate that the improvement is statistically significant for all but the tongue task. This illustrates that the high power of the Bayesian GLM facilitates considering only activations above a scientifically relevant effect size, and that these can be reliably identified in individual subjects. 

\begin{figure}
\centering
    \begin{tabularx}{\textwidth}{|XXX|}
        \multicolumn{1}{c}{\textbf{Classical GLM}} & 
        \multicolumn{1}{c}{\textbf{Classical GLM}} &
        \multicolumn{1}{c}{\textbf{Bayesian GLM}} \\
        \multicolumn{1}{c}{\textbf{(unsmoothed)}} &
        \multicolumn{1}{c}{\textbf{(FWHM = 6mm)}} &
        \multicolumn{1}{c}{\textbf{(unsmoothed)}} \\
        \hline
        \Includegraphics[width=0.3\textwidth]{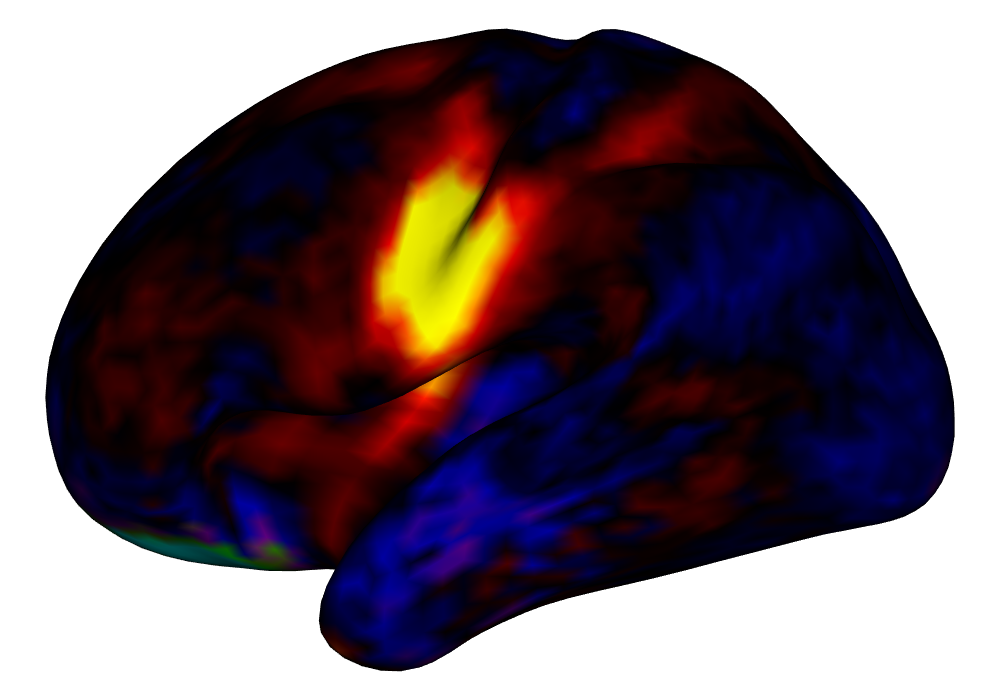} &
        \Includegraphics[width=0.3\textwidth, trim=0 125mm 170mm 0, clip]{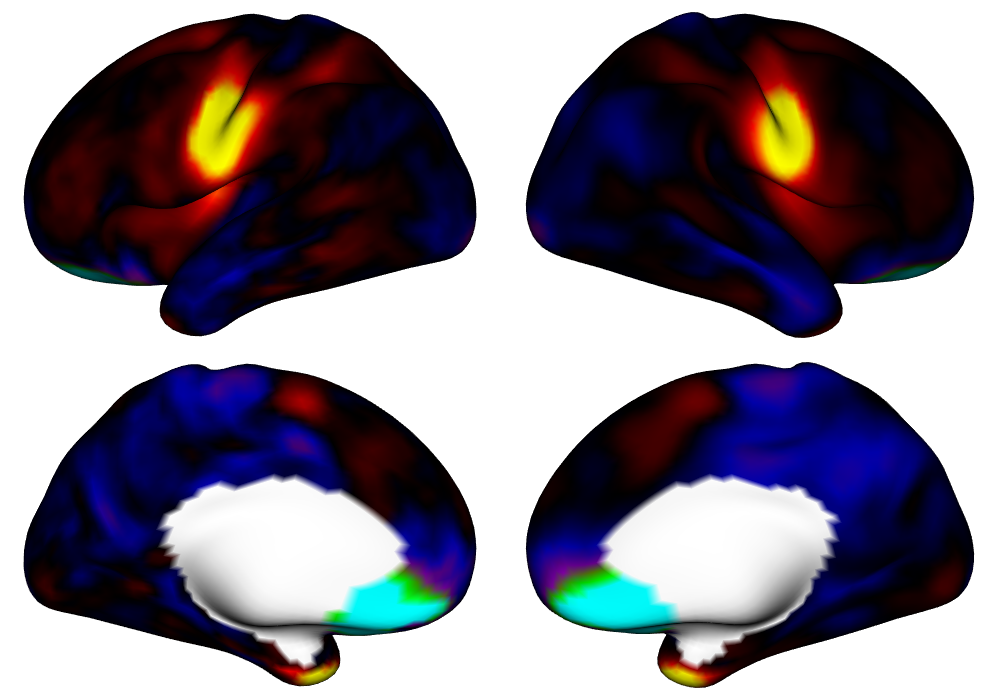} &
        \Includegraphics[width=0.3\textwidth, trim=0 125mm 170mm 0, clip]{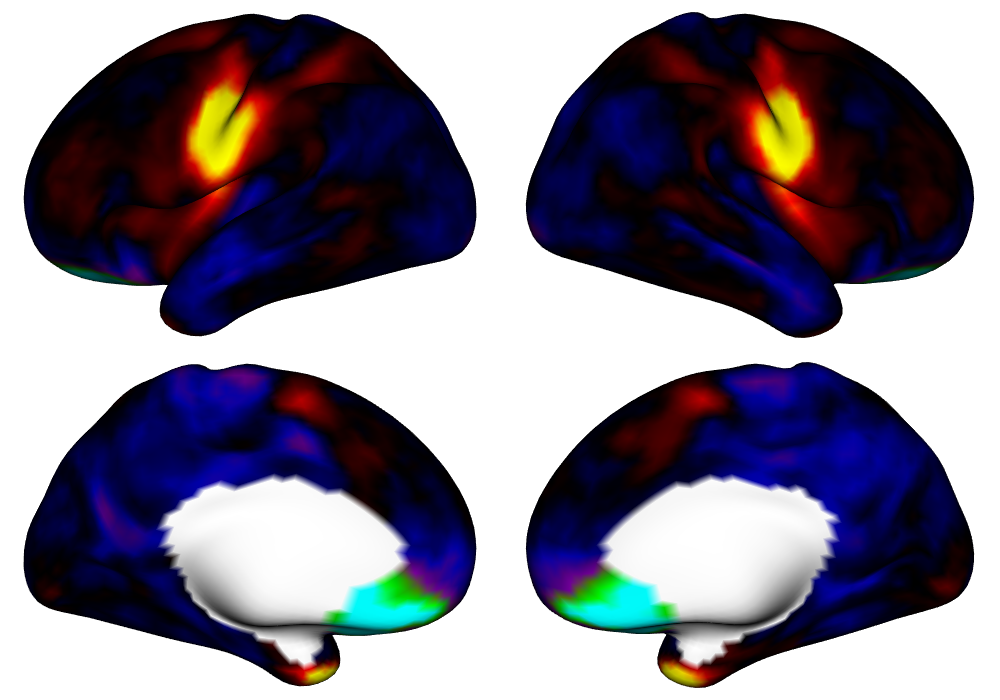} \\
        \multicolumn{3}{|c|}{\Includegraphics[width = 0.3\textwidth]{607_legend_estimate.png}} \\
        \hline
        \Includegraphics[width=0.3\textwidth]{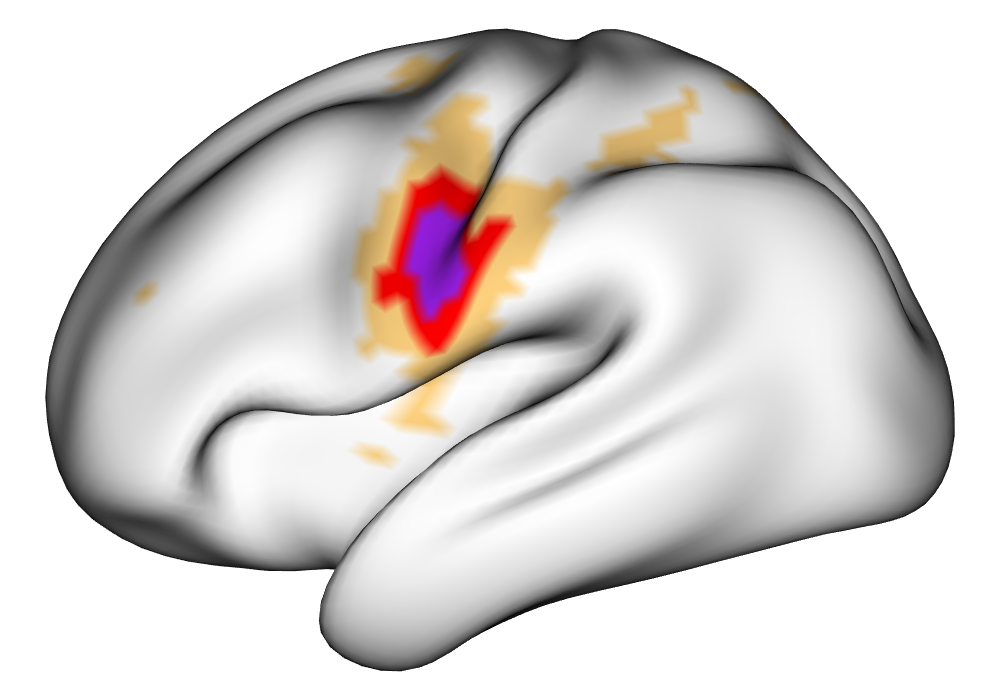} &
        \Includegraphics[width=0.3\textwidth, trim=0 125mm 170mm 0, clip]{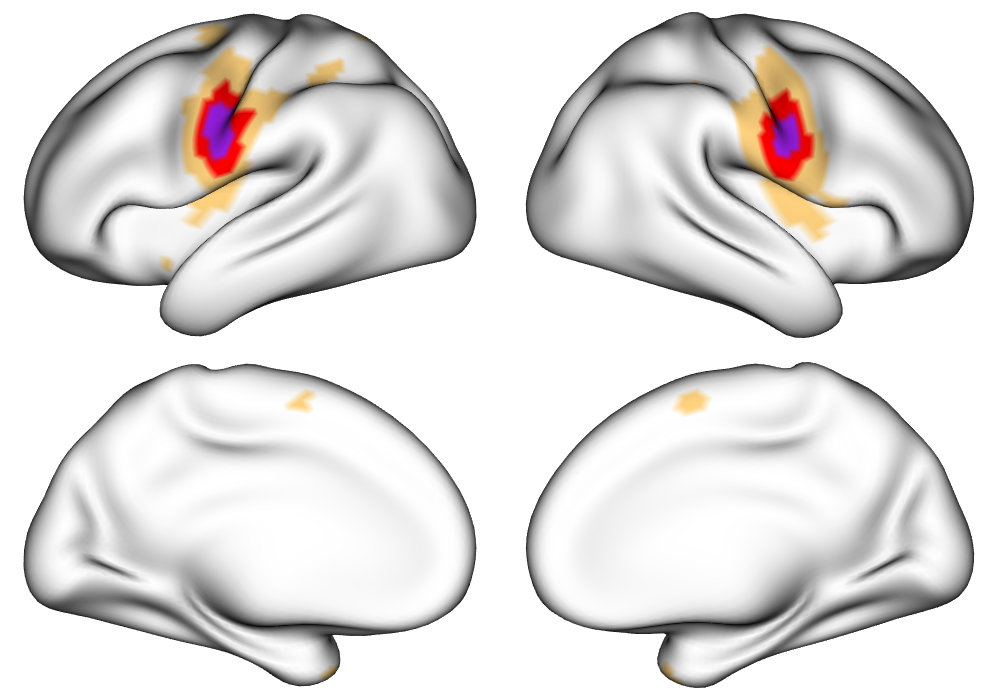} &
        \Includegraphics[width=0.3\textwidth, trim=0 125mm 170mm 0, clip]{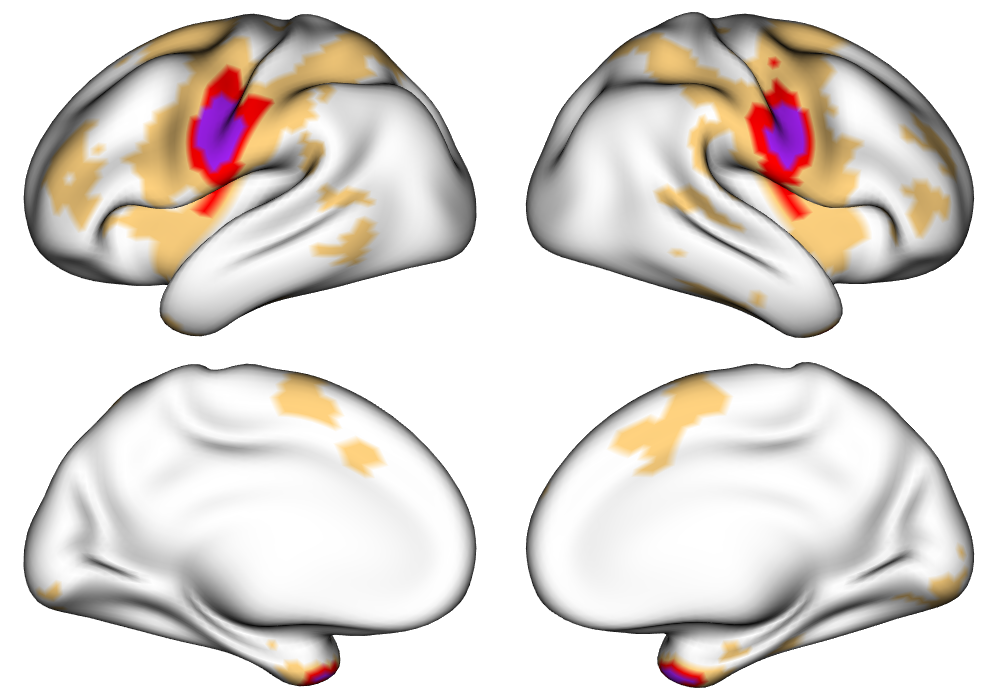} \\ 
        \multicolumn{3}{|c|}{$\gamma =$ \textcolor[HTML]{FFD27F}{$\blacksquare$} 0\% 
          \textcolor[HTML]{FF0000}{$\blacksquare$} 0.5\% 
          \textcolor[HTML]{A020F0}{$\blacksquare$} 1\% } \\
        \hline
    \end{tabularx}
\caption{\textbf{Group-average estimates of activation amplitude (top) and areas of activation (bottom) for the tongue task.} Results are based on the average across 45 subjects. Areas of activation remain smaller in the classical GLM versus the Bayesian GLM (even with smoothed data), suggesting reduced power to detect activations, even at the standard classical GLM hypothesis testing threshold of $\gamma=0\%$.}
\label{fig:group_estimates}
\end{figure}

\begin{figure}
\begin{subfigure}[b]{0.49\textwidth}        \includegraphics[width=\textwidth]{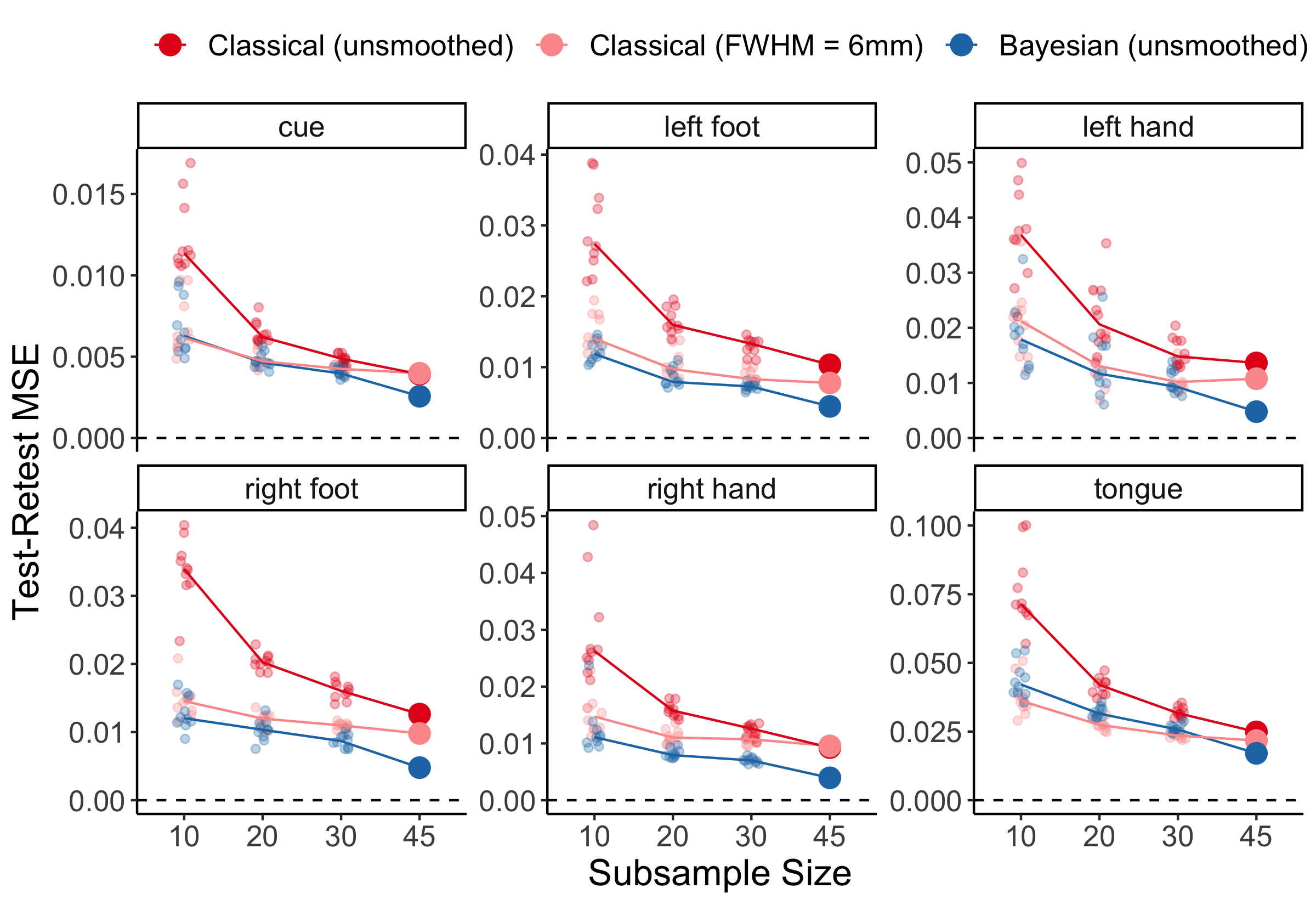}
\caption{Test-retest mean squared error (MSE)}
\end{subfigure}
\begin{subfigure}[b]{0.49\textwidth}
\includegraphics[width=\textwidth]{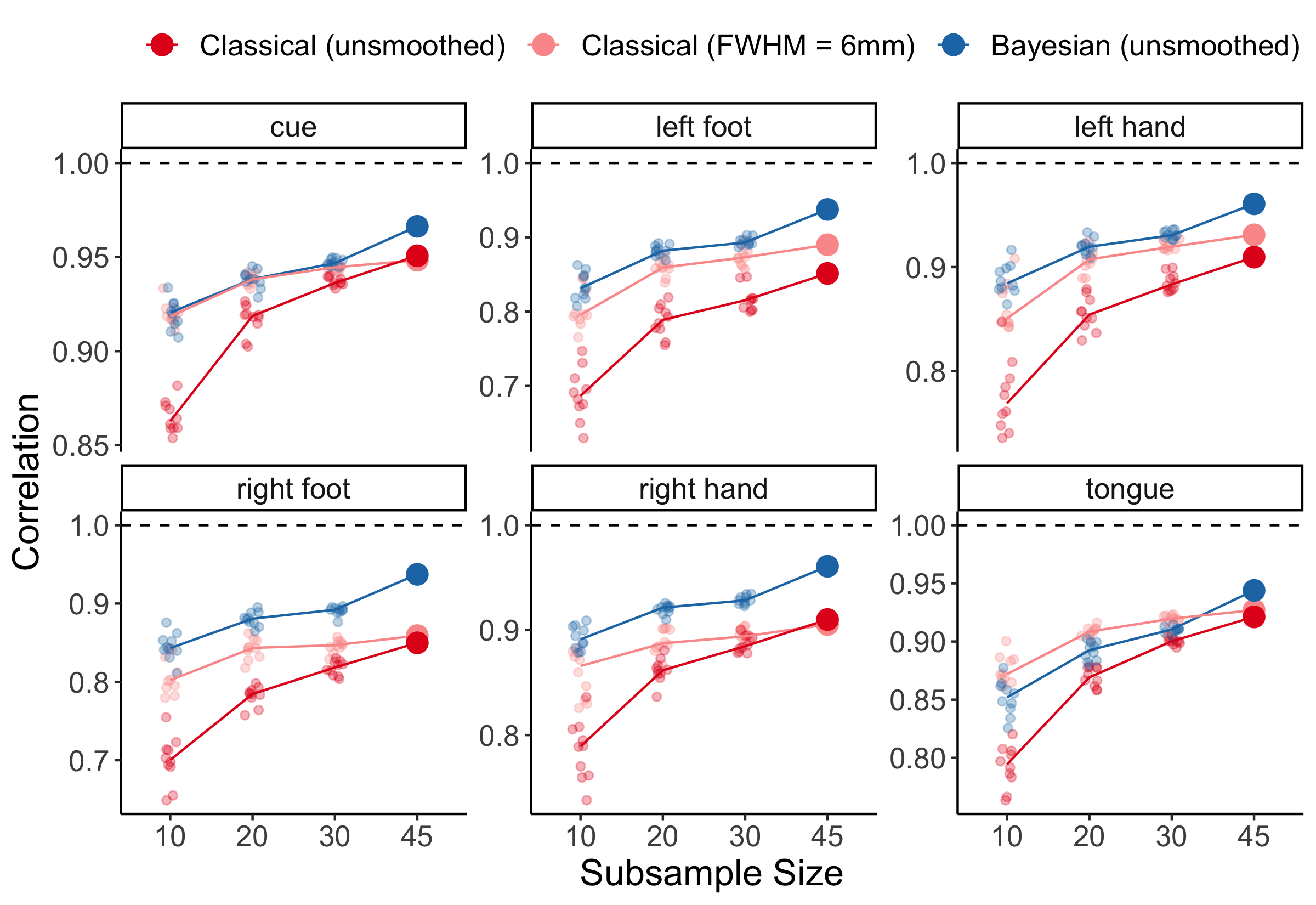} 
\caption{Test-retest correlation}
\end{subfigure}
    \caption{\textbf{Test-retest reliability of group-level estimates of activation amplitude, in terms of test-retest MSE and correlation.} Note that we used the \textit{classical} GLM visit 2 group-level amplitude estimates found using unsmoothed data as a noisy but unbiased proxy for the truth to compute MSE and correlation.  This avoids any bias in favor of the Bayesian GLM or smoothing, but will tend to result in inflated MSE and underestimated correlation for both GLMs.  Both panels show that the classical and Bayesian GLMs become more reliable as sample size increases. The Bayesian GLM produces more reliable group-level estimates of activation compared with the classical GLM using smoothed data in all but the tongue task, where they have similar performance. For the full sample of $45$ subjects, the Bayesian GLM outperforms the classical GLM (using smoothed or unsmoothed data) for all tasks.}
    \label{fig:group_mse_cor}
\end{figure}

\subsection{Group-level estimates and areas of activation}
\label{sec:group_est_and_act}

The surface-based spatial Bayesian GLM can also produce group-level estimates and areas of activation in a computationally efficient way.  Here, we assess the reliability and power of the group-level Bayesian GLM in comparsion with the classical GLM.

\textbf{Figure \ref{fig:group_estimates}} displays group-average estimates of activation amplitude and areas of activation for the tongue task, based on all 45 subjects (\textbf{Appendix Figures \ref{fig:alltasks_group_est}} and \textbf{\ref{fig:alltasks_group_act}} show group-average estimates and activations for all tasks). The estimates of activation are visually similar for the classical and Bayesian GLMs.  However, the size of activations are substantially larger with the Bayesian GLM at every activation threshold. This suggests that the Bayesian GLM has higher power to detect activations even in group analysis with a moderate sample size. In fact, with the Bayesian GLM at the $0\%$ threshold we see large areas of the cortex being detected as statistically significant. These include areas of small effect size, as seen in the estimates of activation amplitude in dark red.  This is similar to the known phenomenon of areas of small effect size being identified as statistically significant in very large group studies using the classical GLM. Due to the greatly increased power of the Bayesian GLM, the issue of small but statistically significant effect sizes may arise even with moderate sample sizes. This illustrates the importance of specifying a threshold above which activations are scientifically meaningful. Here, for example, adopting an activation threshold of $0.5\%$ produces areas of activation that closely mimic the peak areas of activation seen in the amplitude maps in yellow and bright red.  

\textbf{Figure \ref{fig:group_mse_cor}} displays two measures of test-retest reliability for the group-level estimates of activation amplitude: mean squared error (MSE) and correlation.  Both measures are based on using the visit 2 \textit{classical} GLM estimates of activation amplitude based on unsmoothed data, providing a noisy but unbiased proxy for the unknown true activation amplitudes and avoiding any bias in favor of the Bayesian GLM or smoothing.  Note that this will tend to result in somewhat pessimistic measures of reliability, e.g. higher MSE and lower correlation, for both GLMs. Yet even so, the Bayesian GLM approaches perfect reliability (MSE of $0$; correlation of $1$) as sample size increases. For instance, the Bayesian GLM estimates of activation amplitude achieve test-retest correlation of approximately $0.95$ across all tasks in the full sample of $n=45$ subjects.  Overall, the Bayesian GLM produces more reliable group-average estimates of activation amplitude across nearly all settings (sample sizes and tasks).   

\textbf{Figure \ref{fig:group_mse_cor}} additionally shows that the Bayesian GLM achieves small-sample reliability similar to or better than the reliability achieved by the classical GLM at different sample sizes. For example, the Bayesian GLM test-retest reliability with a sample of $n=20$ is generally similar to that of the classical GLM with $n=45$, more than double the sample size.  This illustrates that the group-level Bayesian GLM can extract more reliable measures of population activation from smaller samples compared with the classical GLM.  Given the high cost of collecting larger samples, this illustrates an important benefit of the Bayesian GLM: it is able to extract more information from a sample, rivaling the benefit of doubling the sample size.

\textbf{Appendix Figure \ref{fig:group_power}} examines the power of the Bayesian GLM for different sample sizes. The size of group-level activations are shown as a function of sample size. Using an activation threshold of $0\%$ signal change, the size of both the Bayesian and classical GLM activations grows with increasing sample size. In the case of the Bayesian GLM these areas are quite large for some tasks, and many of these locations exhibit small effect size which may not be of scientific interest. This illustrates the importance of considering effect size when identifying areas of activation, especially when power is high as in the Bayesian GLM.  However, when considering activations above $0.5\%$ signal change the size of Bayesian activations is virtually flat as sample size grows. This illustrates that the Bayesian GLM has high power to detect areas that activate above $0.5\%$ signal change, even in very small samples. The size of classical GLM activations is much smaller and continues to grow nearly linearly with increasing sample size, suggesting that the classical GLM is underpowered to detect these effects, even with moderately sized samples.  

One interesting feature seen in \textbf{Appendix Figure \ref{fig:group_power}} is high variance in the number of activations associated with the tongue task for the Bayesian GLM when $n=10$. This likely stems from two sources. First, the tongue task has larger active regions than the other motor tasks, and similar \textit{relative} changes to the number of active locations will be more apparent when considering the \textit{number} of detected activations. Second, the subject-level activations for the tongue task exhibit highly individualized areas of activation, which vary in size across subjects (see Figure \ref{fig:tongue_est_subject}). Smaller samples may not be representative of populations in terms of their activations, leading to variance across samples. As the sample size increases, the amount of variability tends to decrease and population-level inference becomes more consistent between samples. 

\section{Discussion}
\label{sec:Discussion}

The surface-based spatial Bayesian general linear model (GLM) leverages information shared between neighboring locations on the cortical surface to improve the accuracy of task amplitude estimates and to increase power to detect significant activations.  We analyze test-retest motor task fMRI data from 45 subjects in the Human Connectome Project (HCP).  Our findings establish that surface-based spatial Bayesian modeling produces reliable subject-level and group-average estimates of activation amplitude and highly consistent subject-level areas of activation.  We also observe a major gain in power over the classical GLM to detect activations in individuals and across groups of subjects.  The Bayesian GLM is computationally efficient at both the subject and group level and is conveniently implemented in the R package \texttt{BayesfMRI}, facilitating the use of this approach to extract accurate and nuanced insights in future task fMRI studies.

\subsection{Unique individual functional topology}

We visualize estimates and areas of activation for several individual subjects to illustrate the effects of smoothness and noise reduction of the Bayesian GLM, but also to show the unique patterns of functional activation we observe in individuals using the Bayesian GLM. More importantly, the Bayesian areas of activation above $0.5\%$ signal change closely resemble the patterns of peak activation seen in the amplitude maps. Areas of activation produced using the classical GLM are not as representative of these patterns, due to lower power at the subject level. We observe the Bayesian areas of activation to be highly similar across visits, suggesting that functional topology is a trait that can be consistently observed in individual subjects.  Indeed, the within-subject test-retest overlap of activations is high in the Bayesian GLM achieving Dice coefficients as high as 0.7 for the left and right hand tasks.  The ability to detect and quantify unique patterns of individual functional topology with relatively little data (e.g., four $12$-second blocks of each motor task) is a valuable product of surface-based spatial Bayesian modeling.  Such subject-level measures could be used to enhance understanding of differences in task performance across subjects, the manifestations of development or aging on functional topology, or the effects of disease progression or treatment on functional engagement.

\subsection{Universally beneficial for subject-level analysis}

We assess the ability of the Bayesian GLM to produce reliable subject-level estimates of activation on average across subjects (using ICC) and in individual subjects (using test-retest MSE and correlation).  We show a substantial increase in the number of brain locations exhibiting at least ``fair’’ or ``good’’ ICC in certain tasks. This illustrates that on average across subjects, the Bayesian estimates are more reflective of reliable features of individual subjects. Furthermore, analyzing the test-retest reliability at the individual subject level, we find that the Bayesian GLM produces more reliable estimates of activation across \textit{every} subject included in our analysis.

The HCP includes two runs of motor task data (plus an additional two runs for the 45 subjects in the HCP retest dataset). This may not be the case for many more typical task fMRI studies, where often a single run may be collected for each subject.  Therefore, we also examine the performance of the Bayesian GLM using only a single run from each subject.  The benefits of the Bayesian GLM are quite apparent for single-run data, producing larger areas of activation at the individual level. The areas of activation in individual subjects, though somewhat smaller than those based on both runs, are already reflective of unique patterns of functional topology.

\subsection{High group-level power in small samples}

A unique feature of this spatial Bayesian GLM is its ability to be easily extended to group-level analysis in a computationally efficient way.  Spatial Bayesian modeling is often assumed to be primarily beneficial for subject-level analysis, as the classical GLM tends to produce under-powered areas of activation due to the low signal-to-noise ratio (SNR) in task fMRI data \citep{welvaert2013definition}.  However, we also observe clear benefits of the Bayesian GLM for group-level analysis.  Namely, we observe improved test-retest reliability of group-average estimates of activation amplitude and increased power to detect activations.  Notably, the power of the Bayesian GLM to detect activations above $0.5\%$ signal change is remarkably consistent across sample sizes from $n=10$ to $n=45$. This illustrates that the Bayesian GLM has high power to detect activations above a scientifically meaningful effect size even in small samples. 

\subsection{Efficient Bayesian computation}

While the benefits of spatial Bayesian modeling for task fMRI analysis have been long recognized \citep{zhang2014spatio,zhang2015bayesian,zhang2016spatiotemporal,guhaniyogi2017bayesian,spencer2020joint}, previous methods for volumetric fMRI were constrained by high computational demands.  Analyzing cortical surface data has the dual benefit of leveraging scientifically relevant spatial dependencies along the cortical surface and of dramatically reducing dimensionality, facilitating efficient computation. The surface-based spatial Bayesian GLM also leverages recent advances in Bayesian computation and spatial statistics to maximize both computational efficiency and accuracy in model estimation.  Areas of activation are based on the joint posterior distribution using an efficient Bayesian computation approach, which maximizes power to detect activations \citep{bolin2018Rexcursions}.  

It is important to note that the Bayesian GLM takes substantially more computation time, memory and processing requirements compared to a massive univariate approach. Indeed, such approaches were originally designed for maximal computational efficiency, given the much more limited computing power available in the early days of task fMRI.  Today, statistical and computational advances make it quite feasible to use more sophisticated techniques to extract more accurate and nuanced information from task fMRI studies.  In our analysis, model estimation per hemisphere requires approximately 6.5 minutes for each individual subject and approximately 3 hours for group-level analysis with $n=45$.  While this certainly represents a greater investment of time than the classical GLM, it is a small fraction of the time and resources already invested in experimental design, participant recruitment, data collection, and data processing.  Therefore, the benefits of the Bayesian GLM are likely worth the computational tradeoff. 

\subsection{Software implementation}

The surface-based spatial Bayesian GLM is implemented in the R package \texttt{BayesfMRI}, which is designed to be maximally convenient from a user perspective.  The main function in \texttt{BayesfMRI} can be used to directly analyze surface data in CIFTI and GIFTI format and performs all processing steps described in this paper, including resampling, scaling, nuisance regression and prewhitening using a high-order, spatially varying AR process.  Integration with the \texttt{ciftiTools} R package \citep{pham2021ciftiTools} allows for direct visualization of estimates and areas of activation in R, as well as the ability to write out results in CIFTI or GIFTI format.

\subsection{Study limitations}

This study is subject to several important limitations.  First, here we analyze data from the young adult HCP, a large repository containing high-quality fMRI data acquired using multi-band techniques optimized for high cortical SNR.  Our findings are therefore reflective of the HCP acquisition and processing and the study population. In other contexts, our findings would surely be somewhat different. However, given the quality of HCP data, particularly on the cortical surface, it represents something of a best-case scenario for the classical GLM.  In higher-noise data, surface-based spatial Bayesian modeling may prove to be even more beneficial.

Second, our analyses are based on test-retest data in lieu of information on the ground truth of task activation at the individual and group level.  Furthermore, for analyses using MSE and correlation, we utilize the \textit{classical GLM} estimates of activation using unsmoothed data as a noisy but unbiased proxy for the ground truth to avoid bias in favor of the Bayesian GLM. Our measures of reliability are therefore also subject to noise. Although this results in an imperfect assessment of reliability, the consistency of our results across subjects, tasks and samples provides clear evidence of the benefits of the Bayesian GLM.

Finally, here we limit our analysis to the cortical surface, excluding subcortical and cerebellar regions.  These areas represent great scientific interest and importance.  Given the low SNR in the subcortex, particularly in HCP-style multiband data, spatial Bayesian modeling may be particularly beneficial for these regions.  While the current implementation of the Bayesian GLM considered here is limited to the cortical surface, extension to subcortical and cerebellar regions is an important area for future research.

\subsection{Future work}

In ongoing and future work, we plan several extensions and improvements to the surface-based spatial Bayesian GLM.  First, we are developing an alternative empirical Bayesian computation approach using expectation-maximization (EM) for greater computational efficiency and flexibility.  Second, we plan to extend the spatial Bayesian GLM to subcortical and cerebellar regions, which are also of interest in the neuroscience research community. Third, we plan on investigating the value of using the full-resolution cortical surface data in conjunction with a mesh that uses fewer points than data locations to see if there is an inferential benefit to using the full-resolution data over using data resampled to a computationally feasible resolution. Finally, we plan to directly integrate HRF derivatives into the Bayesian model to fully account for variability in the temporal properties of the hemodynamic response.  These improvements and extensions will be incorporated into future versions of the \texttt{BayesfMRI} R package.

Future work should assess the \textit{value} of the improvement in subject-level reliability of the Bayesian GLM.  Does the improvement in reliability lead to better prediction of behavioral measures, such as task performance?  Are the unique functional topologies that we see reflected in subject-level areas of activation reliable enough to serve as a ``fingerprint'', whereby a subject is identifiable based on these patterns?  Future work should also assess the effect of the Bayesian GLM on the scan duration necessary to produce highly reliable estimates and areas of activation. For example, how much data is required at the subject level to produce estimates that are predictive of behavior, or functional topologies that are identifiable?  These questions are important to address in order to better understand the true benefits of surface-based spatial Bayesian modeling, in terms of extracting not only reliable but informative measures of task activation in individuals, and of reducing the burden of long and repeated scanning sessions to achieve these goals.

\section{Conclusion}

In this study, we assess the reliability of individual and group-average task activations produced by a surface-based spatial Bayesian general linear model (GLM), compared with the classical ``massive univariate’’ GLM.  Based on an analysis of test-retest motor task fMRI data from the Human Connectome Project, we find that surface-based spatial Bayesian modeling produces reliable subject-level and group-average estimates of activation amplitude and highly consistent subject-level areas of activation. Furthermore, the Bayesian GLM has high power to detect activations in individuals and small group studies.  The Bayesian GLM is computationally efficient at both the subject and group level and is conveniently implemented in the R package \texttt{BayesfMRI}.  The ease of implementation makes this powerful method widely accessible.

The code used to perform the analyses and produce the visualizations used in this validation study can be found online via GitHub\footnote{\url{https://github.com/danieladamspencer/BayesGLM_Validation}}.

\section{Acknowledgments}

Data were provided in part by the Human Connectome Project, WU- Minn Consortium (Principal Investigators: David Van Essen and Kamil Ugurbil; 1U54MH091657) funded by the 16 NIH Institutes and Centers that support the NIH Blueprint for Neuroscience Research; and by the McDonnell Center for Systems Neuroscience at Washington University.

The research of Daniel Spencer and Amanda Mejia is partially funded by the National Institute of Biomedical Imaging and Bioengineering (R01EB027119).

\newpage

\bibliographystyle{plainnat}
\bibliography{BayesGLMValidation.bib}

\newpage
\appendix

\renewcommand\thefigure{\thesection.\arabic{figure}}    
\setcounter{figure}{0} 

\section{Midthickness versus spherical surface distances}

One of the advantages to using an SPDE prior is improved accounting for spatial dependence through subject-specific cortical surfaces. This offers an improvement by reducing the distance distortion resulting from using spherical surfaces (see figure \ref{fig:sphere_dist} in the Appendix). 

\begin{figure}[H]
    \centering
    \includegraphics[width=6in]{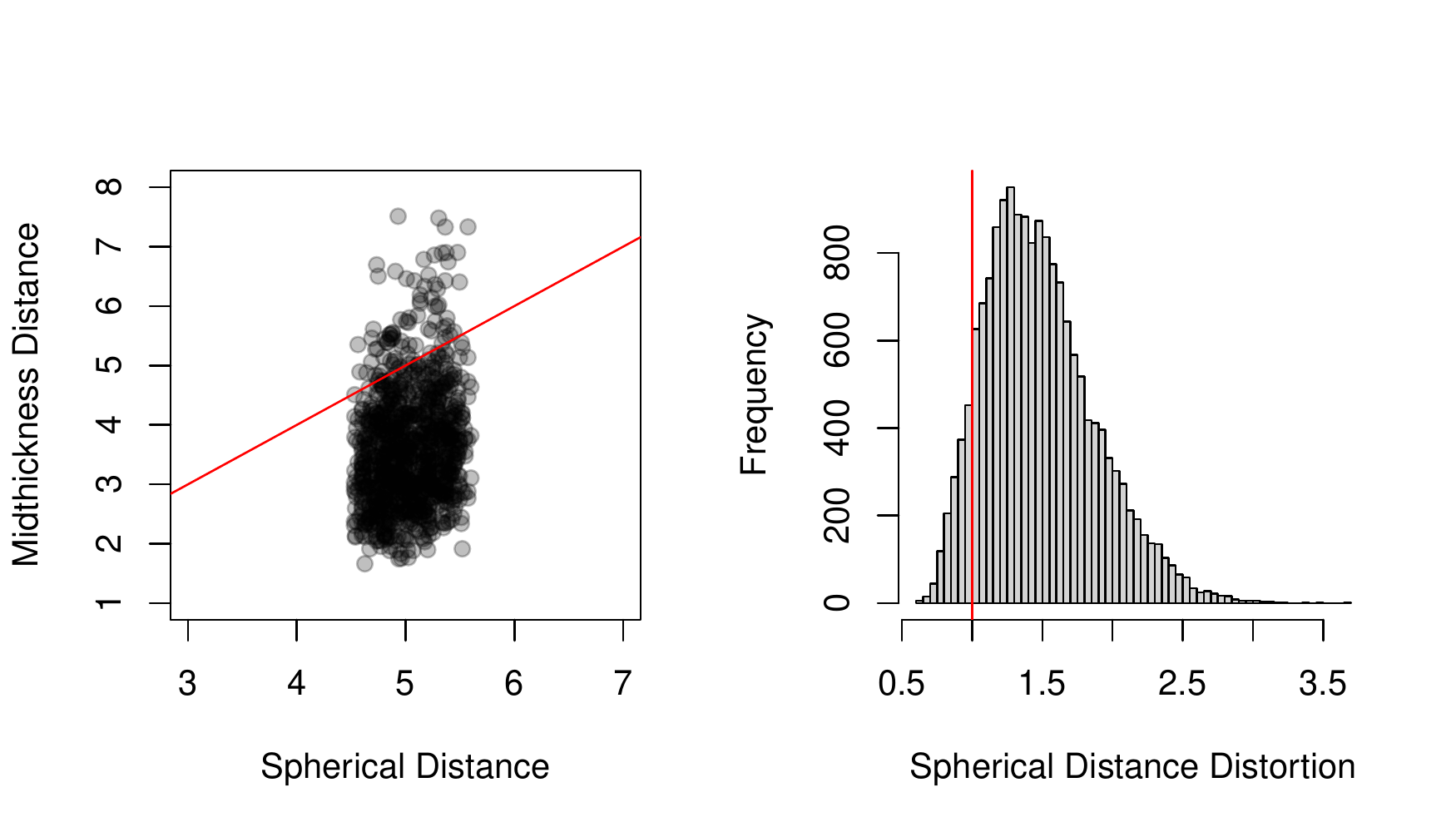}
    \caption{\textbf{Distance distortions between neighboring vertices on spherical surface relative to midthickness surface.} For this analysis, group-average spherical and midthickness 32k surfaces were resampled to 6k using the Connectome Workbench. The distance between each pair of neighbors in the triangular mesh was computed for each surface. On the left-hand plot, a random sample of 1000 neighbors is shown. The red line indicates equality. Distances on the spherical surface are much more uniform and tend to be larger than those on the midthickness surface. The right panel shows the distribution of spherical distance distortions. The red line indicates no distortion. For each pair of neighboring vertices, spherical distance distortion is defined as the ratio of their distance on the spherical surface to their distance on the midthickness surface. Distortions range from approximately $0.5$ (distance is halved on the spherical surface) to $3.5$ (distance is over $3$ times as large on the spherical surface).}
    \label{fig:sphere_dist}
\end{figure}

\begin{figure}[H]
    \centering
    \begin{subfigure}[b]{0.49\textwidth}
    \includegraphics[height=2.3in]{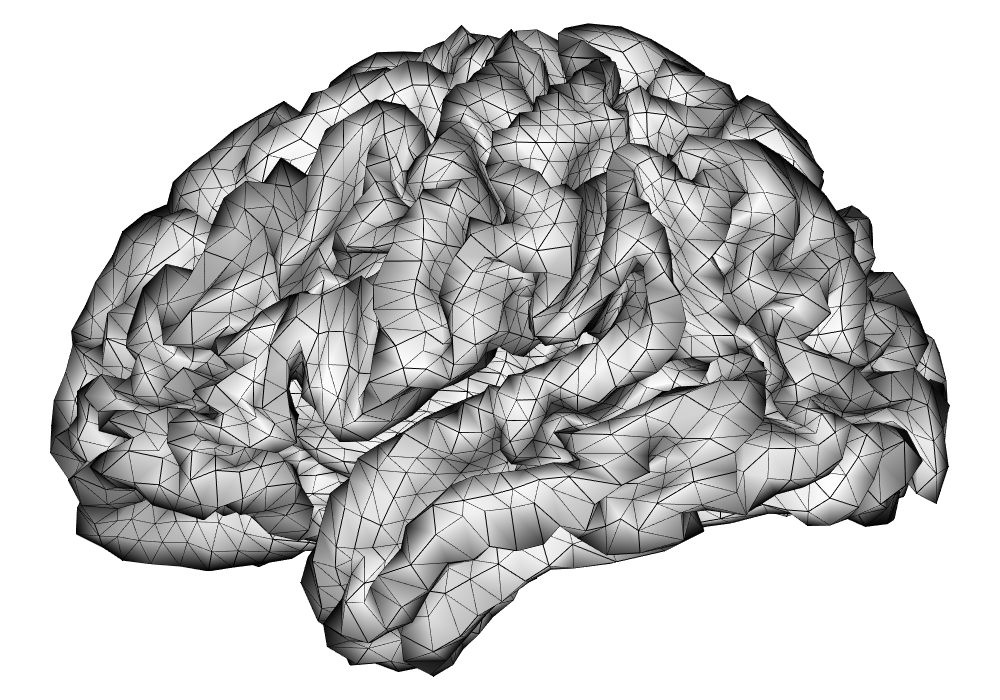}
    \caption{Midthickness Surface (5k vertices)}
    \end{subfigure}
    \begin{subfigure}[b]{0.49\textwidth}
    \includegraphics[height=2.5in]{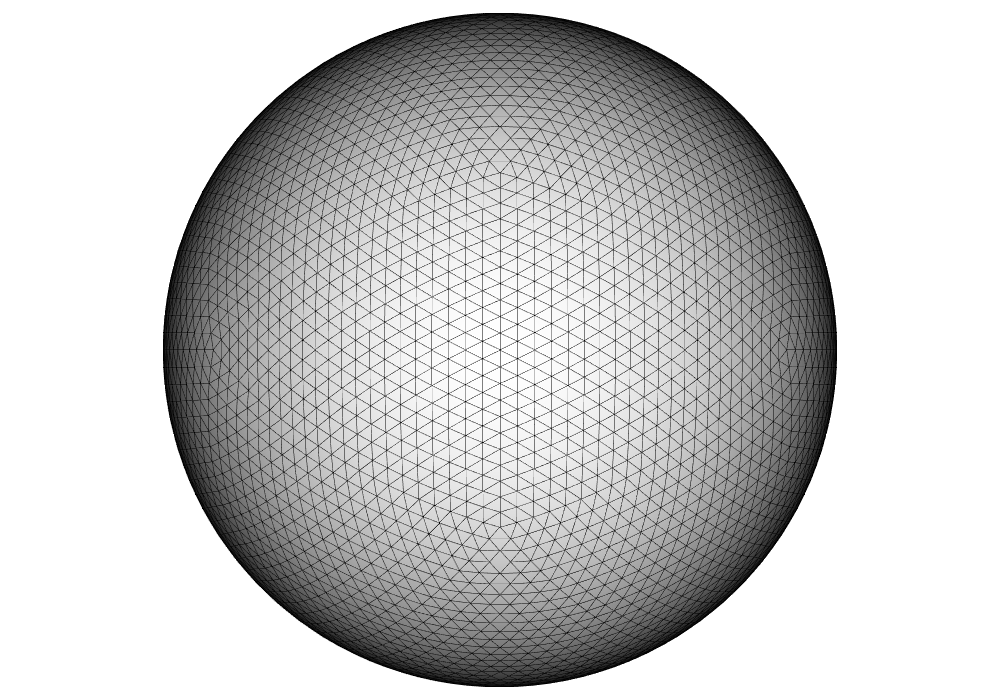}
    \caption{Spherical Surface (5k vertices)}
    \end{subfigure}
    \caption{An example of the triangular mesh structure on the midthickness and spherical surface resampled to a resolution of around 5,000 vertices.}
    \label{fig:cifti_mesh}
\end{figure}

\section{Prewhitening}

\begin{figure}[H]
    \centering
    \begin{subfigure}[b]{0.31\textwidth}
    \includegraphics[height=2.5in]{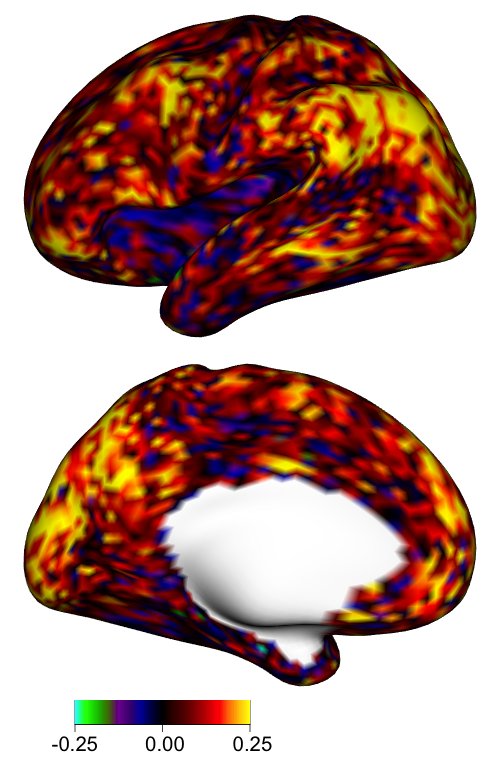}\\
    \caption{AR(1) coefficients}
    \end{subfigure}
    \begin{subfigure}[b]{0.31\textwidth}
    \includegraphics[height=2.5in]{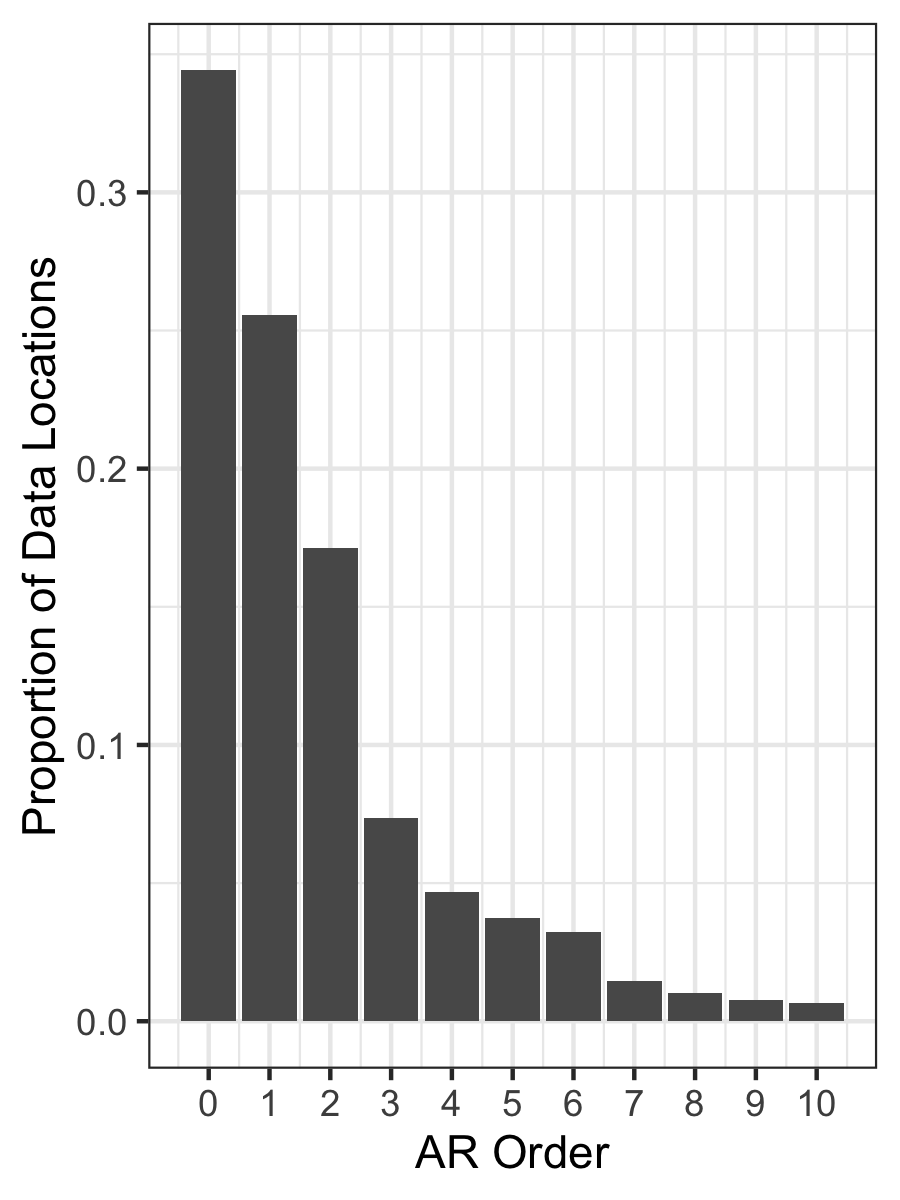}
    \caption{Optimal AR model order}
    \end{subfigure}
    \begin{subfigure}[b]{0.31\textwidth}
    \includegraphics[height=2.5in]{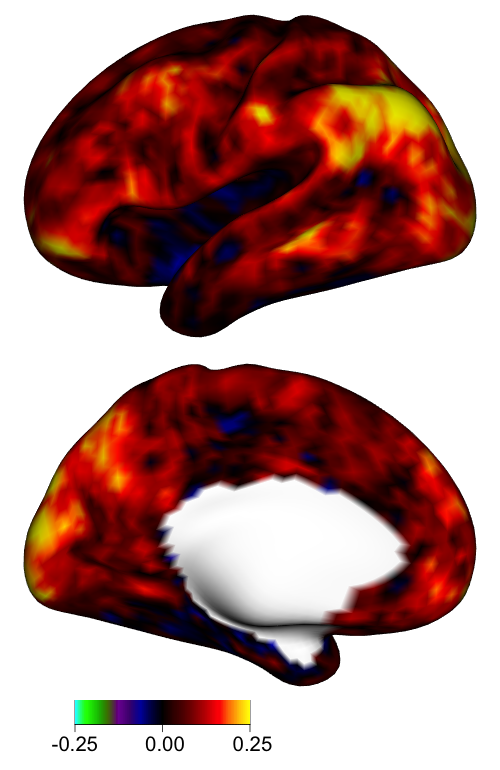}\\
    \caption{Regularized AR(1) coefficients}
    \end{subfigure}
    \caption{\textbf{Illustration of prewhitening procedure.} (a) Left hemisphere AR(1) coefficient estimates from a single subject and run. Systematic spatial variation in the degree of autocorrelation is clearly apparent. (b) Histogram of the optimal AR model order at each vertex, based on AIC.  A low-order AR process (e.g. AR(1) or AR(2)) would fail to fully capture and remove the residual autocorrelation at many locations in the brain.  (c) Left hemisphere estimates of the AR(1) coefficient at each vertex after regularization through averaging over runs and surface smoothing at $5$mm FWHM.}
    \label{fig:prewhitening}
\end{figure}

\section{Choice of multiplicity correction method in the classical GLM}
\label{appendix:multiplecomparisons}

\begin{figure}[H]
    \begin{tabularx}{\textwidth}{c|X|X|X|X|}
        \multicolumn{1}{c}{} &
        \multicolumn{1}{c}{\textbf{FDR}} &
        \multicolumn{1}{c}{\textbf{Bonferroni}} &  \multicolumn{1}{c}{\textbf{Permutations}} & \multicolumn{1}{c}{\textbf{Excursions}} \\ 
        \multicolumn{1}{c}{} &
        \multicolumn{1}{c}{} &
        \multicolumn{1}{c}{(FWER)} &  \multicolumn{1}{c}{(FWER)} & \multicolumn{1}{c}{} \\ \cline{2-5}
        \rotatebox[origin=c]{90}{\textbf{Subject A}\,} &
        \Includegraphics[width=.22\textwidth]{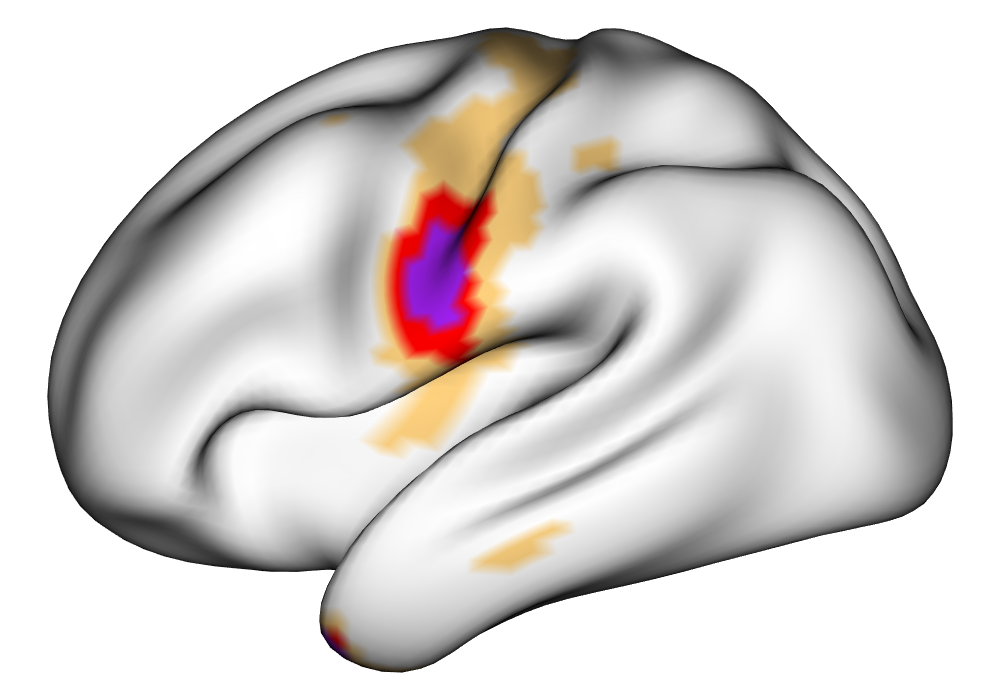} &
        \Includegraphics[width=0.22\textwidth, trim=0 125mm 170mm 0, clip]{607_classical_103818_visit1_tongue_activations.png} &
        \Includegraphics[width=.22\textwidth]{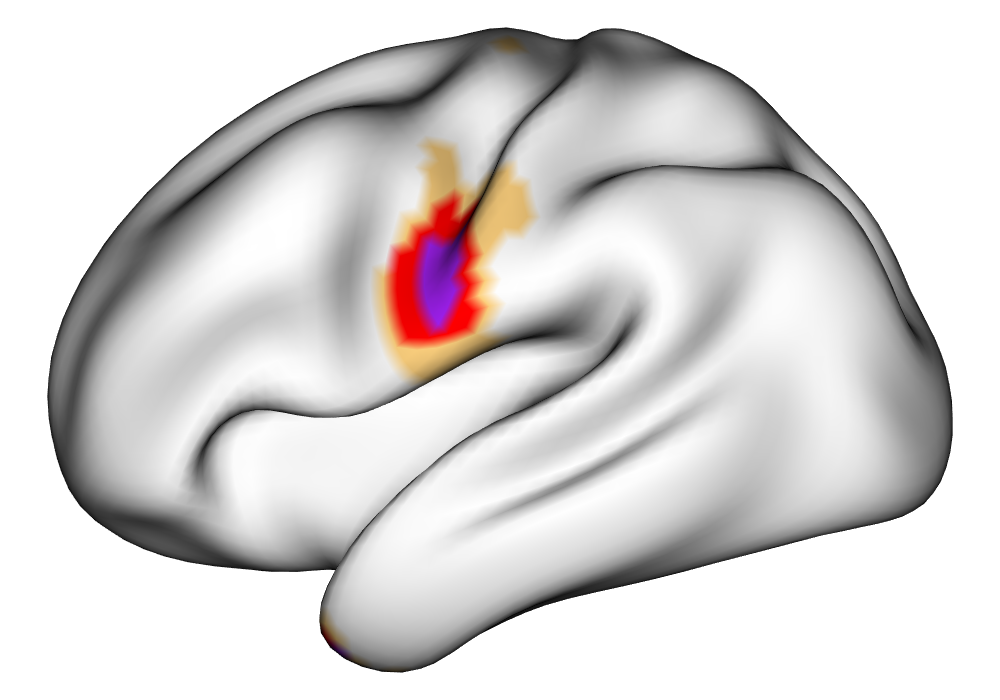} &
        \Includegraphics[width=0.22\textwidth, trim=0 125mm 170mm 0, clip]{607_bayes_103818_visit1_tongue_activations.png} \\ \cline{2-5}
        \rotatebox[origin=c]{90}{\textbf{Subject B}\,} &
        \Includegraphics[width=.22\textwidth]{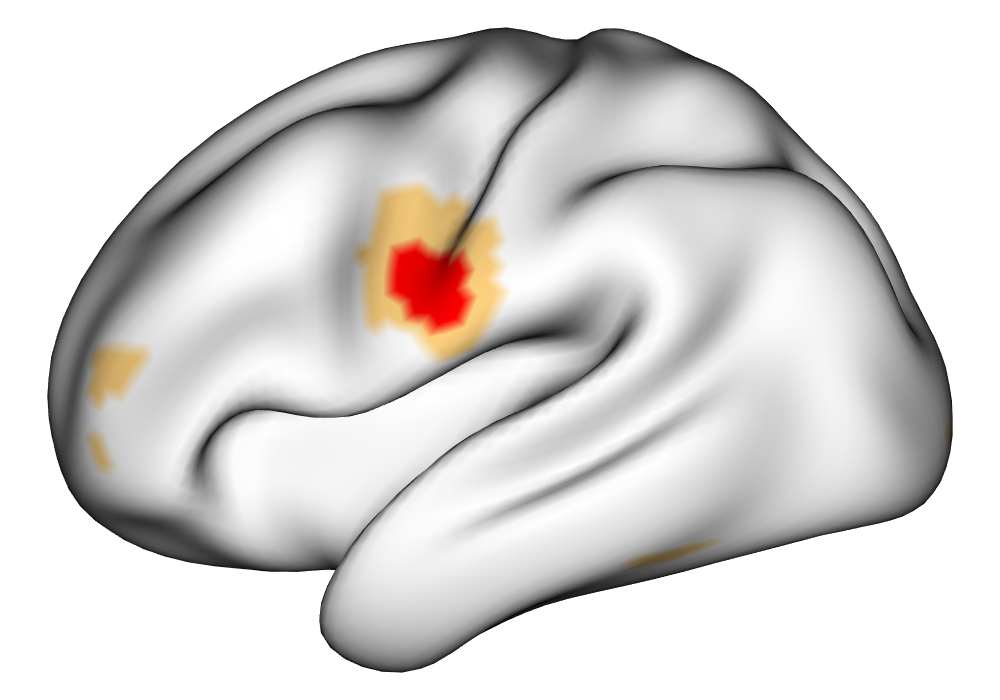} &
        \Includegraphics[width=0.22\textwidth, trim=0 125mm 170mm 0, clip]{607_classical_105923_visit1_tongue_activations.png} &
        \Includegraphics[width=.22\textwidth]{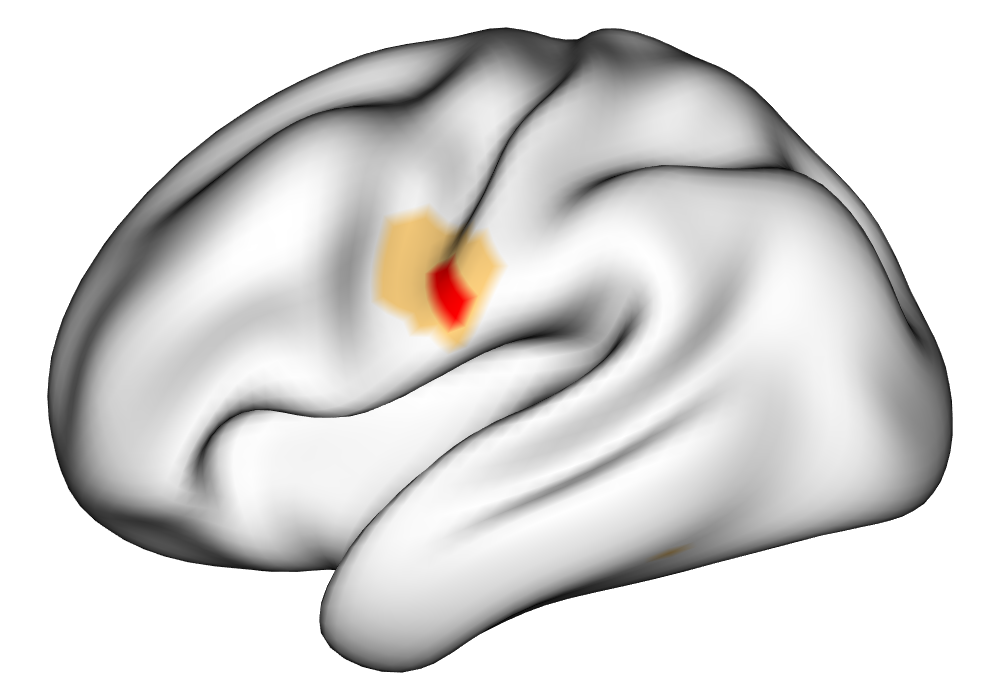} &
        \Includegraphics[width=0.22\textwidth, trim=0 125mm 170mm 0, clip]{607_bayes_105923_visit1_tongue_activations.png} \\ \cline{2-5}
        \rotatebox[origin=c]{90}{\textbf{Subject C}\,} &
        \Includegraphics[width=.22\textwidth]{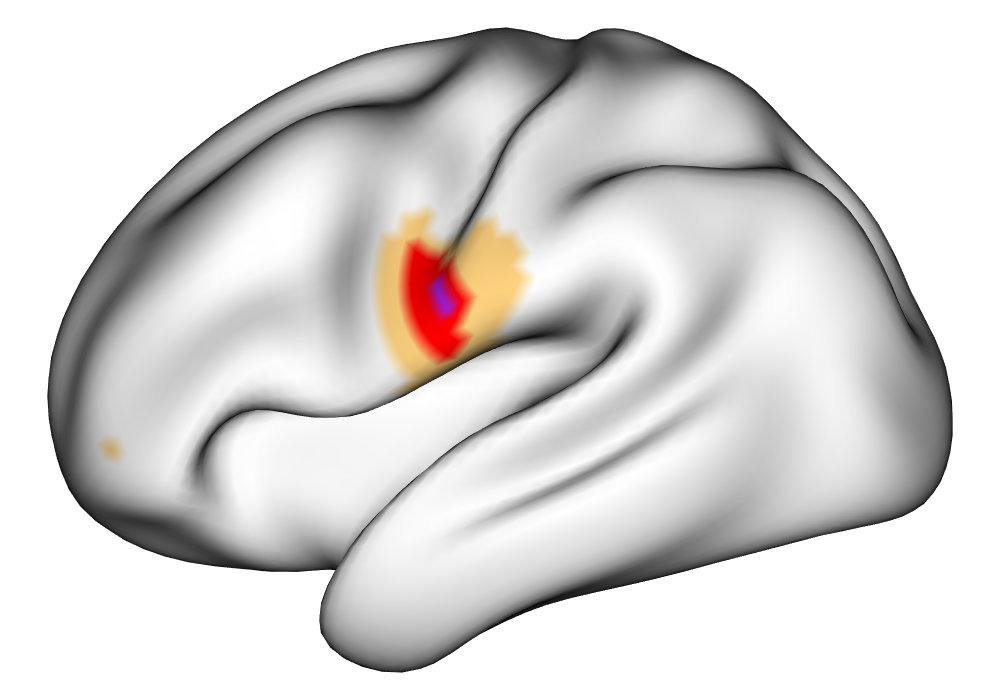} &
        \Includegraphics[width=0.22\textwidth, trim=0 125mm 170mm 0, clip]{607_classical_114823_visit1_tongue_activations.png} &
        \Includegraphics[width=.22\textwidth]{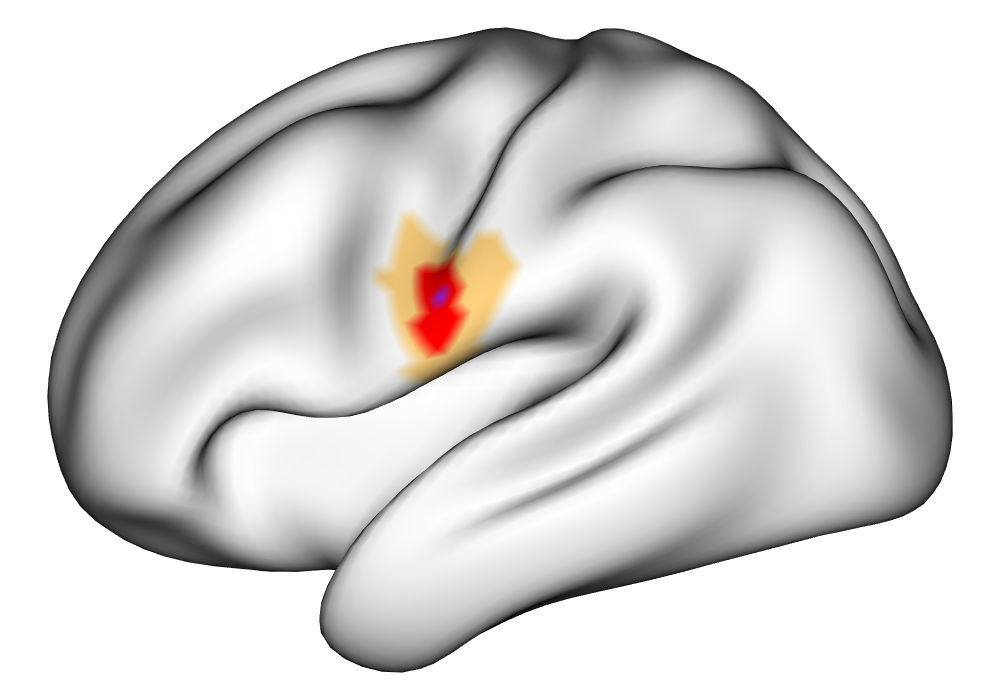} &
        \Includegraphics[width=0.22\textwidth, trim=0 125mm 170mm 0, clip]{607_bayes_114823_visit1_tongue_activations.png} \\ \cline{2-5}
        \multicolumn{1}{c}{} &
        \multicolumn{4}{c}{$\gamma =$ \textcolor[HTML]{FFD27F}{$\blacksquare$} 0\% 
           \textcolor[HTML]{FF0000}{$\blacksquare$} 0.5\% 
           \textcolor[HTML]{A020F0}{$\blacksquare$} 1\%}
    \end{tabularx}
    \caption{A comparison of the activations identified using different multiple comparisons correction methods for the tongue task in a single subject using cs-fMRI data smoothed with a Gaussian kernel with FWHM = 6mm. The Bayesian activations found using the excursions method found using unsmoothed data are shown as a comparison. Bonferroni correction and permutation testing produced similar results, with permutation testing being slightly more conservative.}
    \label{fig:determining_activations}
\end{figure}

In order to compare activations detected between the classical GLM and the Bayesian GLM, the multiplicity correction method to use in the classical GLM is an important question. Here we compare three popular methods: Bonferroni correction, the Benjamini-Hochverg procedure to control the FDR, and nonparametric permutation testing. Both the Bonferroni multiple testing method and the permutations method control the FWER, which is the probability of at least one false positive.  Each method is based on first performing a $t$-test at each vertex.  The test statistic at location $v$ for task $k$ is $t_{v,k}^* = (\hat{\beta}_{v,k} - \gamma) / \text{SE}(\hat{\beta})$. Uncorrected p-values $p_{v,k}$ are based on the $t$ distribution. In the Bonferroni correction, p-values are multiplied by the number of locations $V$ to produce corrected p-values $p_{v,k}^{\text{Bonferroni}}$. These corrected p-values are then compared to a significance level $\alpha$, where any locations in which $p_{v,k}^{\text{Bonferroni}} < \alpha$ are determined to be active. The Benjamini-Hochberg procedure first orders the $V$ p-values and determines a location to be active if $\frac{\ell}{V}p_{v,k} < \alpha$ for the $\ell$th lowest p-value. The nonparametric permutation testing method first randomly reorders the (prewhitened) BOLD time series to create $M$ reordered time series, $\mathbf{y}_{v,m}$, for $m = 1,\ldots,M$, with $M$ large. Null hypothesis test statistics $t_{v,k,m}^{\text{null}}$ are found for each location and task for each reordering, which induces no relationship on average between the BOLD signal and the task paradigm. Next, the maximum null test statistic across locations, $t_{k,m}^{\text{null},\text{max}}$, is found for each reordering, and then the ($1 - \alpha$) percentile is found across all $t_{k,m}^{\text{null},\text{max}}$ to produce a test statistic threshold $t_{k}^{\text{threshold}}$ for task $k$. Any location $v$ where $t_{v,k}^* > t_{k}^{\text{threshold}}$ is considered to be active. \textbf{Appendix Figure \ref{fig:determining_activations}} shows the activation map for a single subject for the tongue task under these three multiple corrections methods. We find that the permutations method is slightly more conservative than the Bonferroni method. As expected, the Benjamini-Hochberg method is much less conservative, as it controls the FDR. Since Bonferroni correction was generally not more conservative than permutation testing, and since FWER control is analogous to the excursions method used in the spatial Bayesian GLM, we adopt Bonferroni correction to identify activations in the classical GLM.

\section{Effect of resampling and smoothing on results}
\label{app:resampling_smoothing}

The spatial Bayesian GLM presented here is generally applied to data that has been resampled (interpolated) to reduce computational demands.  This comes with a tradeoff in terms of resolution of the inference on the latent task activation fields. However, it is important to note that typical levels of spatial smoothing actually result in greater interpolation (and hence more loss of fine spatial detail) than resampling, as illustrated by \cite{mejia2020bayesian}. Since in the spatial Bayesian GLM presented here we perform resampling but no smoothing, finer details are preserved in the underlying data relative to the traditional classical GLM approach, which is based on full-resolution but smoothed data.

\begin{figure}[H]
    \centering
    \begin{tabularx}{\textwidth}{c|X|X|X|}
        \multicolumn{1}{c}{} & \multicolumn{1}{c}{\textbf{Bayesian}} & \multicolumn{1}{c}{\textbf{Classical}} & \multicolumn{1}{c}{\textbf{Classical (Smoothed)}} \\
        \cline{2-4}
        \rotatebox[origin=c]{90}{\textbf{5k Vertices}} &
        \Includegraphics[width=0.3\textwidth]{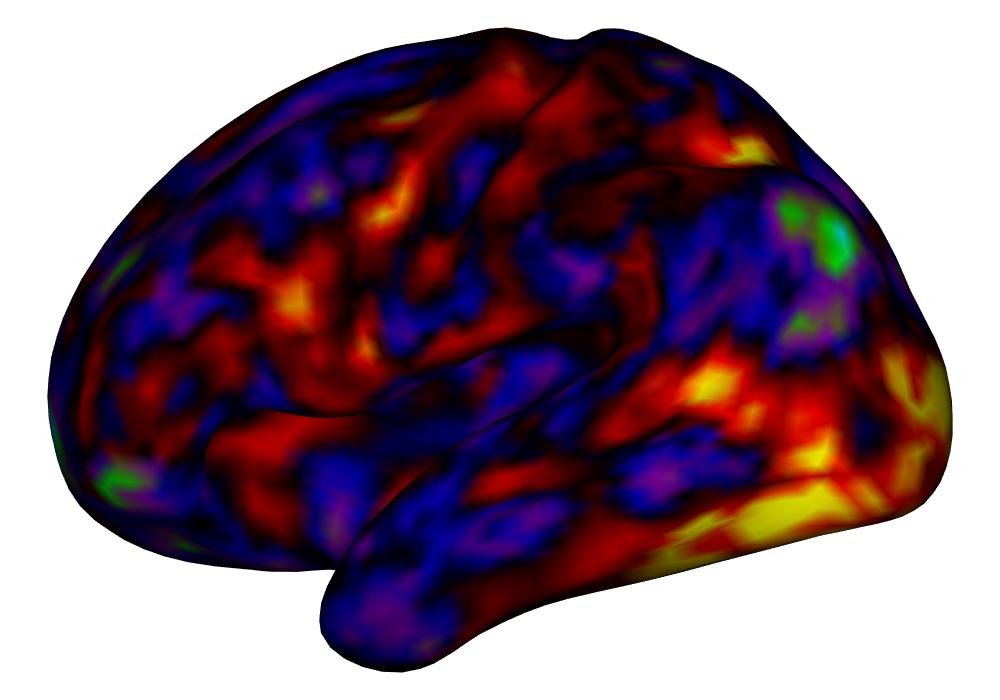} &
        \Includegraphics[width=0.3\textwidth]{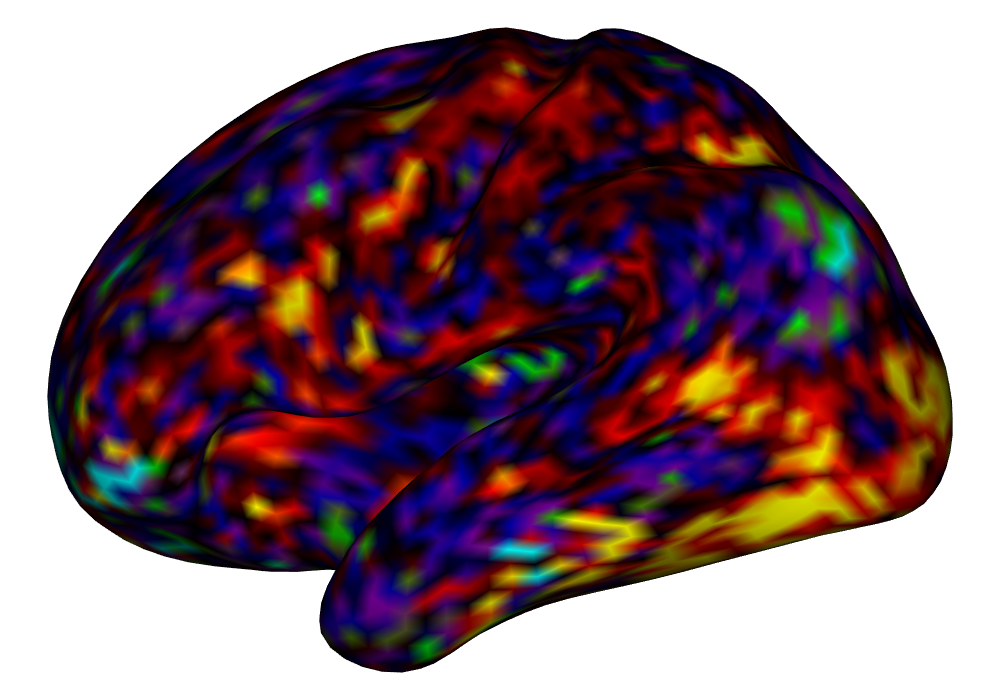} &
        \Includegraphics[width=0.3\textwidth]{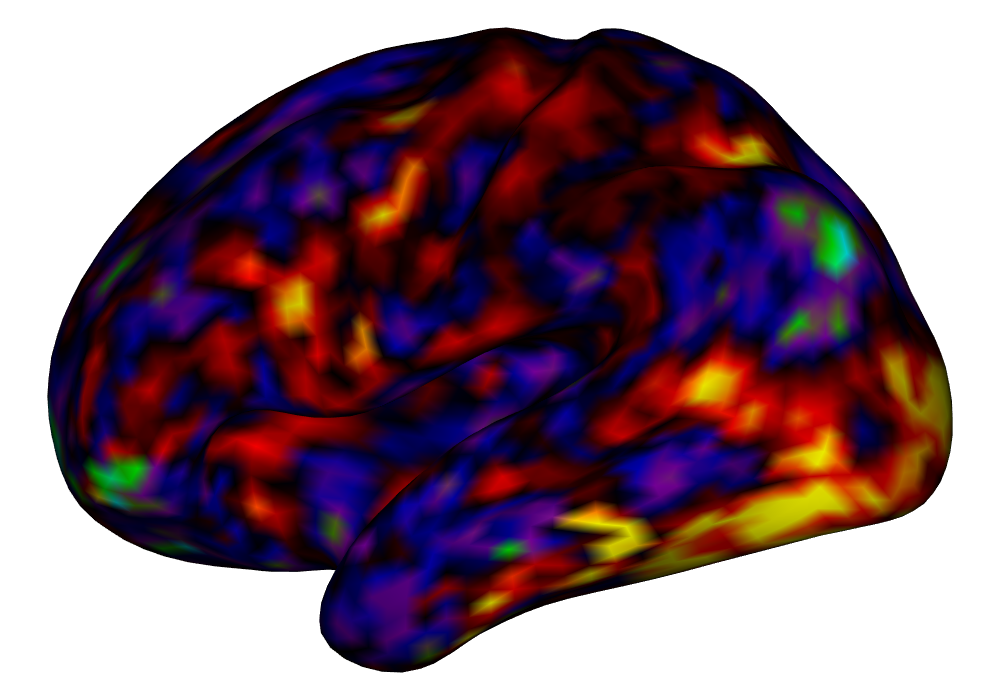} \\
        \cline{2-4}
        \rotatebox[origin=c]{90}{\textbf{32k Vertices}} &
        \Includegraphics[width=0.3\textwidth]{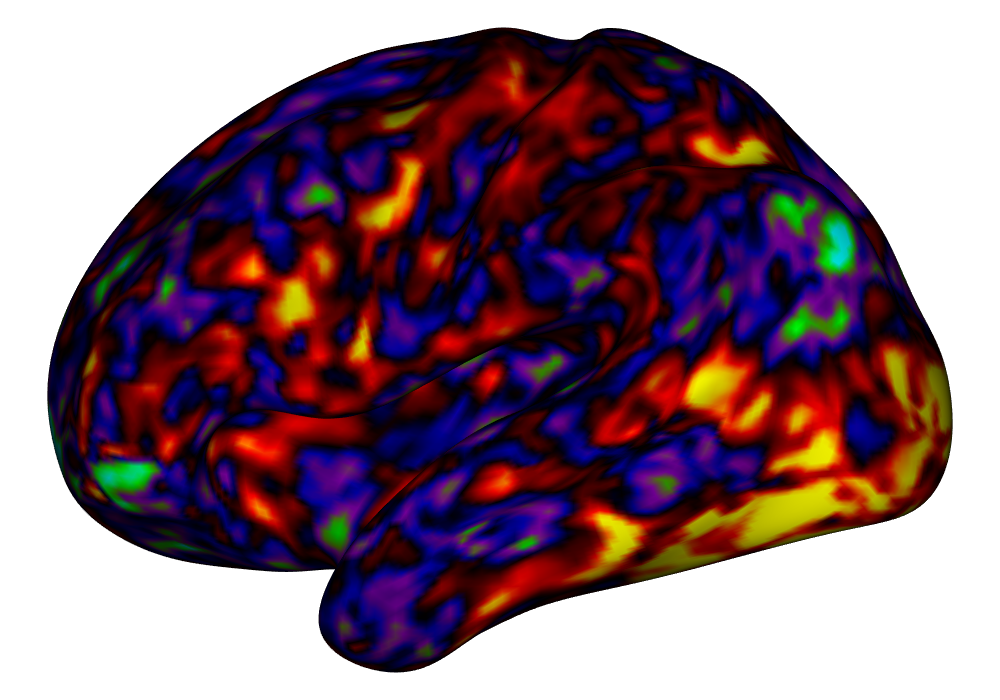} &
        \Includegraphics[width=0.3\textwidth]{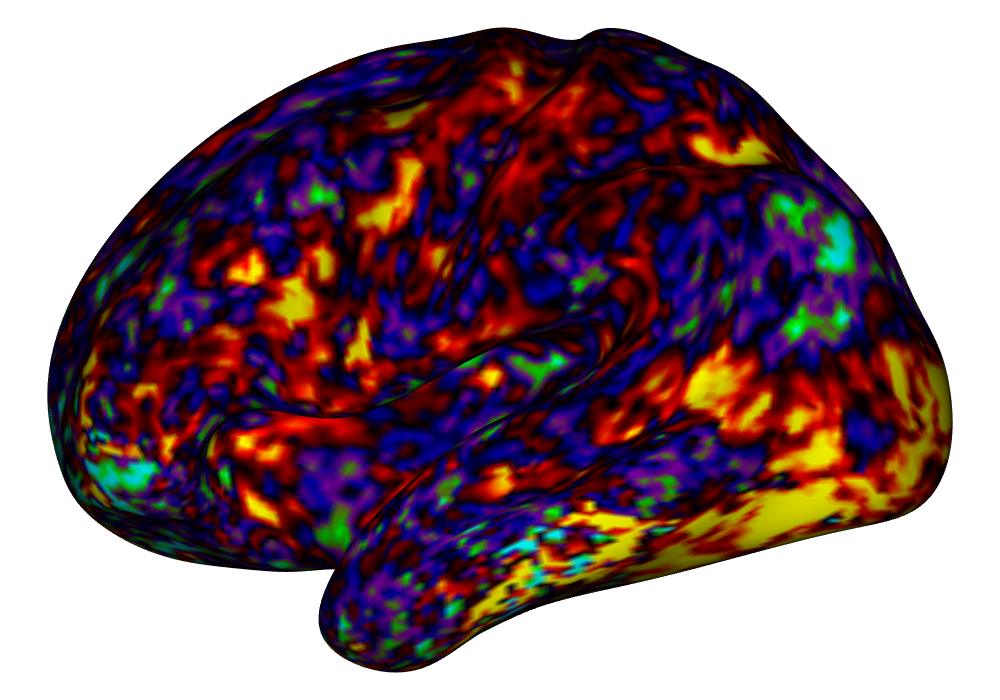} &
        \Includegraphics[width=0.3\textwidth]{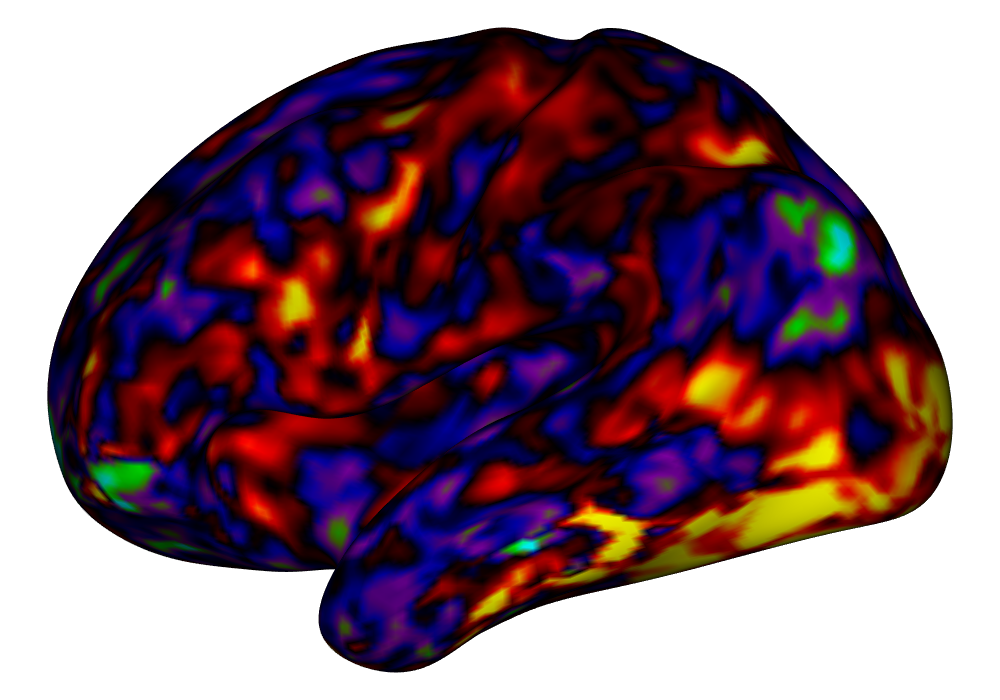} \\
        \cline{2-4}
        \multicolumn{1}{c}{} &
        \multicolumn{3}{c}{\Includegraphics[width = 0.4\textwidth]{607_legend_estimate.png}}
    \end{tabularx}
    \caption{A subject-level comparison of the results of the gambling task analysis at different resolutions. Smoothing was performed using a Gaussian kernel with a full-width half-maximum}
    \label{fig:compare_smooth_resample}
\end{figure}

In order to investigate the effects of resampling, we examine the HCP Gambling task. While there are three tasks in the experiment (for winning, losing, and neutral events), all three events show similar activation patterns \citep{barch2013function}. Thus, combining the three tasks into a single ``gambling event" task can be used to infer which parts of the brain are associated with participation in a gambling task. To ease the computational burden, only one run is considered in this analysis. Smoothing done for the classical analysis was performed using a Gaussian kernel with a full-width half-maximum (FWHM) of 6mm. The results of these analyses can be seen in \textbf{Figure \ref{fig:compare_smooth_resample}}. Resampling indeed produces a smoothing effect on the estimates for both the classical and the Bayesian GLM.  In the classical GLM, the effect of smoothing (32K) is more dramatic than the effect of resampling (5K), which illustrates that the resampling we perform results in less interpolation than standard spatial smoothing.  Considering the Bayesian estimates, while the resampled 5K results are smoother than the full-resolution 32K results, the primary areas of activation remain well-preserved.

In conclusion, resampling provides important computational advantages, allowing the SBSB GLM to be fit when there are more tasks, and longer and/or multiple runs.  However, the amount of resampling required will vary based on computational resources, the number of tasks, and the duration and quantity of runs per subject.  The amount of resampling should therefore be minimized to preserve the maximum amount of spatial detail.

\begin{figure}[H]
    \centering
    \includegraphics[width=0.8\textwidth]{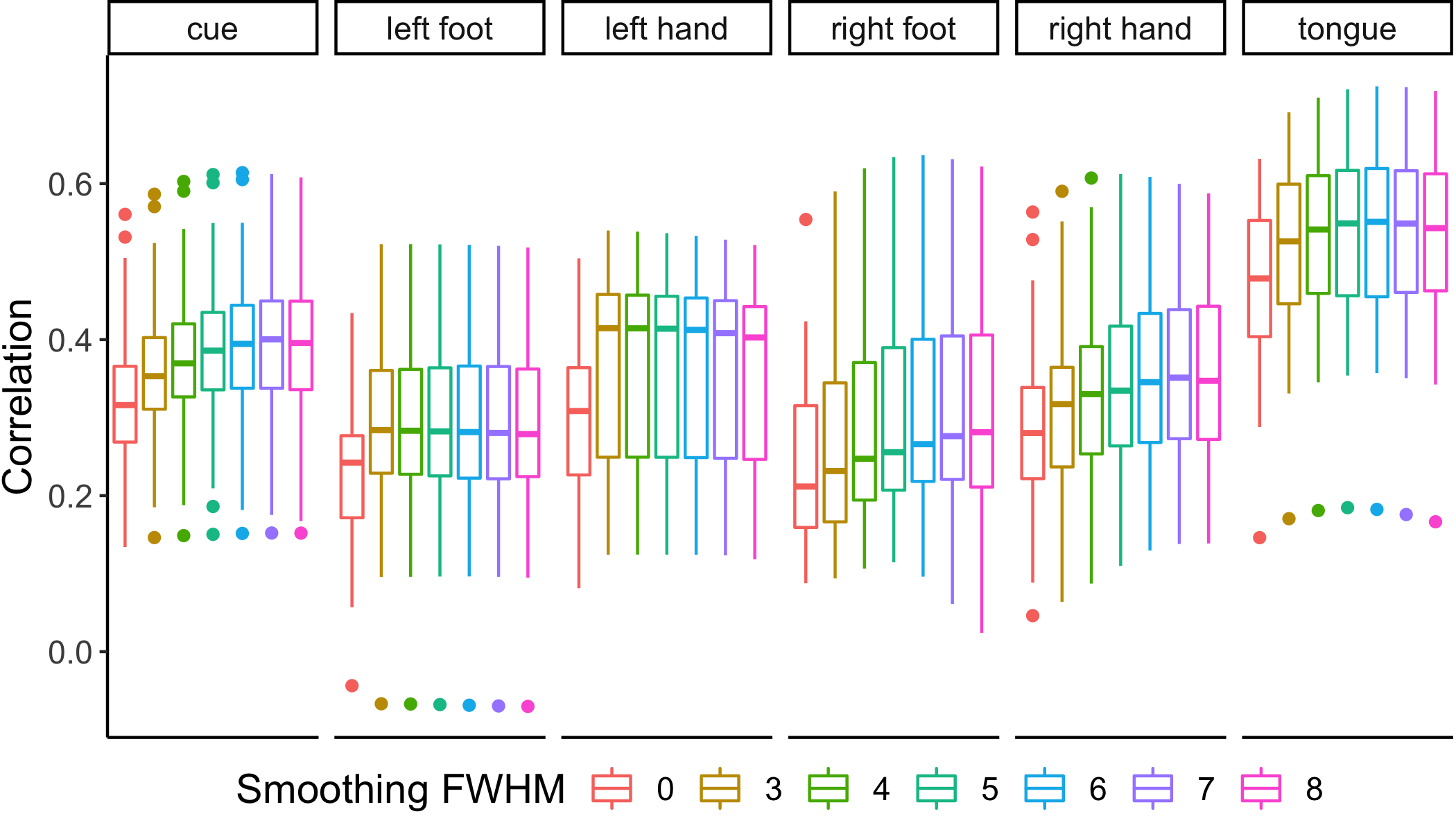}
    \caption{\textbf{Comparison of the test-retest correlations for the classical results.} Each of the different smoothing kernel full-width half-maxima (FWHM) in millimeters data were fit using the classical GLM for all 45 subjects using both sessions of the first visit data. The correlations with the estimates from the classical GLM using unsmoothed data from the second visit across all tasks are used as a basis of comparison to see which smoothing kernel achieved the optimal balance between smoothing out spurious activation signals and reducing the peak estimates from truly active signals. }
    \label{fig:smooth_cor_plot}
\end{figure}

\textbf{Figure \ref{fig:smooth_cor_plot}} shows the test-retest reliability of subject-level classical GLM estimates using different smoothing kernels.  Though the optimal degree of smoothing varies by task, a 6mm FWHM kernel is near optimal across all tasks.  Therefore, we adopt this smoothing level in the classical GLM analyses presented in the paper.
    
\section{Additional reliability analysis figures}
\label{app:additional_results_figures}

\begin{figure}[H]
	\begin{tabularx}{\textwidth}{c|X|X|}
		\multicolumn{1}{c}{} & \multicolumn{1}{c}{\textbf{Classical GLM}} & \multicolumn{1}{c}{\textbf{Bayesian GLM}}  \\ \cline{2-3}
		\rotatebox[origin=c]{90}{\textbf{Subject A}} & \Includegraphics[width=0.45\textwidth, trim=0 125mm 0 0, clip]{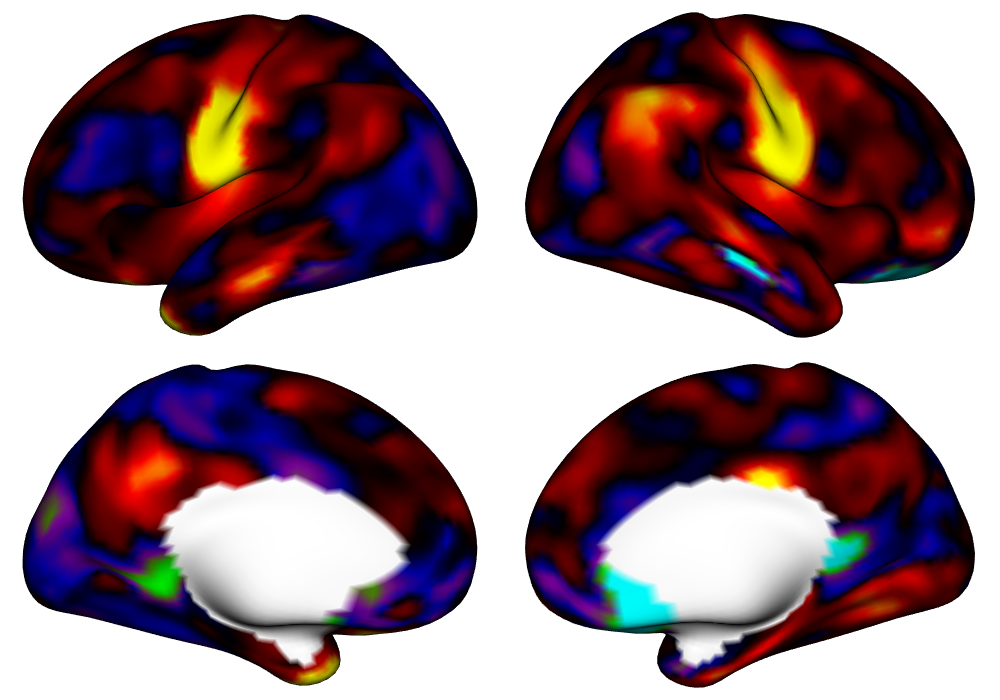} &
		\Includegraphics[width=0.45\textwidth, trim=0 125mm 0 0, clip]{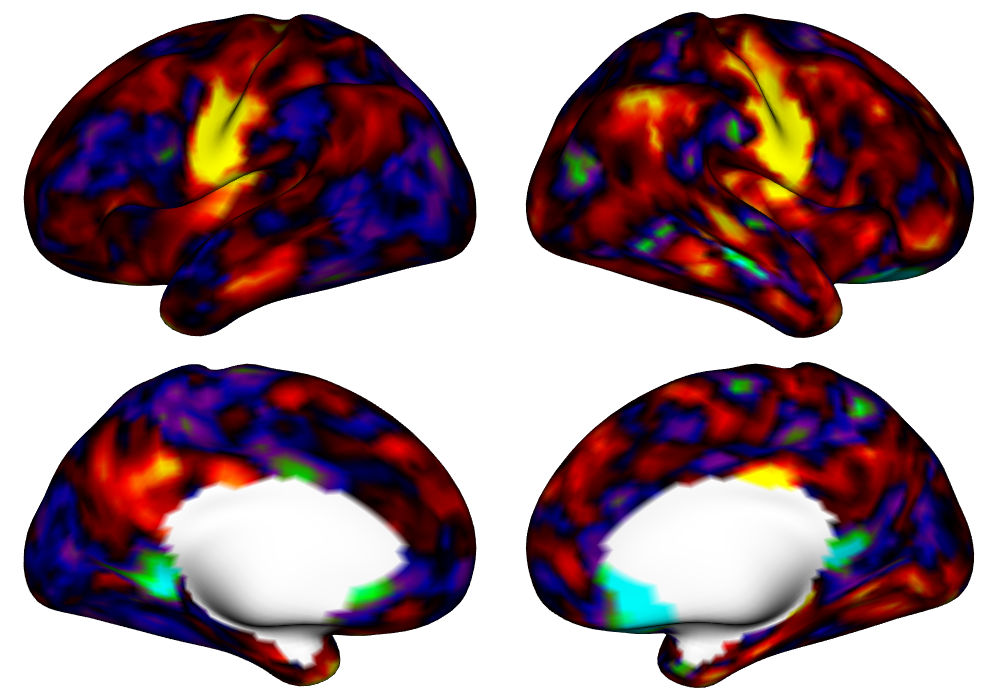} \\ \cline{2-3}
		\rotatebox[origin=c]{90}{\textbf{Subject B}} & \Includegraphics[width=0.45\textwidth, trim=0 125mm 0 0, clip]{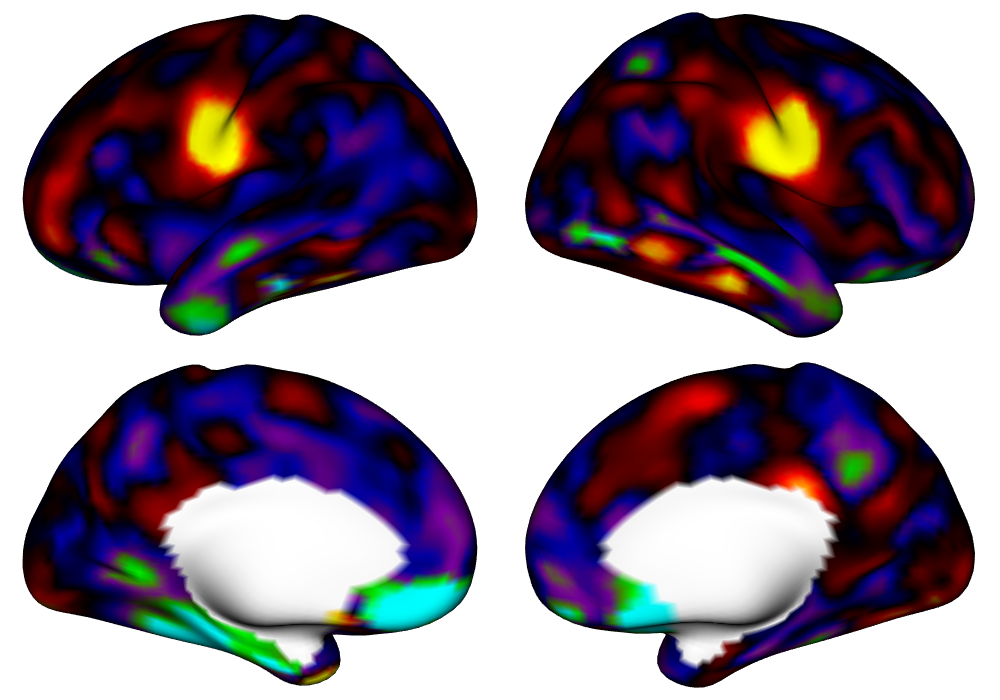} &
		\Includegraphics[width=0.45\textwidth, trim=0 125mm 0 0, clip]{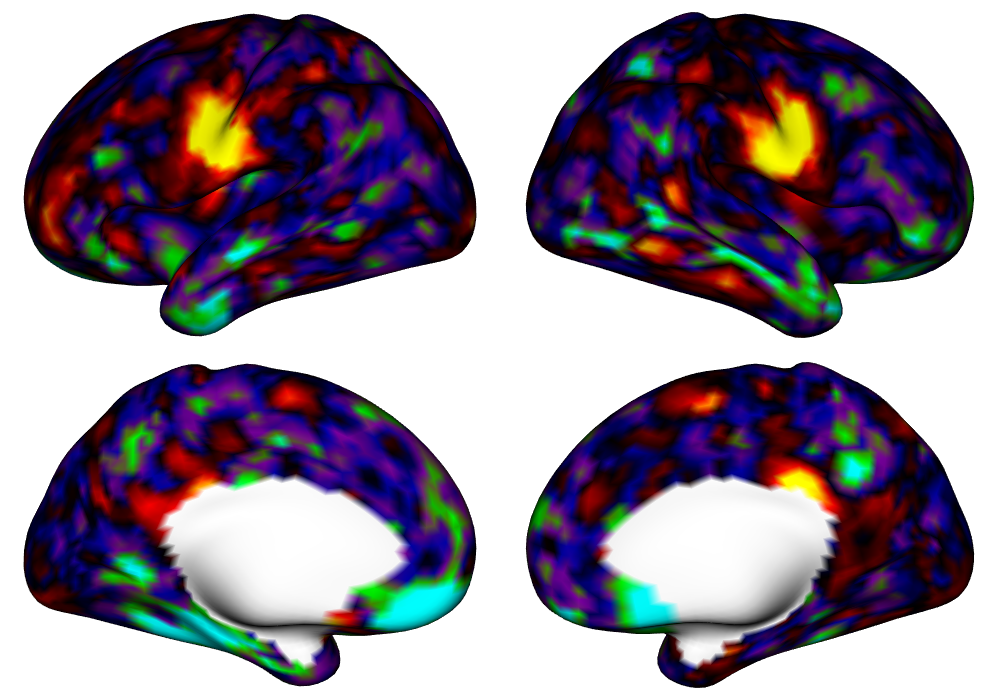} \\ \cline{2-3}
		\rotatebox[origin=c]{90}{\textbf{Subject C}} & \Includegraphics[width=0.45\textwidth, trim=0 125mm 0 0, clip]{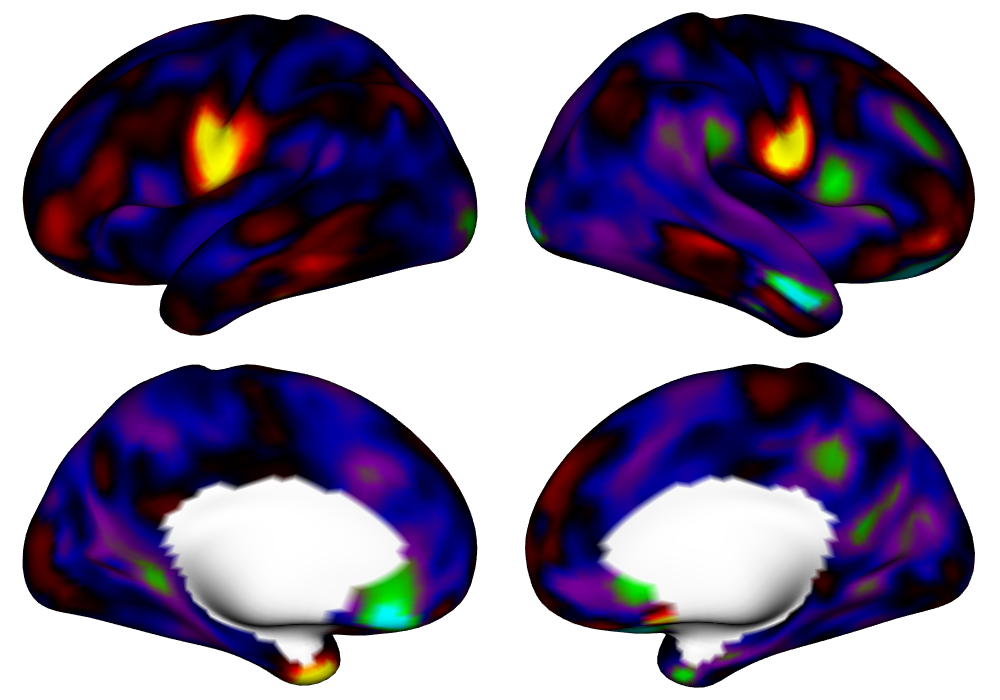} &
		\Includegraphics[width=0.45\textwidth, trim=0 125mm 0 0, clip]{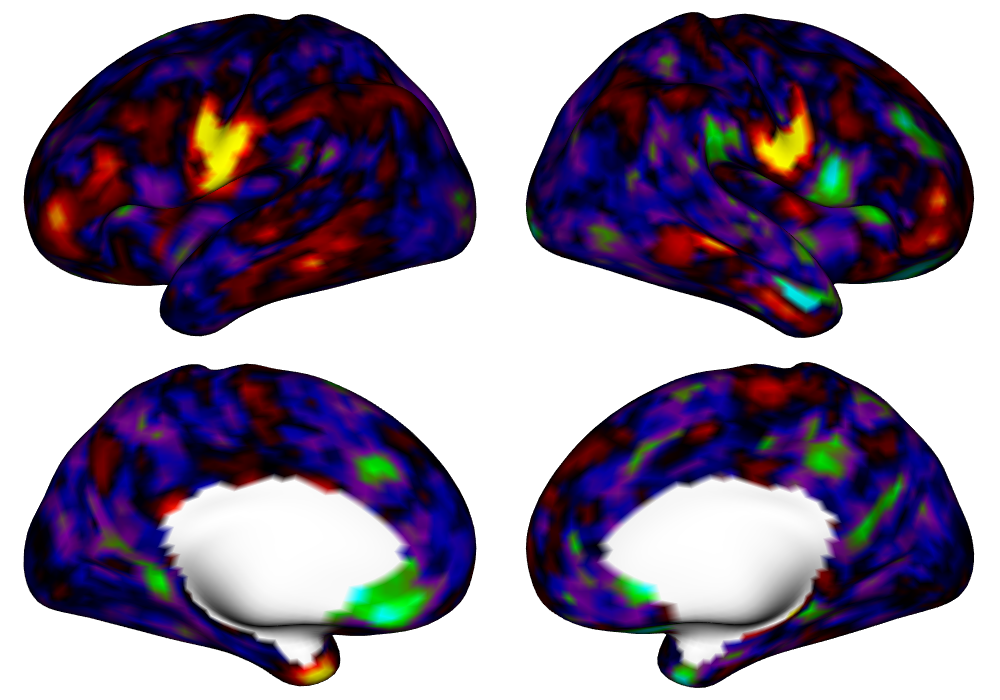} \\ \cline{2-3}
	\end{tabularx}
\caption{Single run (LR) amplitude estimation results for subjects A, B, and C.}
\label{fig:sessionLR_subjectA_estimates}
\end{figure}

\begin{figure}[H]
    \centering
	\begin{tabularx}{0.9\textwidth}{c|X|X|X|X|}
		\multicolumn{1}{c}{} & \multicolumn{2}{c}{\textbf{Classical GLM}} & \multicolumn{2}{c}{\textbf{Bayesian GLM}}  \\ 
		\cline{2-5}
		\rotatebox[origin=l]{90}{\qquad \qquad \textbf{Visual Cue}} & \multicolumn{2}{c|}{\Includegraphics[width=0.405\textwidth]{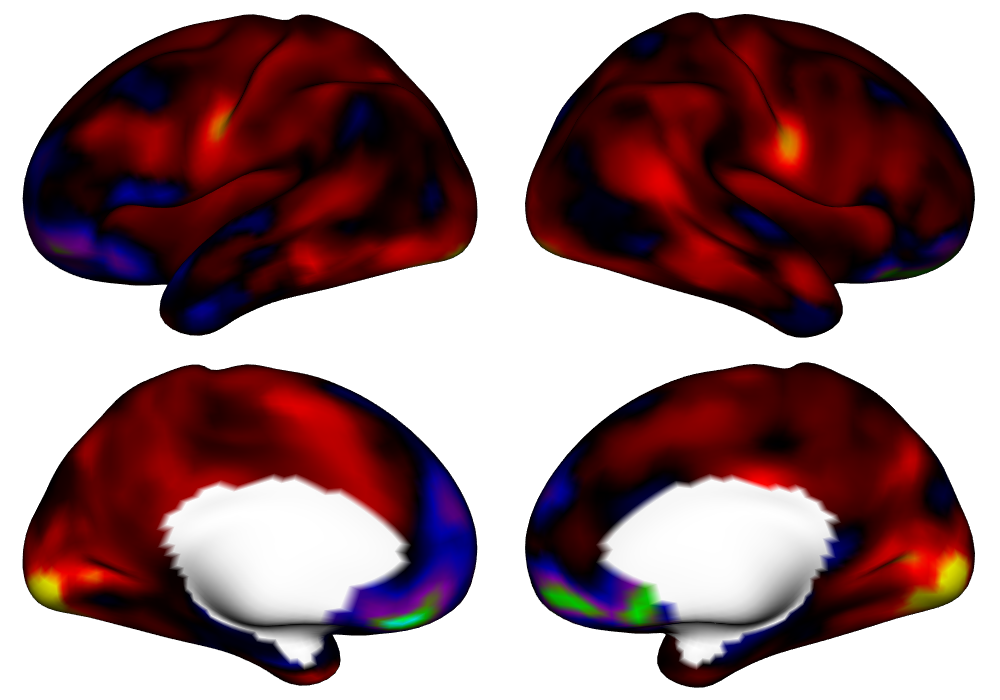}} &
		\multicolumn{2}{c|}{\Includegraphics[width=0.405\textwidth]{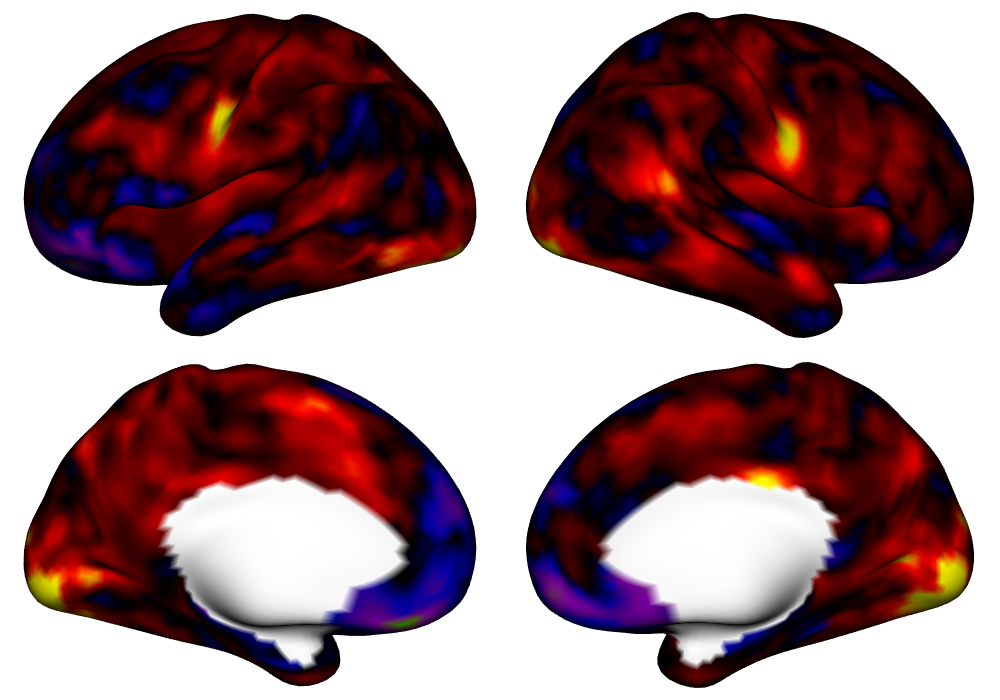}} \\ \cline{2-5}
		\rotatebox[origin=l]{90}{\qquad \qquad \quad \textbf{Tongue}} &
		\multicolumn{2}{c|}{\Includegraphics[width=0.405\textwidth]{607_classical_103818_visit1_sessionLR_tongue_estimate.png}} &
		\multicolumn{2}{c|}{\Includegraphics[width=0.405\textwidth]{607_bayes_103818_visit1_sessionLR_tongue_estimate.png}} \\ \cline{2-5}
		\rotatebox[origin=c]{90}{\textbf{Foot}} & \Includegraphics[width=0.198\textwidth]{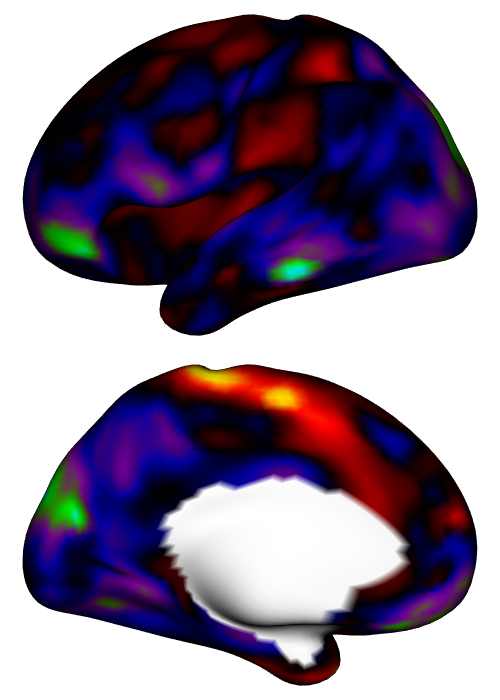} &
		\Includegraphics[width=0.198\textwidth]{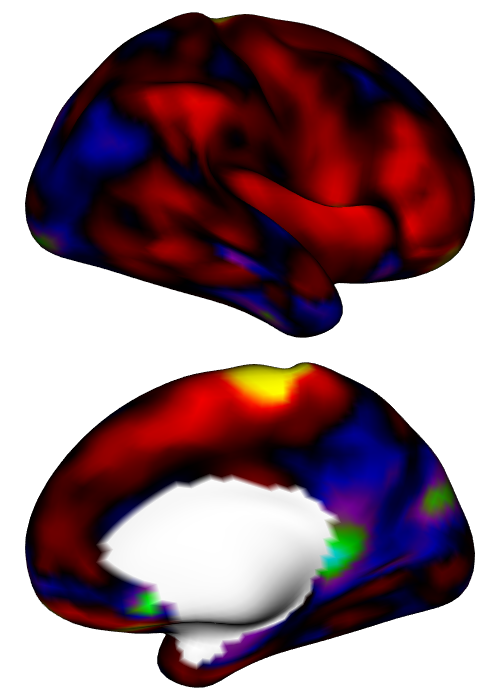} &
		\Includegraphics[width=0.198\textwidth]{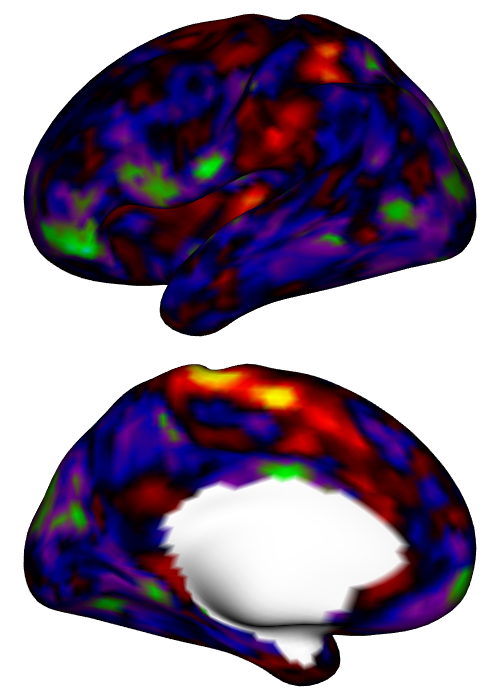} &
		\Includegraphics[width=0.198\textwidth]{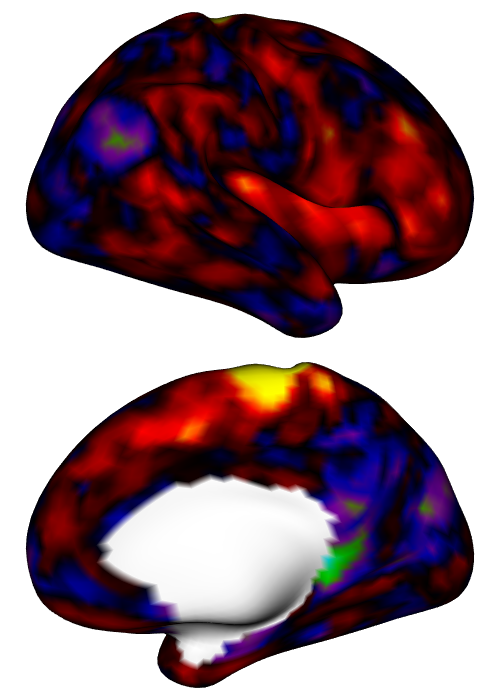} \\ 
		\cline{2-5}
		\rotatebox[origin=c]{90}{\textbf{Hand}} & 
		\Includegraphics[width=0.198\textwidth]{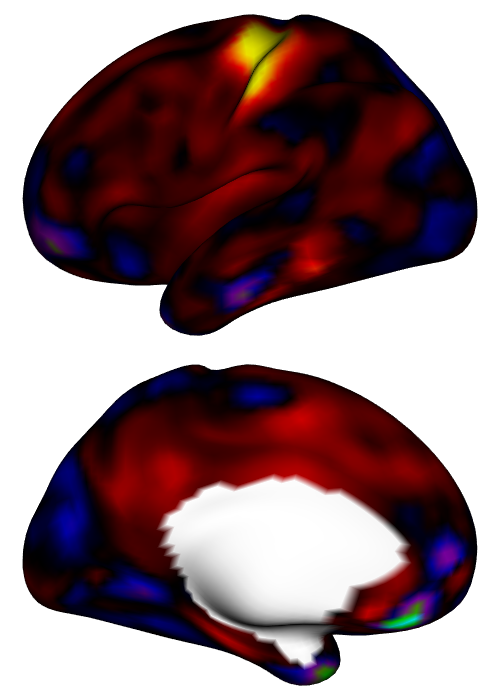} &
		\Includegraphics[width=0.198\textwidth]{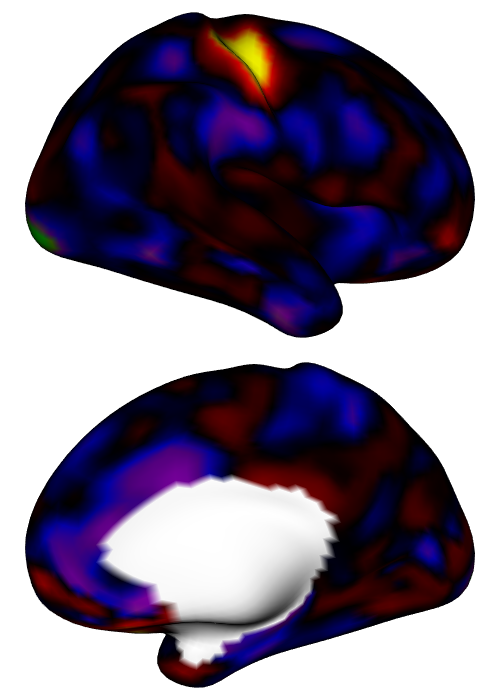} &
		\Includegraphics[width=0.198\textwidth]{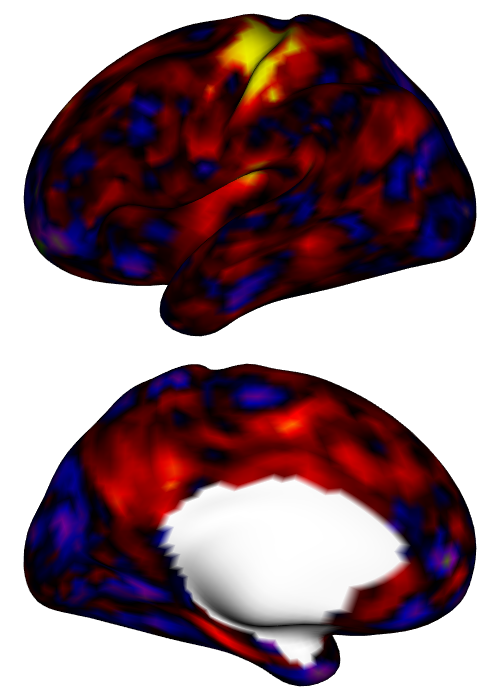} &
		\Includegraphics[width=0.198\textwidth]{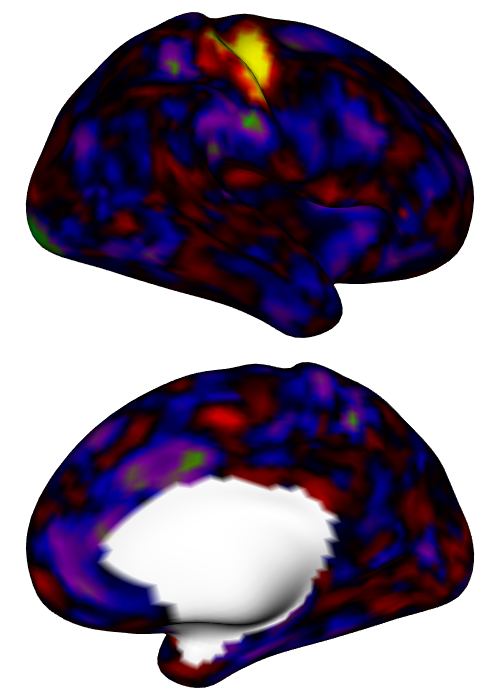} \\ 
		\cline{2-5}
		\multicolumn{1}{c}{} & \multicolumn{4}{c}{\Includegraphics[width = 0.4\textwidth]{607_legend_estimate.png}}
	\end{tabularx}
	\caption{Subject-level estimates of activation for each motor task and the visual cue, in units of local percent signal change, based on a single run (LR). For lateral tasks, only the contralateral hemisphere is displayed.}
	\label{fig:alltasks_subj_est_LR}
\end{figure}

\begin{figure}[H]
    \centering
	\begin{tabularx}{0.9\textwidth}{c|X|X|X|X|}
		\multicolumn{1}{c}{} & \multicolumn{2}{c}{\textbf{Classical GLM}} & \multicolumn{2}{c}{\textbf{Bayesian GLM}}  \\ 
		\cline{2-5}
		\rotatebox[origin=l]{90}{\qquad \qquad \textbf{Visual Cue}} & \multicolumn{2}{c|}{\Includegraphics[width=0.405\textwidth]{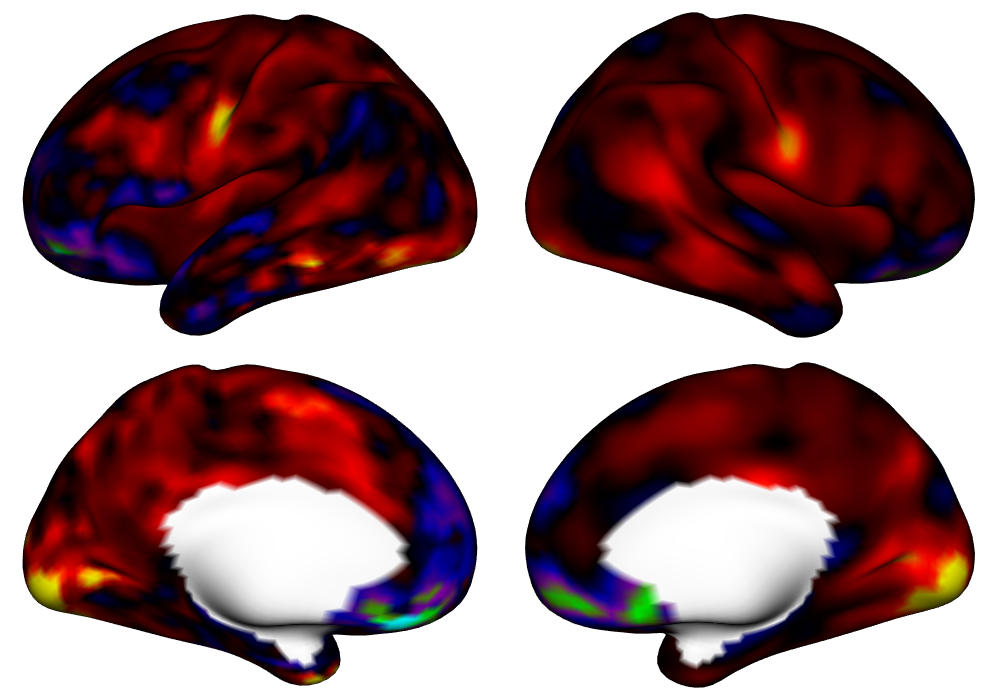}} &
		\multicolumn{2}{c|}{\Includegraphics[width=0.405\textwidth]{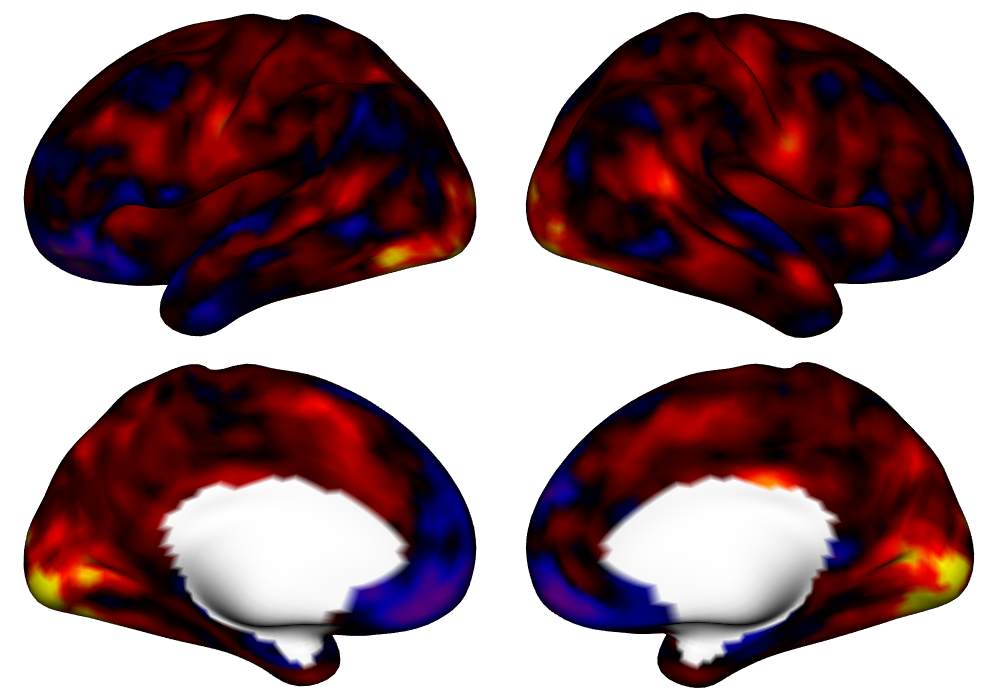}} \\ \cline{2-5}
		\rotatebox[origin=l]{90}{\qquad \qquad \quad \textbf{Tongue}} &
		\multicolumn{2}{c|}{\Includegraphics[width=0.405\textwidth]{600_subject_103818_tongue_classical_estimates.png}} &
		\multicolumn{2}{c|}{\Includegraphics[width=0.405\textwidth]{600_subject_103818_tongue_estimates.png}} \\ \cline{2-5}
		\rotatebox[origin=c]{90}{\textbf{Foot}} & \Includegraphics[width=0.198\textwidth]{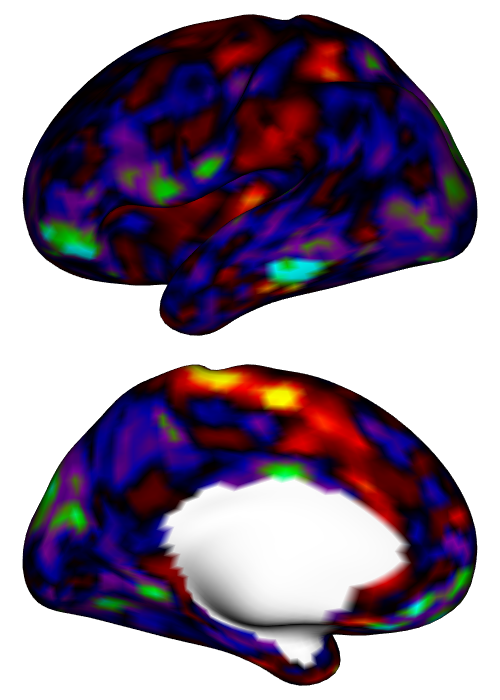} &
		\Includegraphics[width=0.198\textwidth]{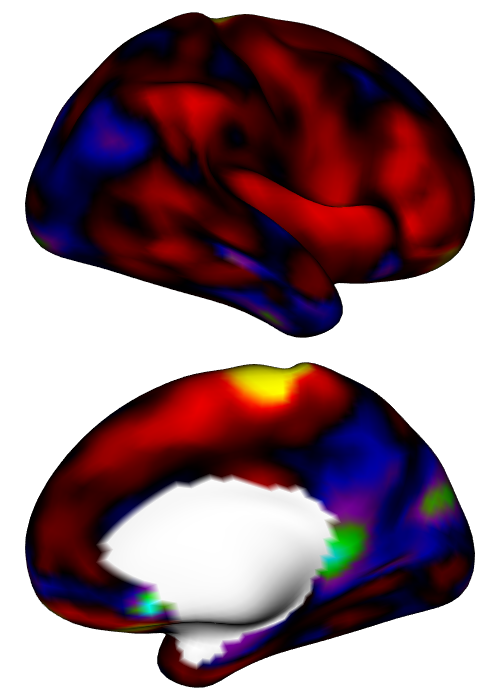} &
		\Includegraphics[width=0.198\textwidth]{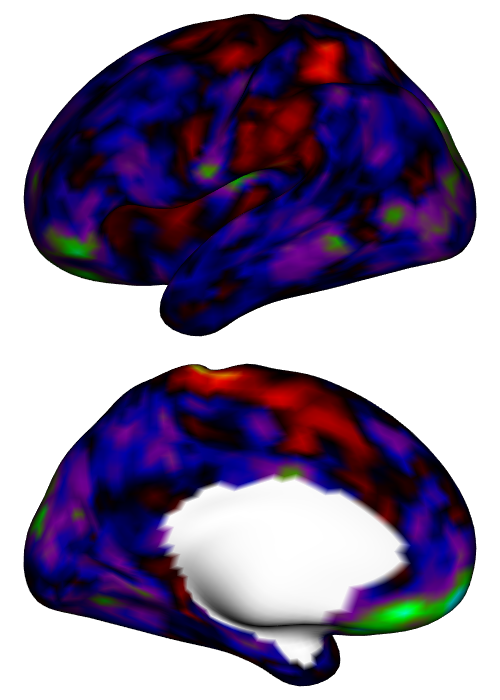} &
		\Includegraphics[width=0.198\textwidth]{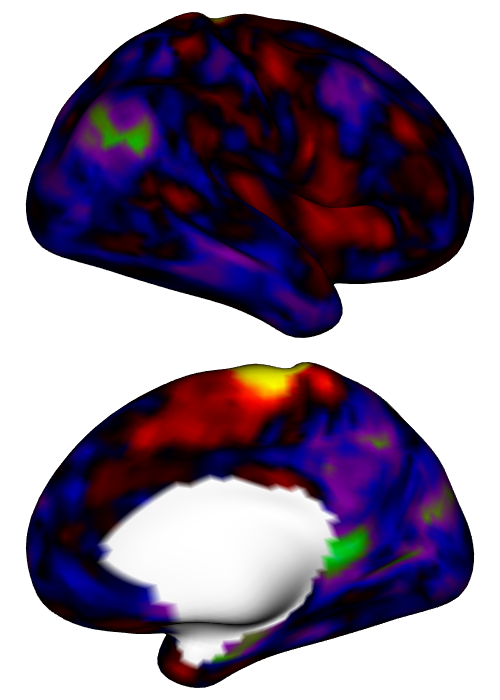} \\ 
		\cline{2-5}
		\rotatebox[origin=c]{90}{\textbf{Hand}} & 
		\Includegraphics[width=0.198\textwidth]{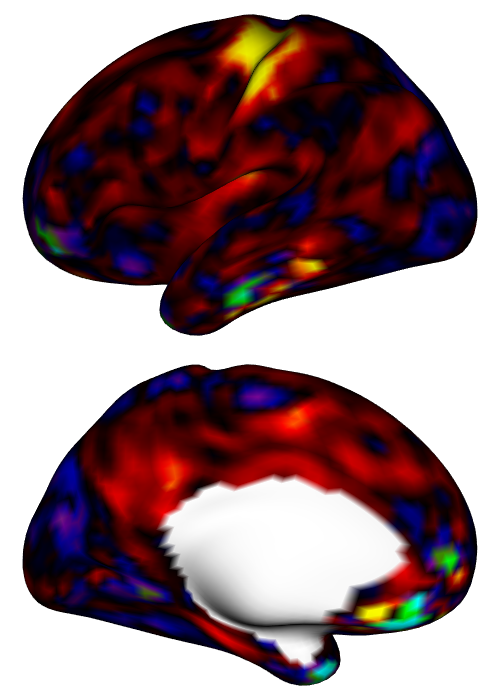} &
		\Includegraphics[width=0.198\textwidth]{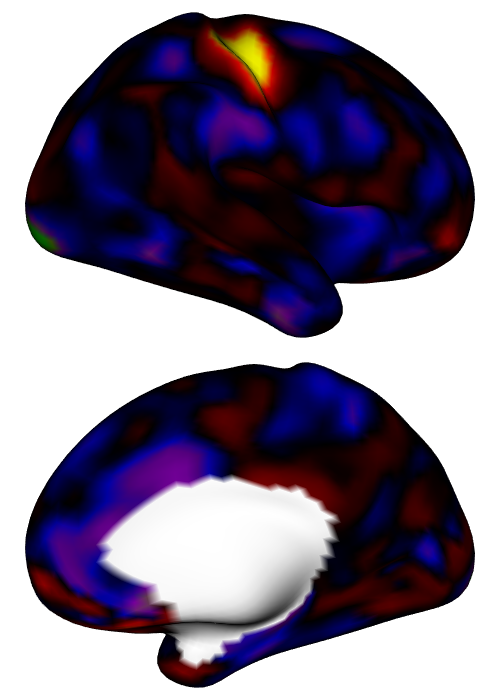} &
		\Includegraphics[width=0.198\textwidth]{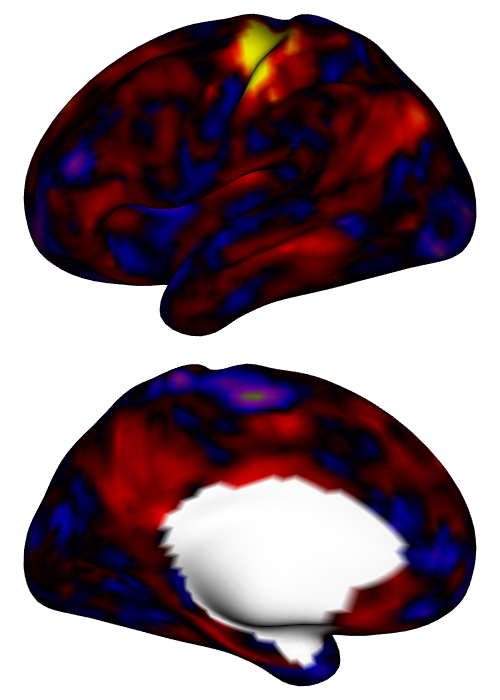} &
		\Includegraphics[width=0.198\textwidth]{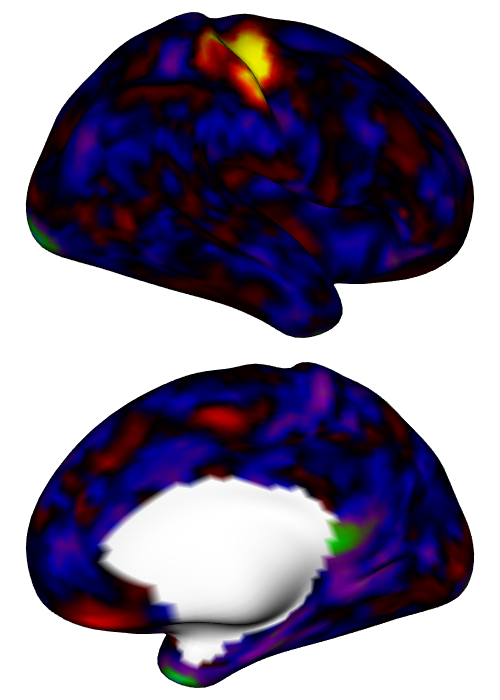} \\ 
		\cline{2-5}
		\multicolumn{1}{c}{} & \multicolumn{4}{c}{	\Includegraphics[width = 0.4\textwidth]{607_legend_estimate.png}}
	\end{tabularx}
	\caption{Subject-level estimates of activation for each motor task and the visual cue, in units of local percent signal change. For lateral tasks, only the contralateral hemisphere is displayed.}
	\label{fig:alltasks_subj_est_avg}
\end{figure}

\begin{figure}[H]
    \centering
	\begin{tabularx}{.9\textwidth}{c|X|X|X|X|}
		\multicolumn{1}{c}{} & \multicolumn{2}{c}{\textbf{Classical GLM}} & \multicolumn{2}{c}{\textbf{Bayesian GLM}}  \\ 
		\cline{2-5}
		\rotatebox[origin=l]{90}{\qquad \qquad \textbf{Visual Cue}} & \multicolumn{2}{c|}{\Includegraphics[width=0.405\textwidth]{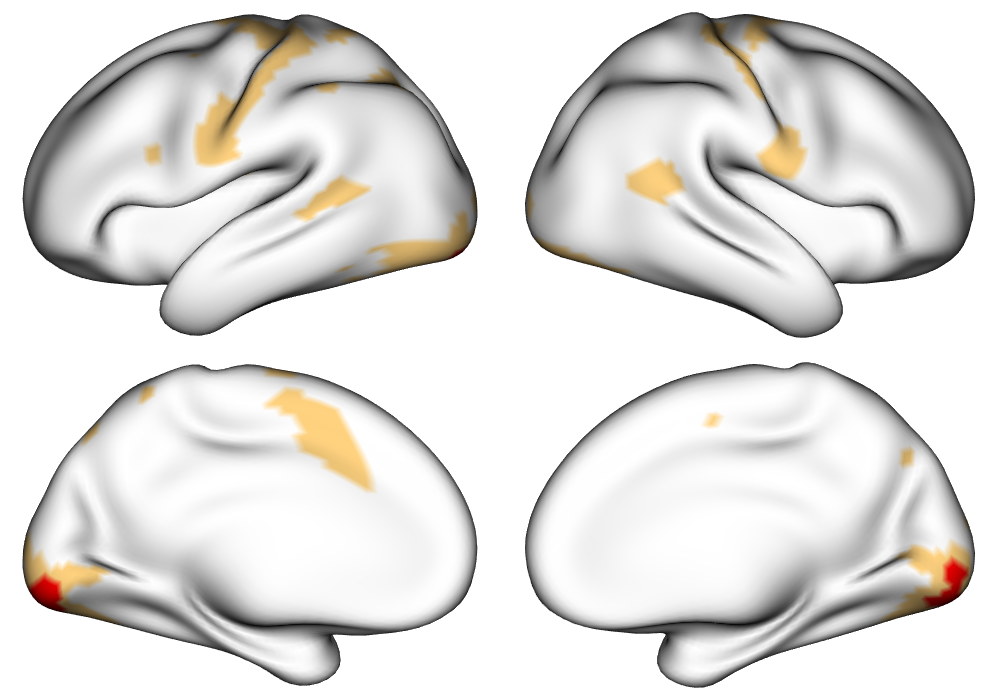}} &
		\multicolumn{2}{c|}{\Includegraphics[width=0.405\textwidth]{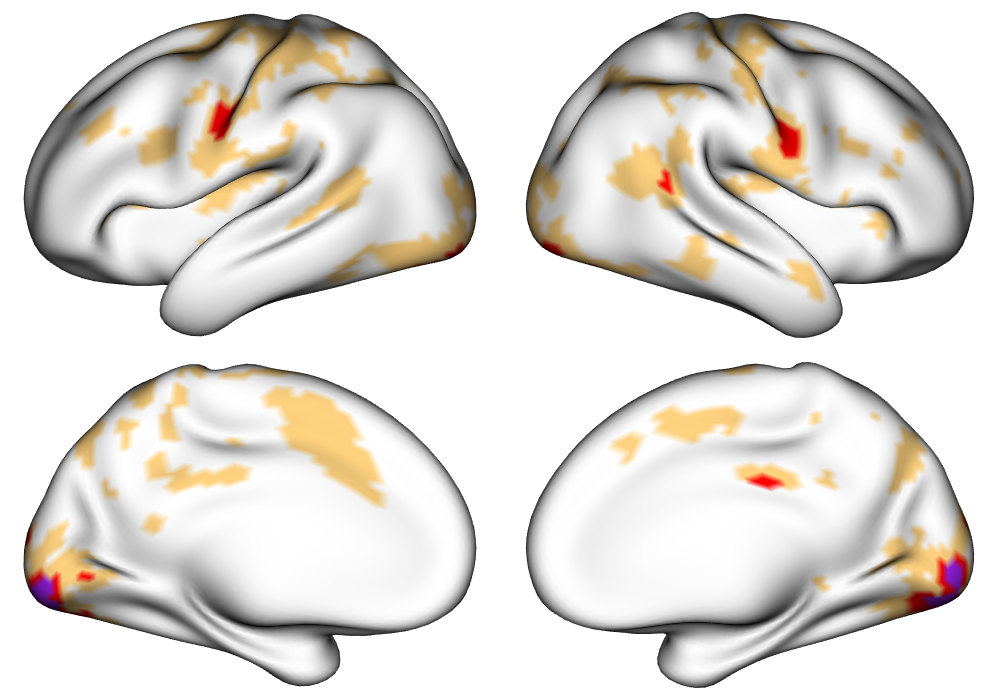}} \\ \cline{2-5}
		\rotatebox[origin=l]{90}{\qquad \qquad \quad \textbf{Tongue}} & \multicolumn{2}{c|}{\Includegraphics[width=0.405\textwidth]{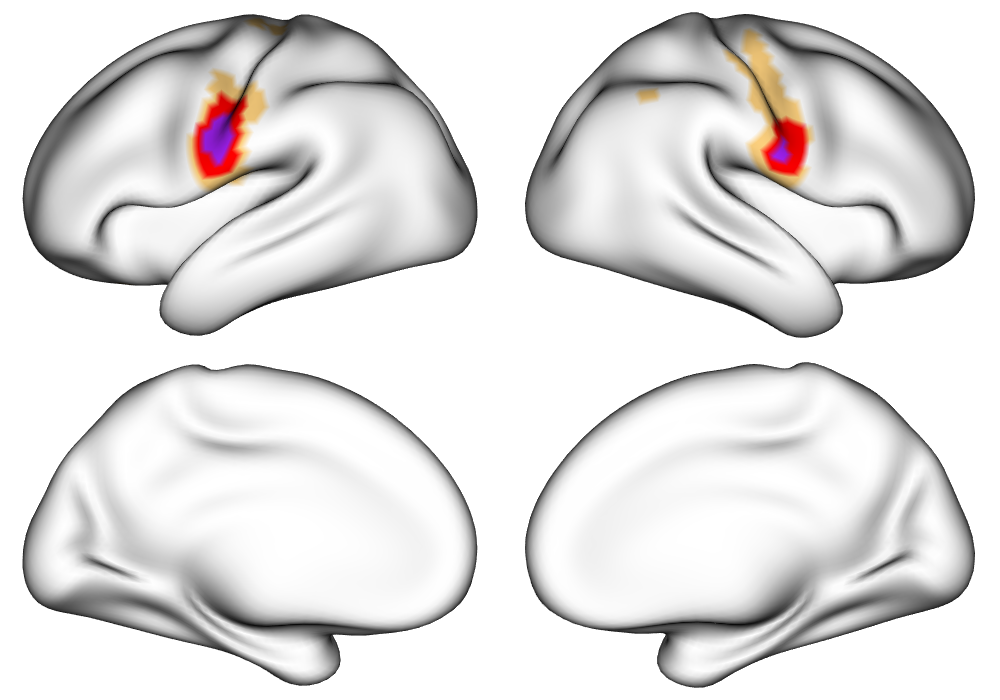}} &
		\multicolumn{2}{c|}{\Includegraphics[width=0.405\textwidth]{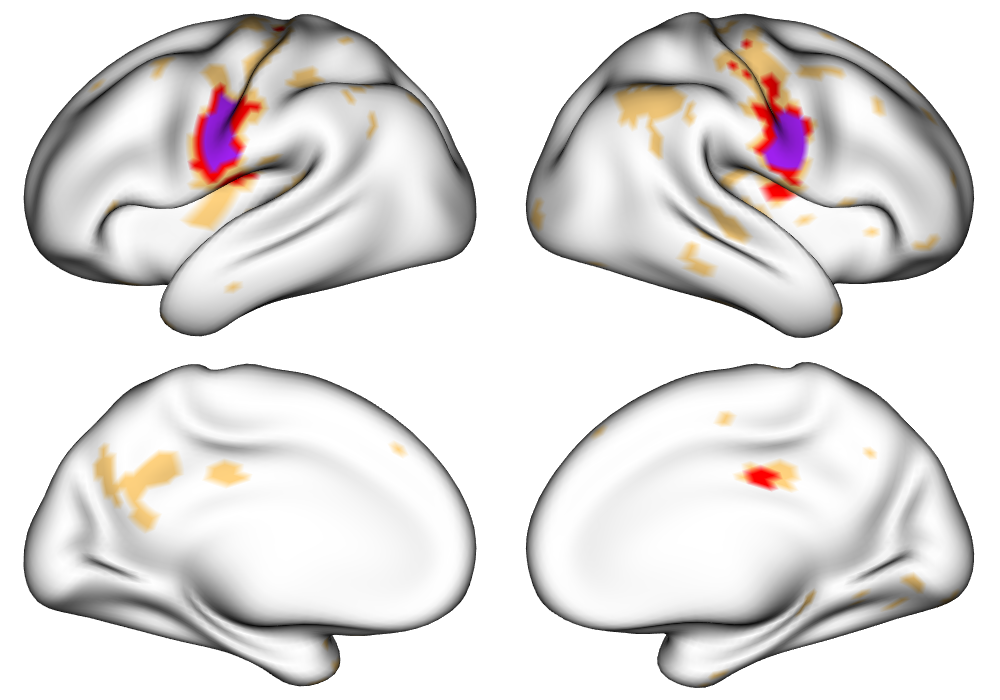}} \\ \cline{2-5}
		\rotatebox[origin=c]{90}{\textbf{Foot}} & \Includegraphics[width=0.198\textwidth]{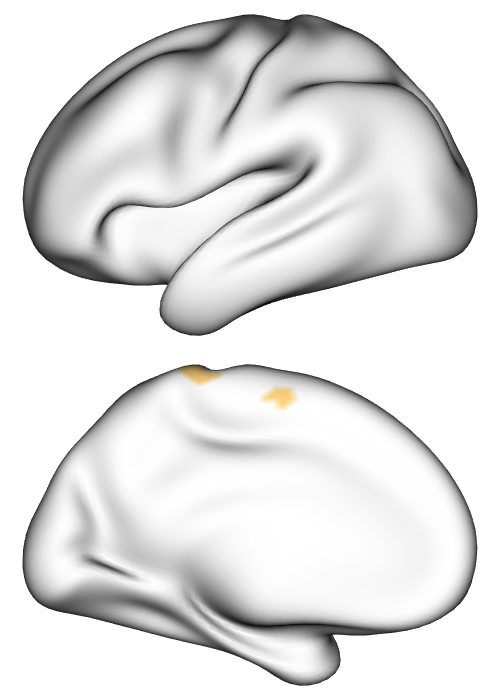} &
		\Includegraphics[width=0.198\textwidth]{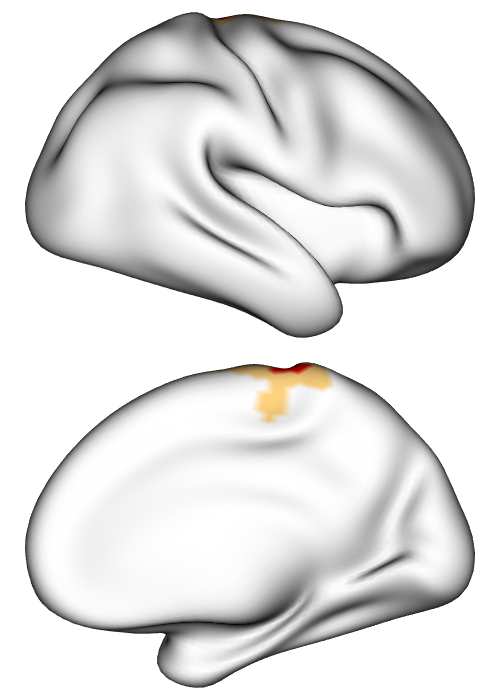} &
		\Includegraphics[width=0.198\textwidth]{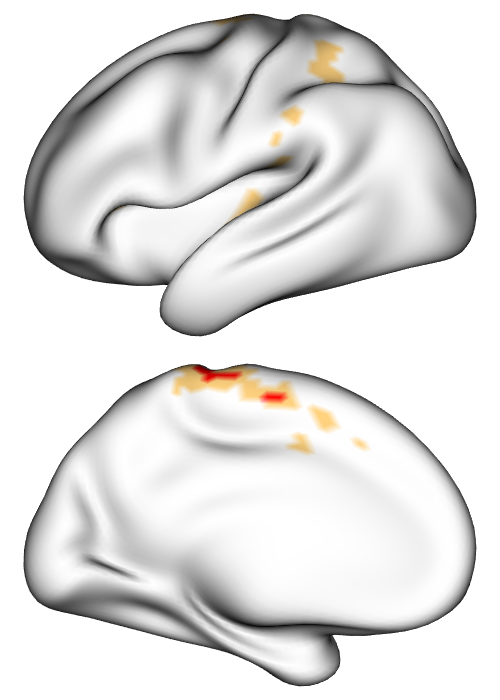} &
		\Includegraphics[width=0.198\textwidth]{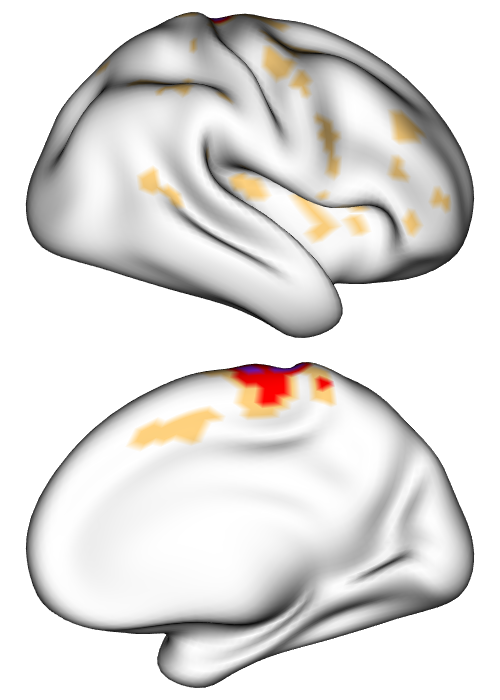} \\ 
		\cline{2-5}
		\rotatebox[origin=c]{90}{\textbf{Hand}} & \Includegraphics[width=0.198\textwidth]{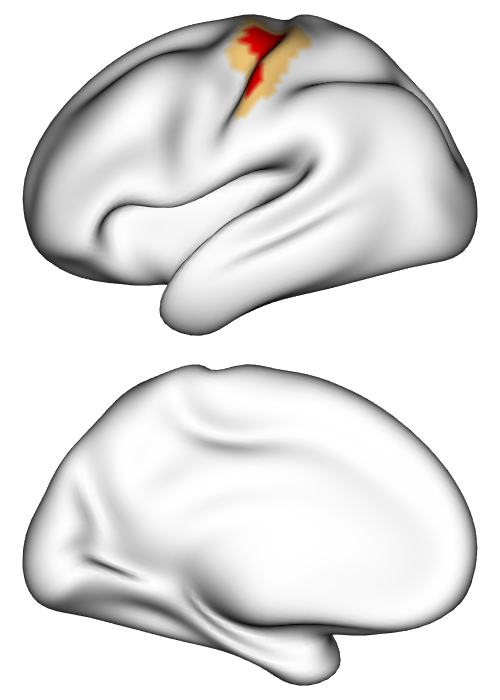} &
		\Includegraphics[width=0.198\textwidth]{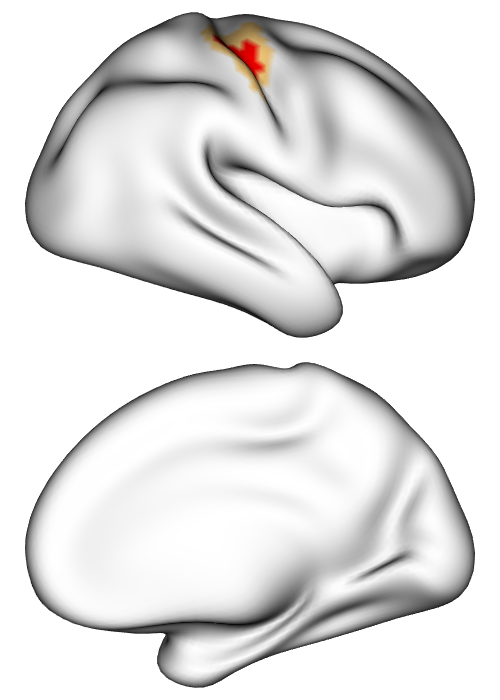} &
		\Includegraphics[width=0.198\textwidth]{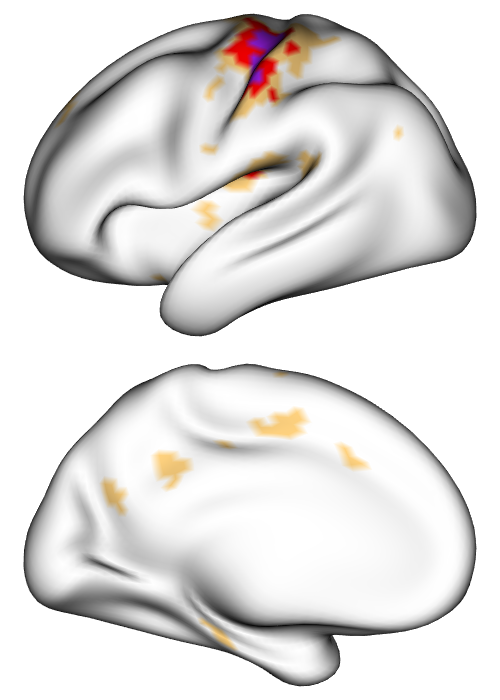} &
		\Includegraphics[width=0.198\textwidth]{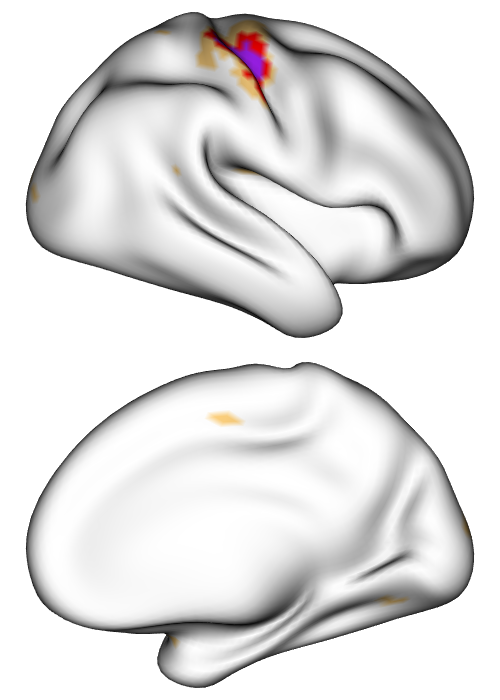} \\ 
		\cline{2-5}
		\multicolumn{1}{c}{} & \multicolumn{4}{c}{$\gamma =$ \textcolor[HTML]{FFD27F}{$\blacksquare$} 0\% 
           \textcolor[HTML]{FF0000}{$\blacksquare$} 0.5\% 
           \textcolor[HTML]{A020F0}{$\blacksquare$} 1\%}
	\end{tabularx}
	\caption{Single-run activation maps for one subject across all six tasks.}
	\label{fig:alltasks_subj_act_LR}
\end{figure}

\begin{figure}[H]
    \centering
	\begin{tabularx}{.9\textwidth}{c|X|X|X|X|}
		\multicolumn{1}{c}{} & \multicolumn{2}{c}{\textbf{Classical GLM}} & \multicolumn{2}{c}{\textbf{Bayesian GLM}}  \\ 
		\cline{2-5}
		\rotatebox[origin=l]{90}{\qquad \qquad \textbf{Visual Cue}} & \multicolumn{2}{c|}{\Includegraphics[width=0.405\textwidth]{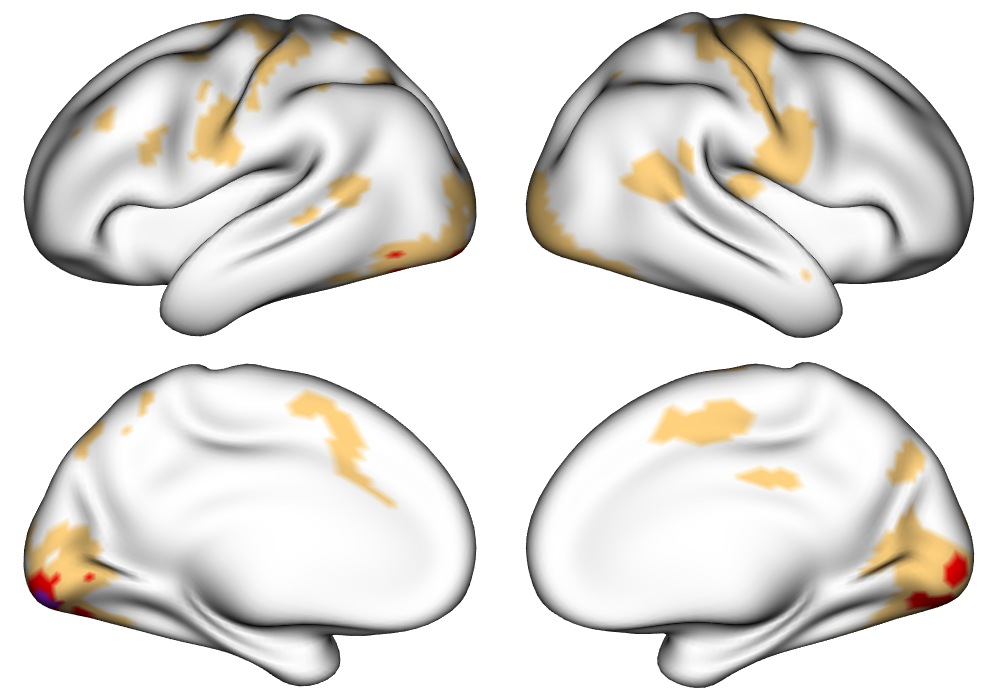}} &
		\multicolumn{2}{c|}{\Includegraphics[width=0.405\textwidth]{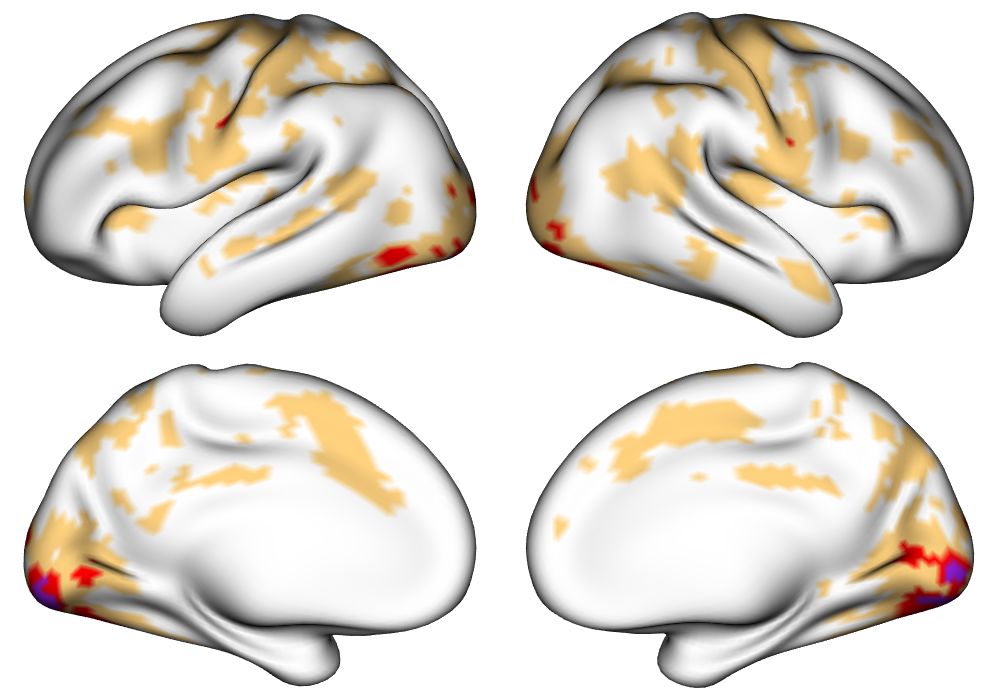}} \\ \cline{2-5}
		\rotatebox[origin=l]{90}{\qquad \qquad \quad \textbf{Tongue}} & \multicolumn{2}{c|}{\Includegraphics[width=0.405\textwidth]{607_classical_103818_visit1_tongue_activations.png}} &
		\multicolumn{2}{c|}{\Includegraphics[width=0.405\textwidth]{607_bayes_103818_visit1_tongue_activations.png}} \\ \cline{2-5}
		\rotatebox[origin=c]{90}{\textbf{Foot}} & \Includegraphics[width=0.198\textwidth]{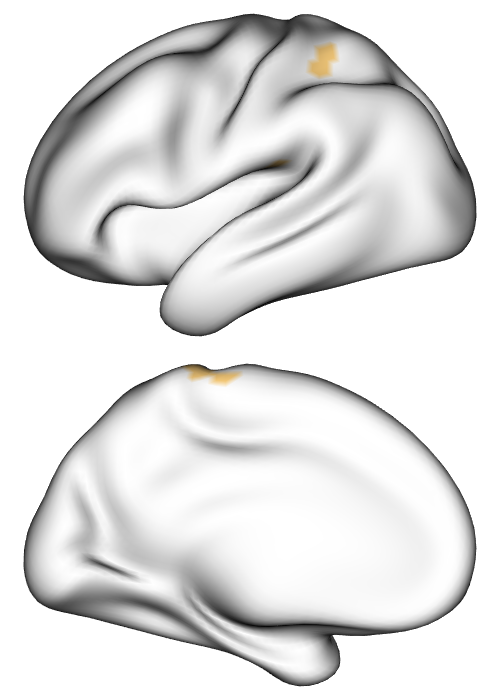} &
		\Includegraphics[width=0.198\textwidth]{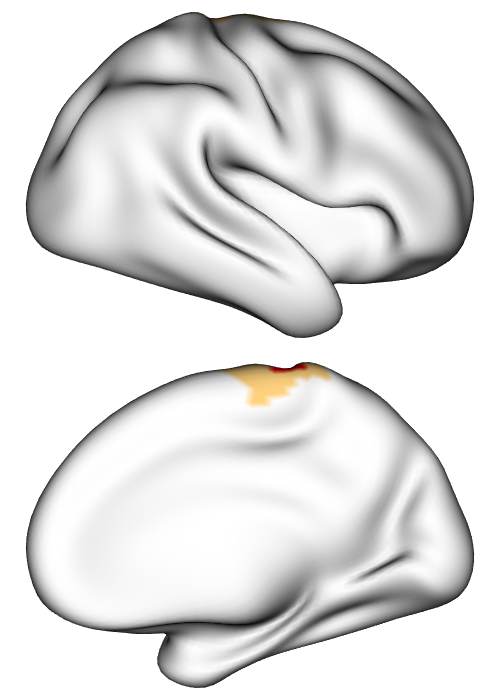} &
		\Includegraphics[width=0.198\textwidth]{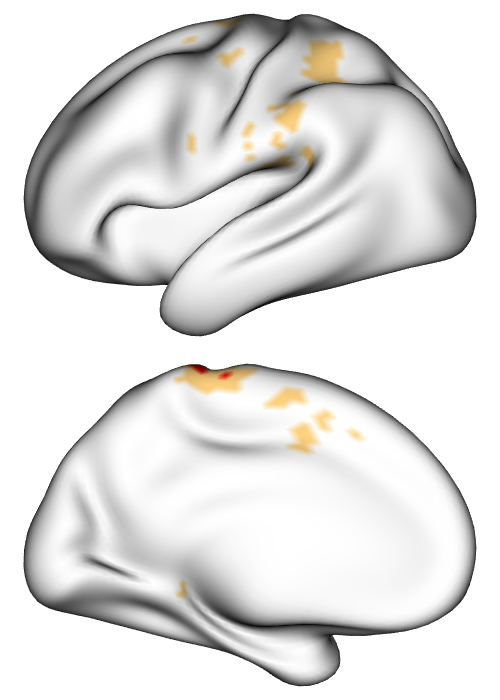} &
		\Includegraphics[width=0.198\textwidth]{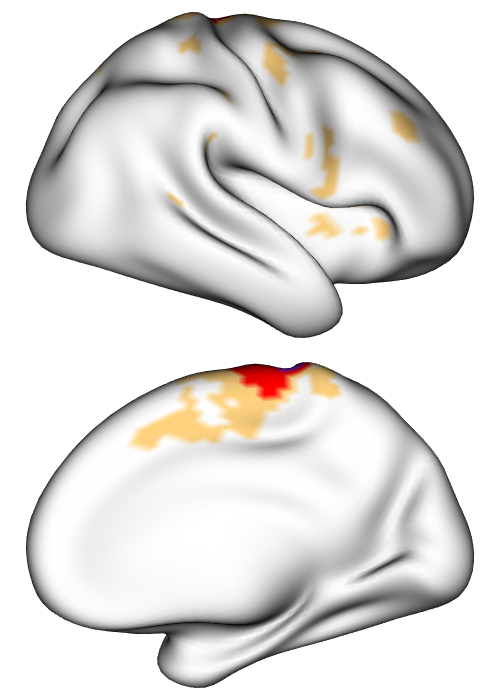} \\ 
		\cline{2-5}
		\rotatebox[origin=c]{90}{\textbf{Hand}} & \Includegraphics[width=0.198\textwidth]{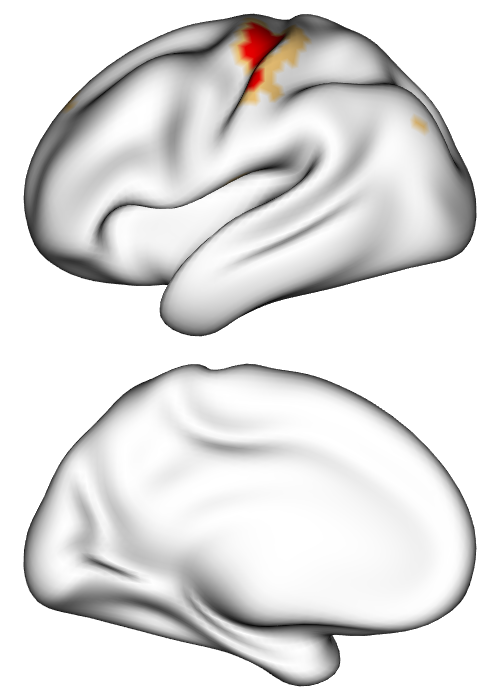} &
		\Includegraphics[width=0.198\textwidth]{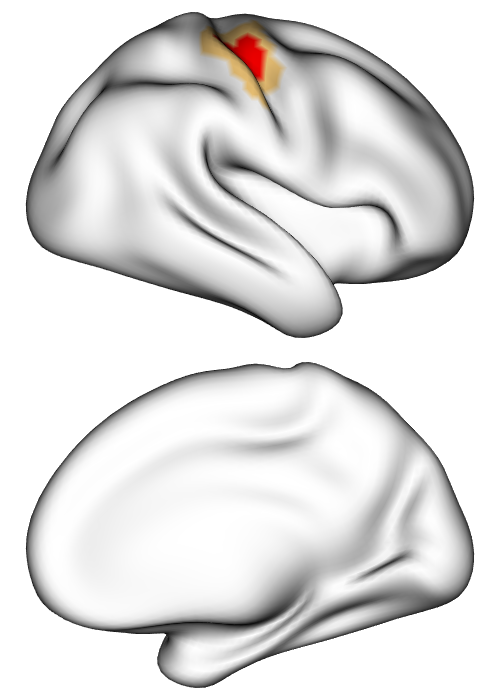} &
		\Includegraphics[width=0.198\textwidth]{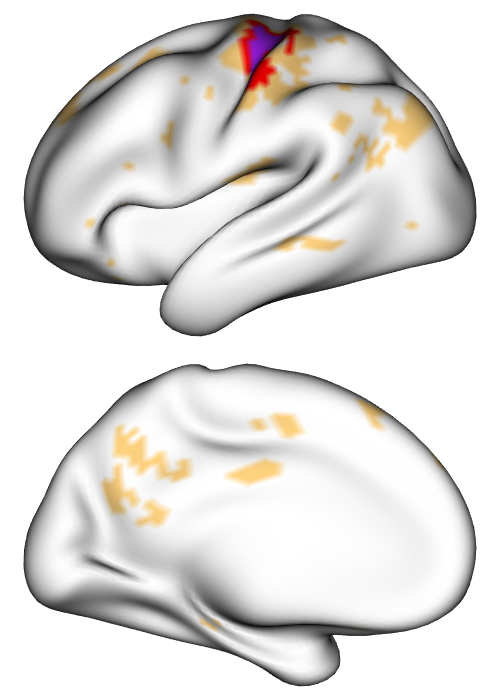} &
		\Includegraphics[width=0.198\textwidth]{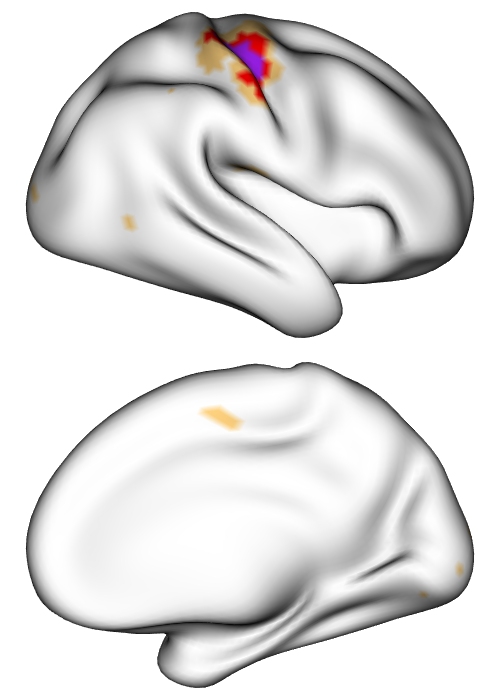} \\ 
		\cline{2-5}
		\multicolumn{1}{c}{} & \multicolumn{4}{c}{$\gamma =$ \textcolor[HTML]{FFD27F}{$\blacksquare$} 0\% 
           \textcolor[HTML]{FF0000}{$\blacksquare$} 0.5\% 
           \textcolor[HTML]{A020F0}{$\blacksquare$} 1\%}
	\end{tabularx}
	\caption{Single-visit (two runs) activation maps for one subject across all six tasks.}
	\label{fig:alltasks_subj_act_avg}
\end{figure}

\begin{figure}[H]
    \centering
	\begin{tabularx}{.9\textwidth}{c|X|X|X|X|}
		\multicolumn{1}{c}{} & \multicolumn{2}{c}{\textbf{$\gamma = 0.5\%$}} & \multicolumn{2}{c}{\textbf{$\gamma = 1\%$}}  \\ 
		\cline{2-5}
		\rotatebox[origin=l]{90}{\qquad \qquad \textbf{Visual Cue}} & \multicolumn{2}{c|}{\Includegraphics[width=0.405\textwidth]{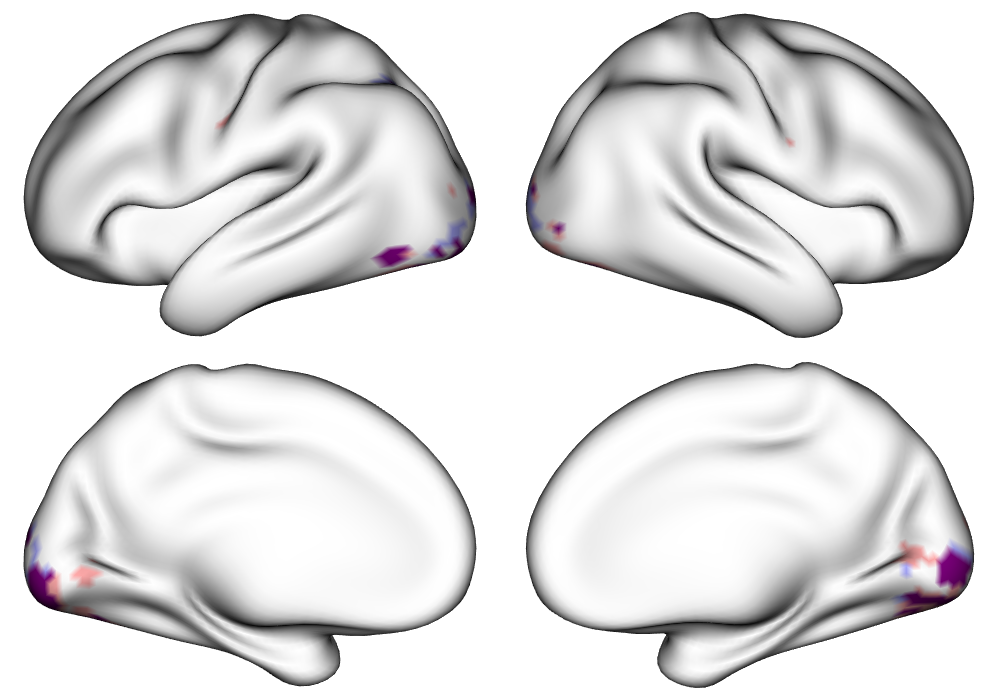}} &
		\multicolumn{2}{c|}{\Includegraphics[width=0.405\textwidth]{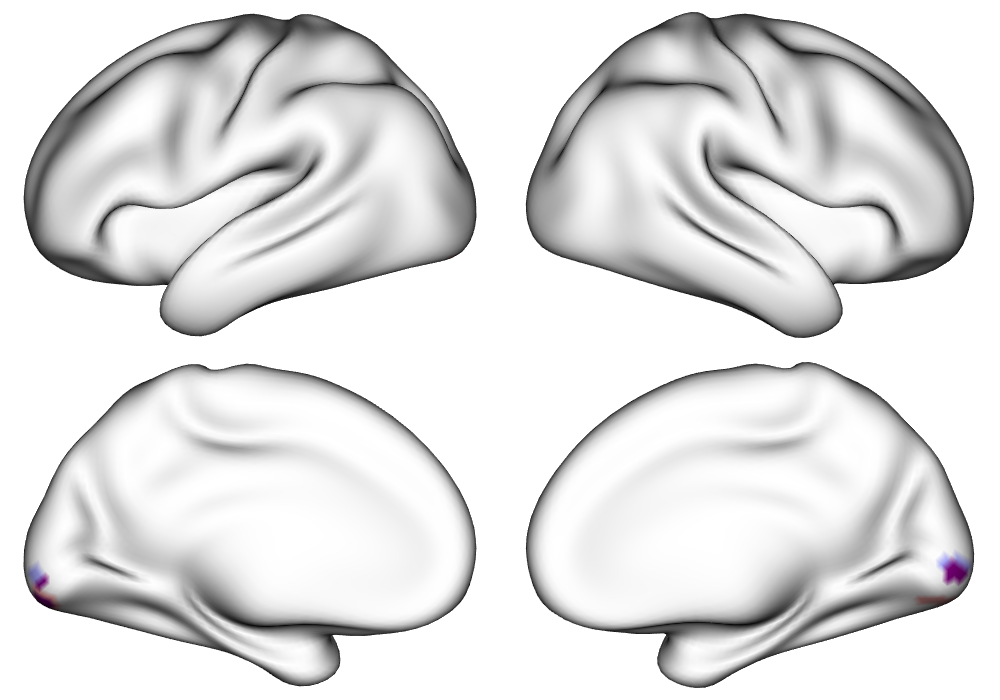}} \\ \cline{2-5}
		\rotatebox[origin=l]{90}{\qquad \qquad \quad \textbf{Tongue}} &
		\multicolumn{2}{c|}{\Includegraphics[width=0.405\textwidth]{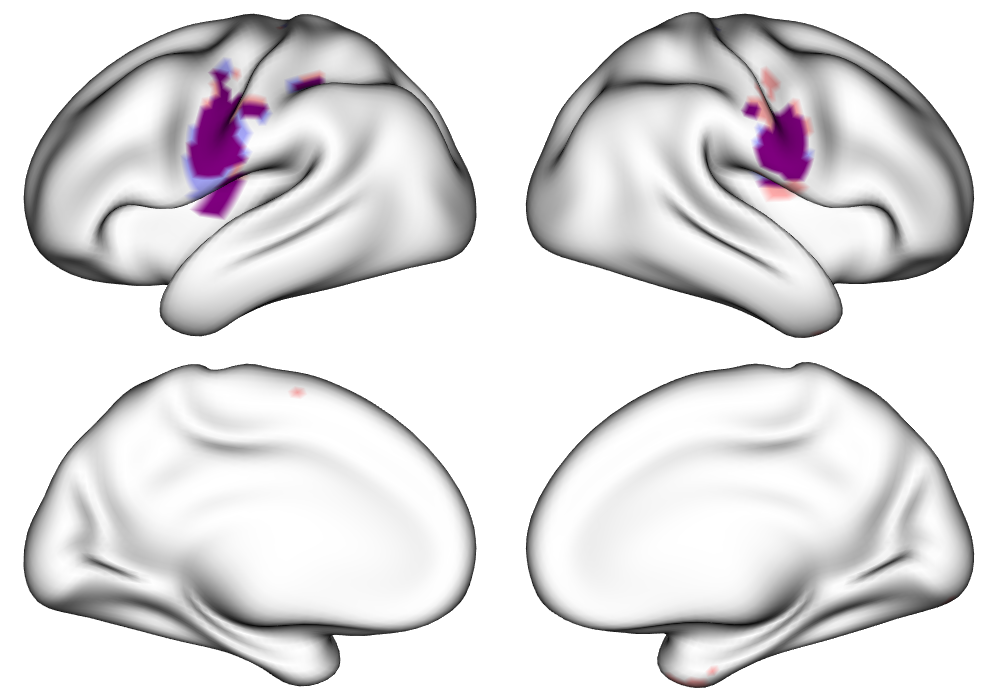}} &
		\multicolumn{2}{c|}{\Includegraphics[width=0.405\textwidth]{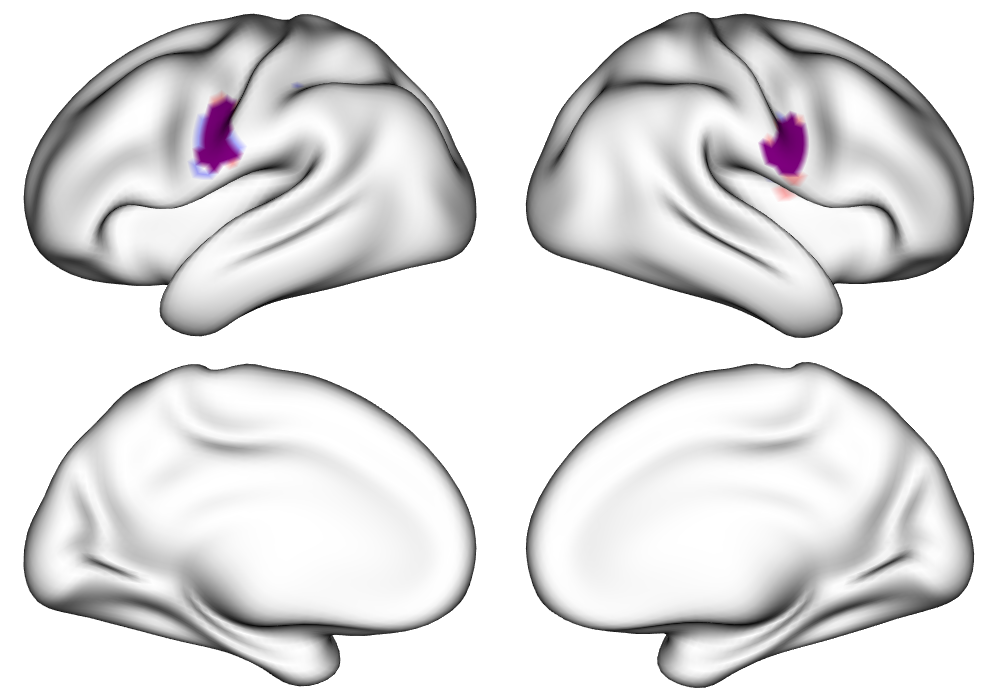}} \\ \cline{2-5}
		\rotatebox[origin=c]{90}{\textbf{Foot}} & \Includegraphics[width=0.198\textwidth]{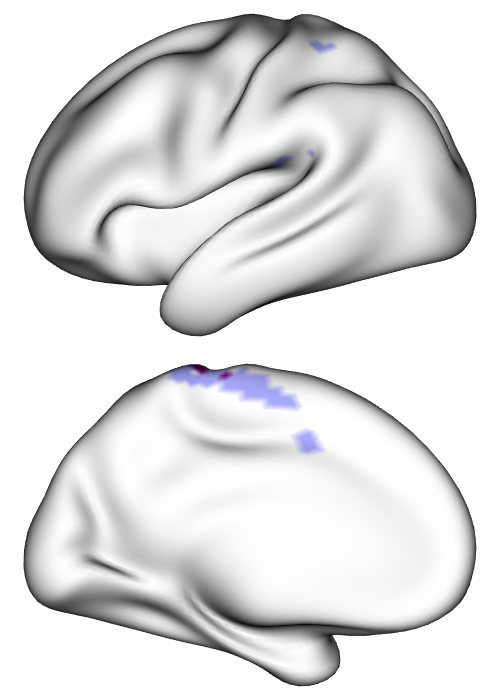} &
		\Includegraphics[width=0.198\textwidth]{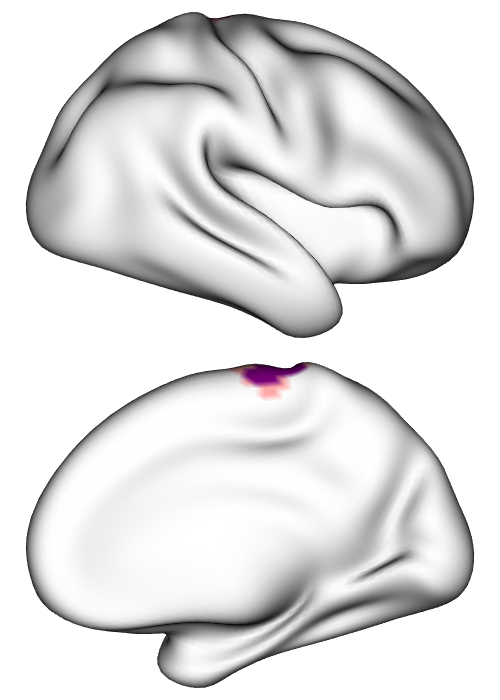} &
		\Includegraphics[width=0.198\textwidth]{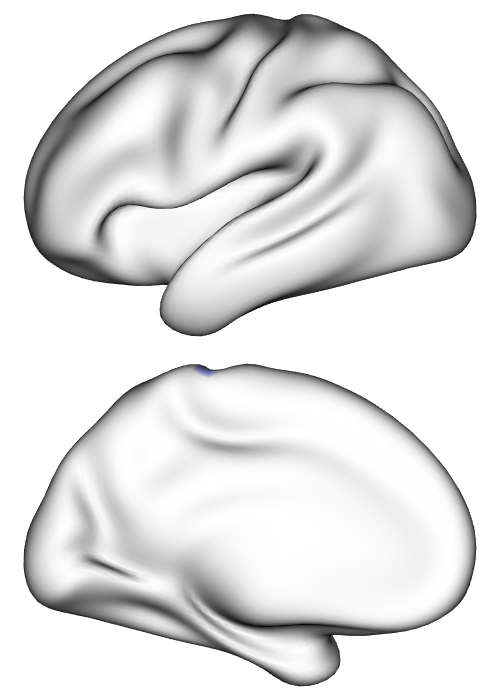} &
		\Includegraphics[width=0.198\textwidth]{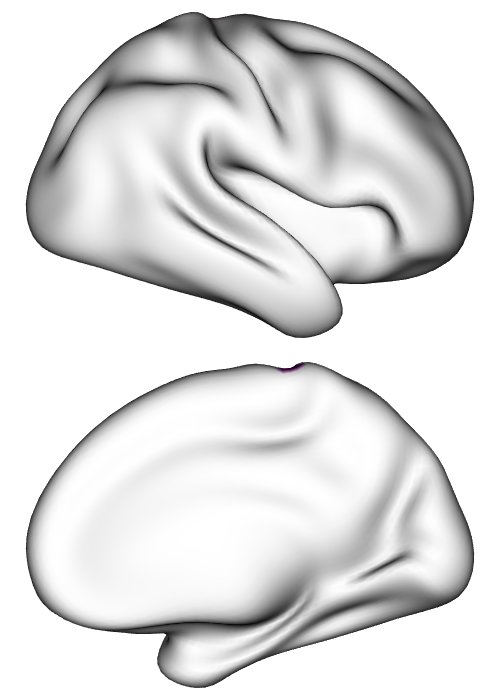} \\ 
		\cline{2-5}
		\rotatebox[origin=c]{90}{\textbf{Hand}} & 
		\Includegraphics[width=0.198\textwidth]{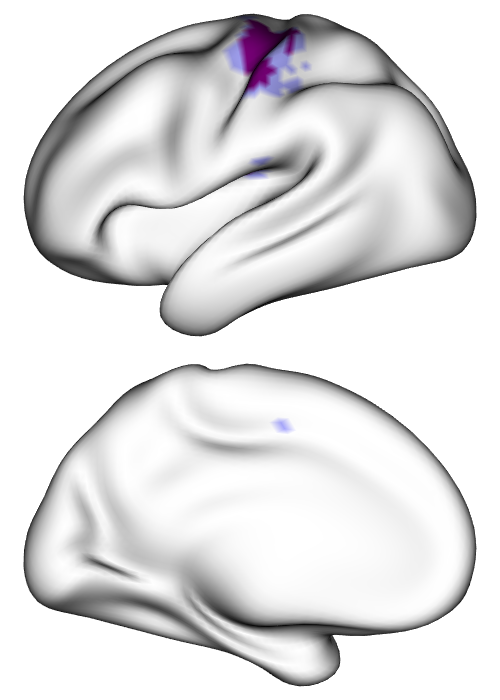} &
		\Includegraphics[width=0.198\textwidth]{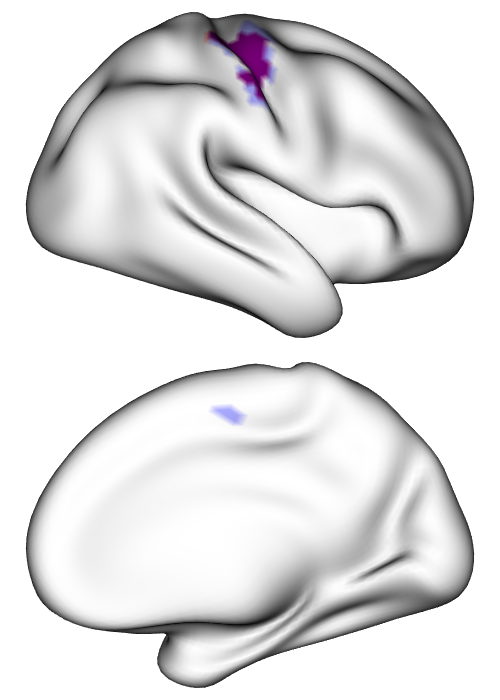} &
		\Includegraphics[width=0.198\textwidth]{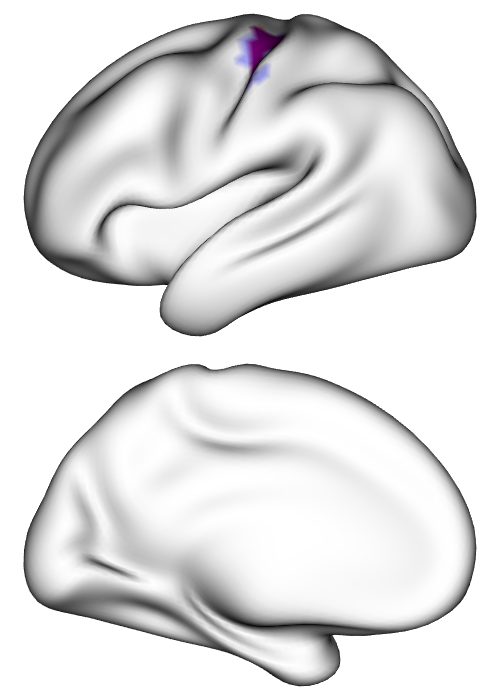} &
		\Includegraphics[width=0.198\textwidth]{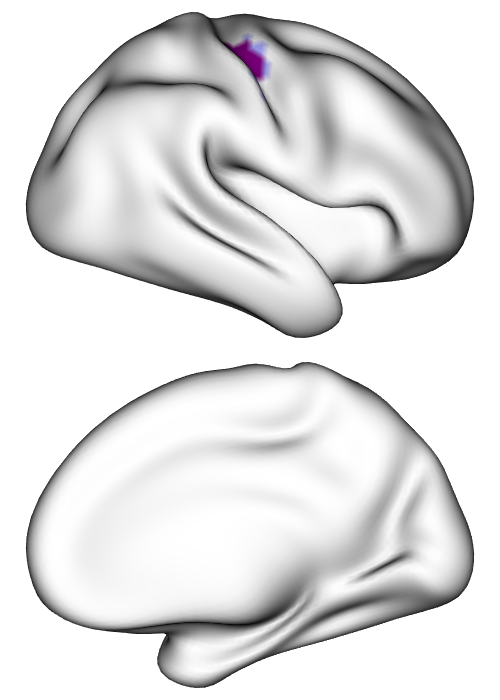} \\ 
		\cline{2-5}
        \multicolumn{1}{c}{} & \multicolumn{4}{c}{Activation detected in \quad \textcolor[HTML]{FFAAAA}{$\blacksquare$} visit 1 \quad
        \textcolor[HTML]{AAAAFF}{$\blacksquare$} visit 2 \quad
        \textcolor[HTML]{7F007F}{$\blacksquare$}  both visits }
	\end{tabularx}
	\caption{Subject-level activations consistently detected across visits for subject A. Areas of activation are found using the Bayesian GLM with activation thresholds $\gamma=(0.5\%,1\%)$ and significance level $\alpha = 0.01$.  Areas of activation are highly consistent within the subject across visits.}
	\label{fig:alltasks_act_single_subject_two_visits}
\end{figure}

\begin{figure}[H]
\centering
\begin{subfigure}[b]{0.39\textwidth}
    \includegraphics[width=1\textwidth]{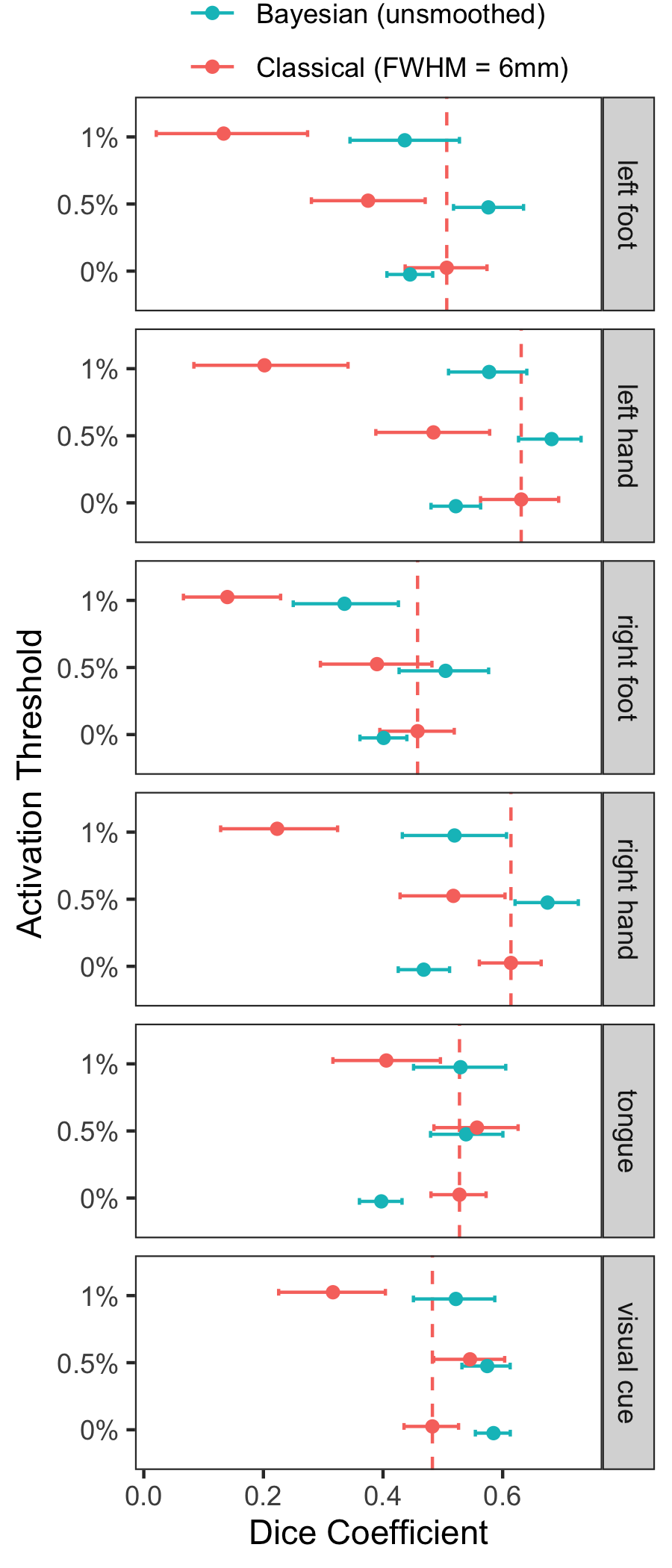} 
\caption{Average Dice overlap}
\end{subfigure}
\begin{subfigure}[b]{0.59\textwidth}
        \includegraphics[width=1\textwidth]{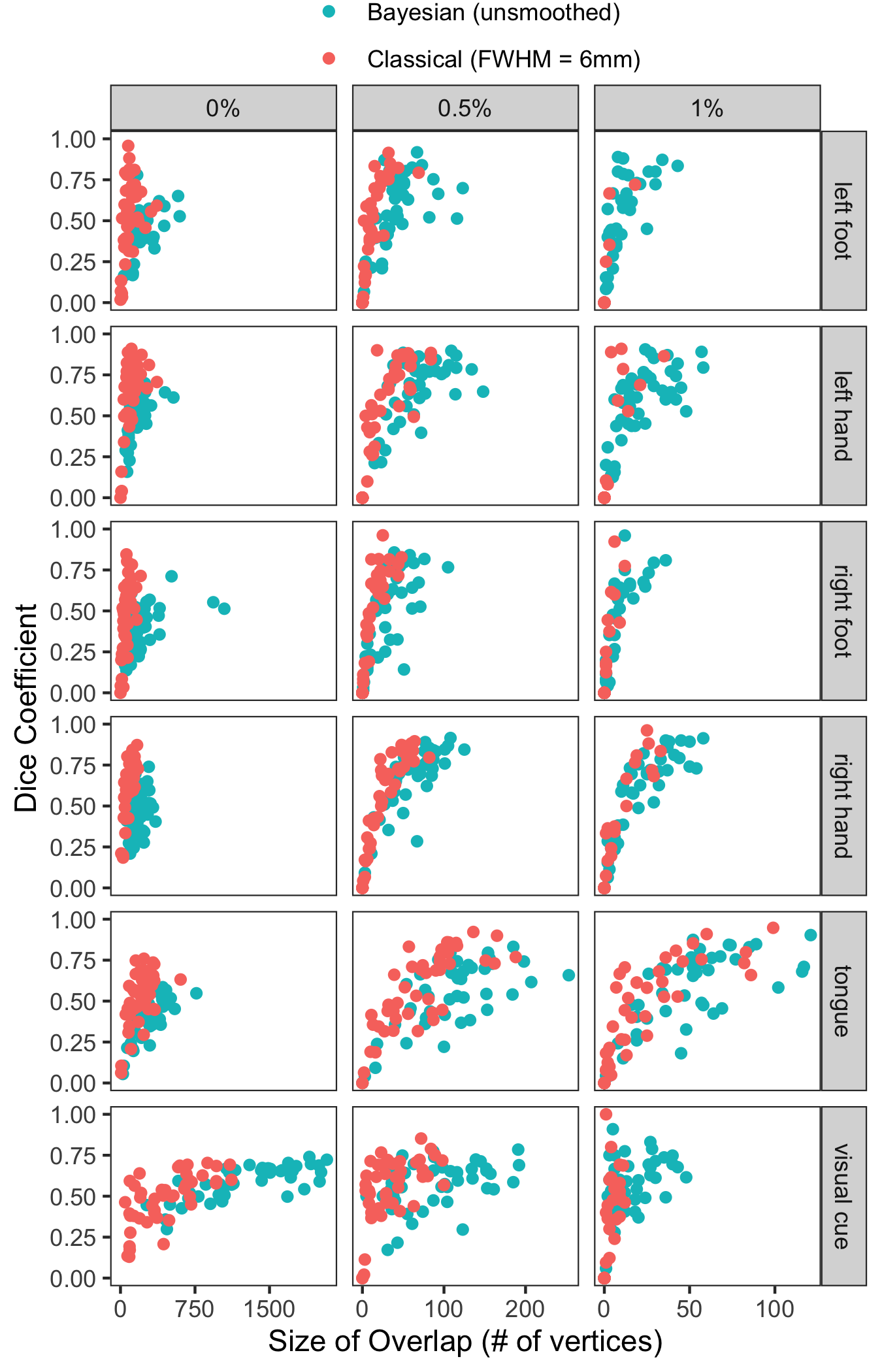}
\caption{Size of activation versus Dice overlap}
\end{subfigure}
\caption{\textbf{Test-retest reliability of subject-level areas of activation.} Bayesian GLM areas of activation are based on the joint posterior distribution of activation amplitude across all vertices. Classical GLM areas of activation are based on performing a hypothesis test at every location and controlling the FWER. For both GLMs, the significance level is $\alpha=0.01$ within each hemisphere. \textbf{(a)} The average Dice test-retest overlap of areas of activation across all subjects, with 95\% bootstrap confidence intervals. For the classical GLM, the most reliable areas of activation are typically produced using activation threshold $\gamma=0\%$, corresponding to a traditional hypothesis-testing approach; this is treated as the benchmark and is indicated with a vertical line. For the Bayesian GLM, an activation threshold of $\gamma=0.5\%$ tends to produce the most reliable results, which significantly outperforms the classical GLM benchmark for all tasks. \textbf{(b) }Size of activation (overlap across both visits) versus Dice overlap.  The Bayesian GLM produces areas of activation that tend to be both larger and more reliable. This illustrates that the Bayesian GLM benefits from both a gain in power, producing larger areas of activation, and a gain in reliability.}
    \label{fig:dice_improvements}
\end{figure}

\begin{figure}[H]
    \centering
	\begin{tabularx}{.9\textwidth}{c|X|X|X|X|}
		\multicolumn{1}{c}{} & \multicolumn{2}{c}{\textbf{Classical GLM}} & \multicolumn{2}{c}{\textbf{Bayesian GLM}}  \\ 
		\cline{2-5}
		\rotatebox[origin=l]{90}{\qquad \qquad \textbf{Visual Cue}} & \multicolumn{2}{c|}{\Includegraphics[width=0.405\textwidth]{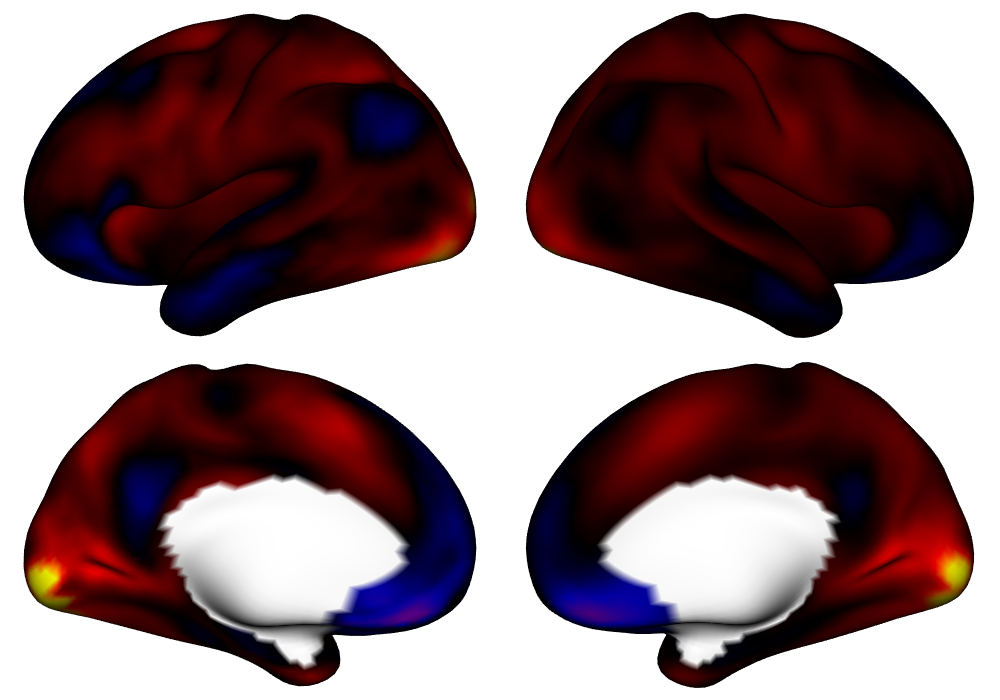}} &
		\multicolumn{2}{c|}{\Includegraphics[width=0.405\textwidth]{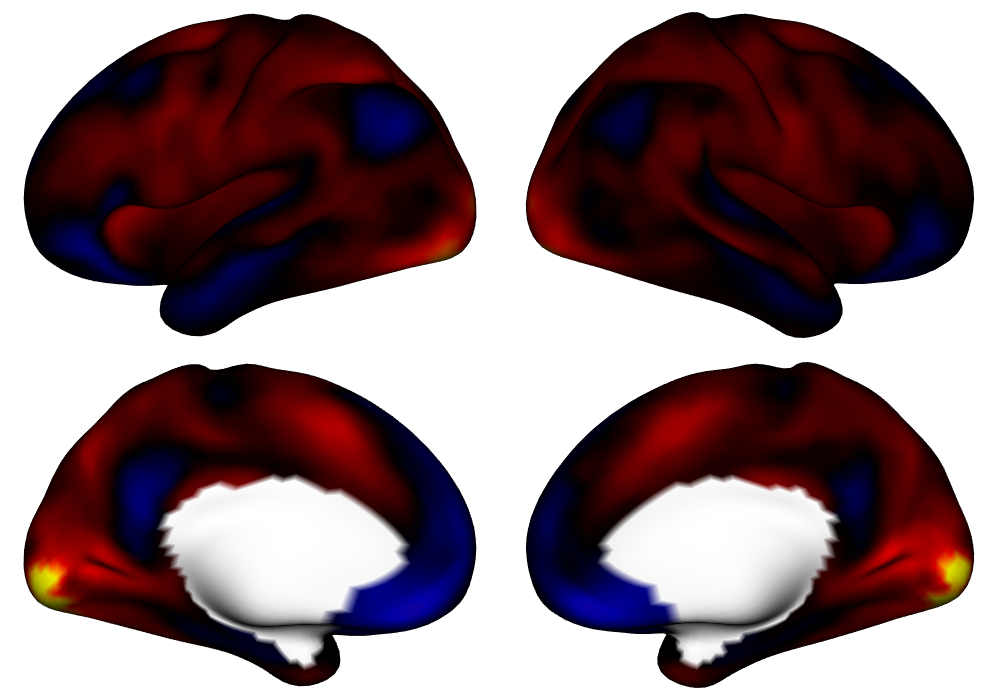}} \\ \cline{2-5}
		\rotatebox[origin=l]{90}{\qquad \qquad \quad \textbf{Tongue}} &
		\multicolumn{2}{c|}{\Includegraphics[width=0.405\textwidth]{607_group_classical_tongue_estimate.png}} &
		\multicolumn{2}{c|}{\Includegraphics[width=0.405\textwidth]{607_group_bayes_tongue_estimate.png}} \\ \cline{2-5}
		\rotatebox[origin=c]{90}{\textbf{Foot}} & \Includegraphics[width=0.198\textwidth]{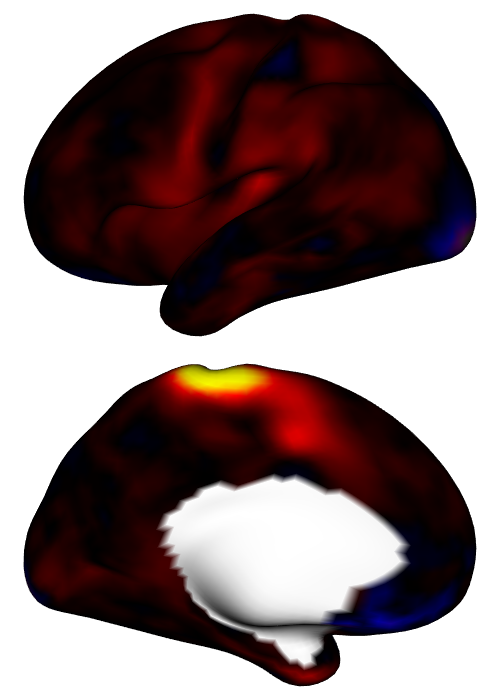} &
		\Includegraphics[width=0.198\textwidth]{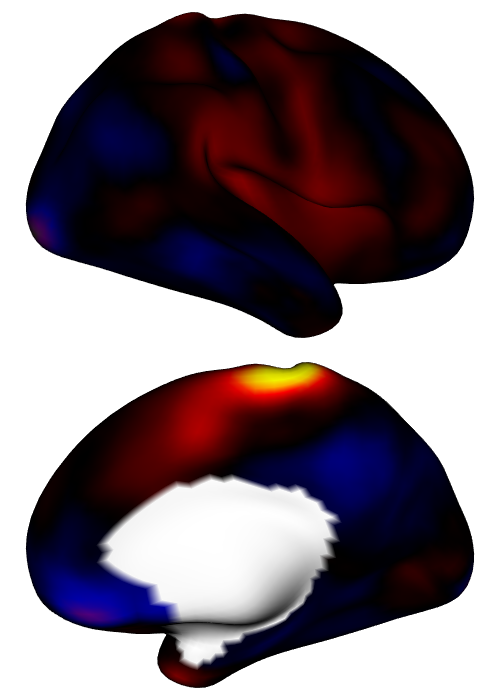} &
		\Includegraphics[width=0.198\textwidth]{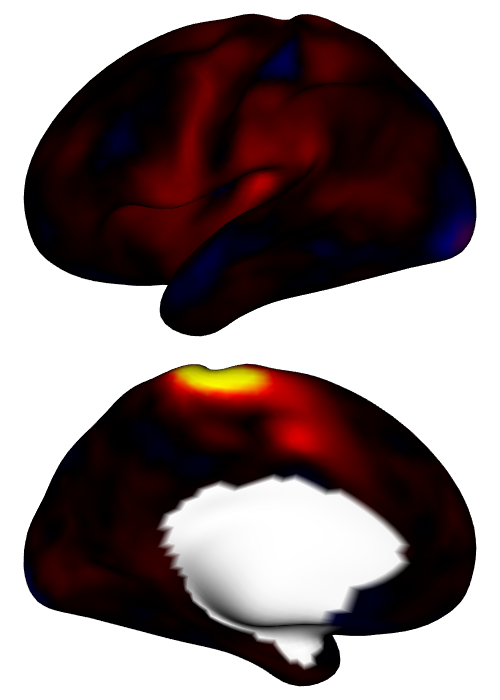} &
		\Includegraphics[width=0.198\textwidth]{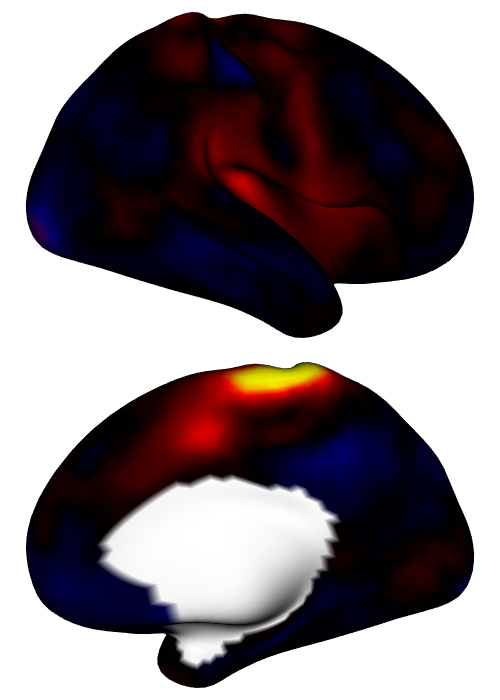} \\ 
		\cline{2-5}
		\rotatebox[origin=c]{90}{\textbf{Hand}} & 
		\Includegraphics[width=0.198\textwidth]{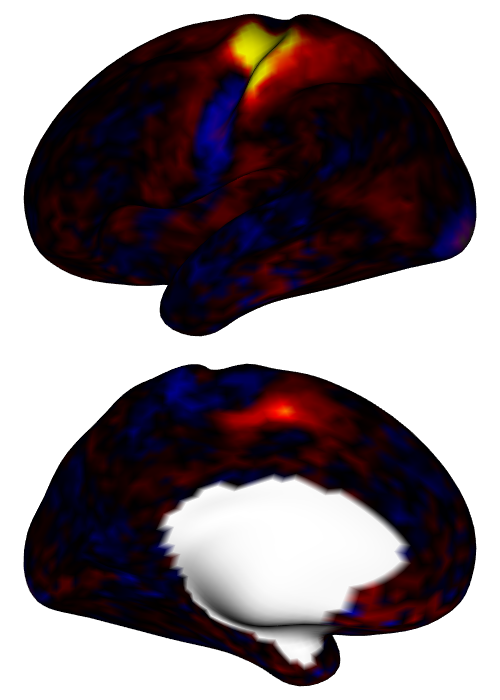} &
		\Includegraphics[width=0.198\textwidth]{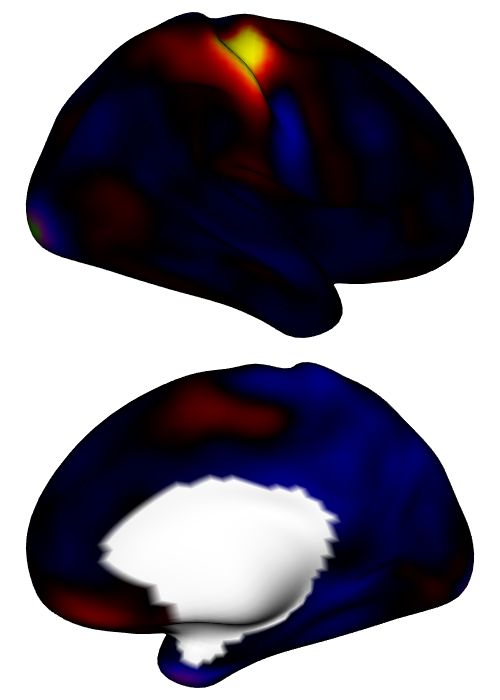} &
		\Includegraphics[width=0.198\textwidth]{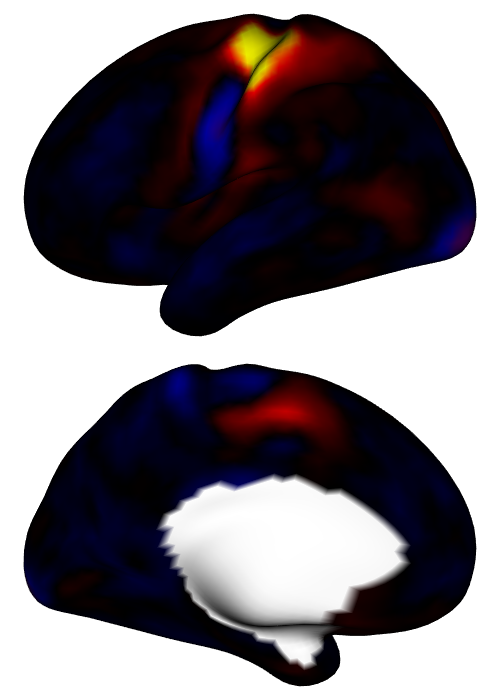} &
		\Includegraphics[width=0.198\textwidth]{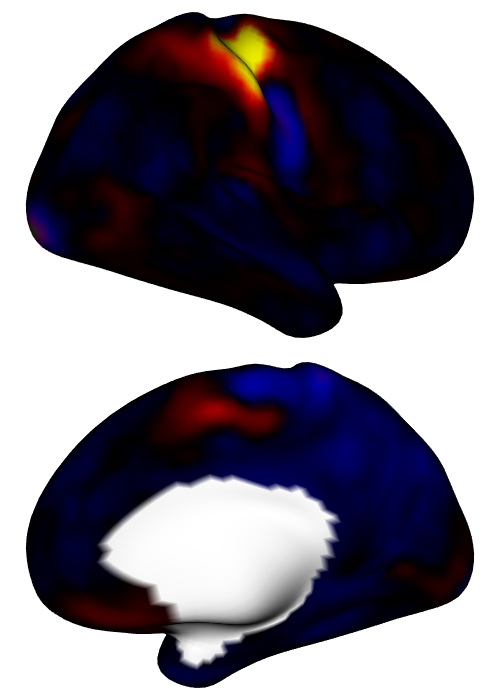} \\ 
		\cline{2-5}
		\multicolumn{1}{c}{} & \multicolumn{4}{c}{\Includegraphics[width = 0.4\textwidth]{607_legend_estimate.png}}
	\end{tabularx}
	\caption{Group-level estimates of activation for each motor task and the visual cue, in units of local percent signal change, based on the average across all subjects using the test data. For lateral tasks, only the contralateral hemisphere is displayed.}
	\label{fig:alltasks_group_est}
\end{figure}

{\setlength{\extrarowheight}{5pt}}

\begin{figure}[H]
    \centering
	\begin{tabularx}{.9\textwidth}{c|X|X|X|X|}
		\multicolumn{1}{c}{} & \multicolumn{2}{c}{\textbf{Classical GLM}} & \multicolumn{2}{c}{\textbf{Bayesian GLM}}  \\
		\cline{2-5} 
		\rotatebox[origin=l]{90}{\qquad \qquad \textbf{Visual Cue}} & \multicolumn{2}{c|}{\Includegraphics[height=4.45cm]{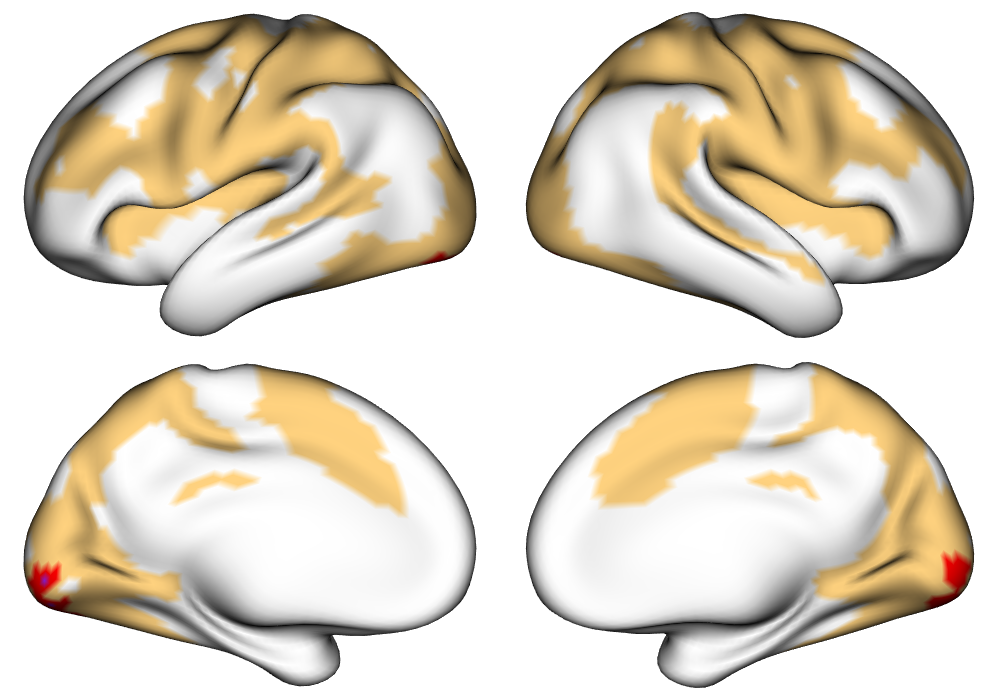}} &
		\multicolumn{2}{c|}{\Includegraphics[height=4.45cm]{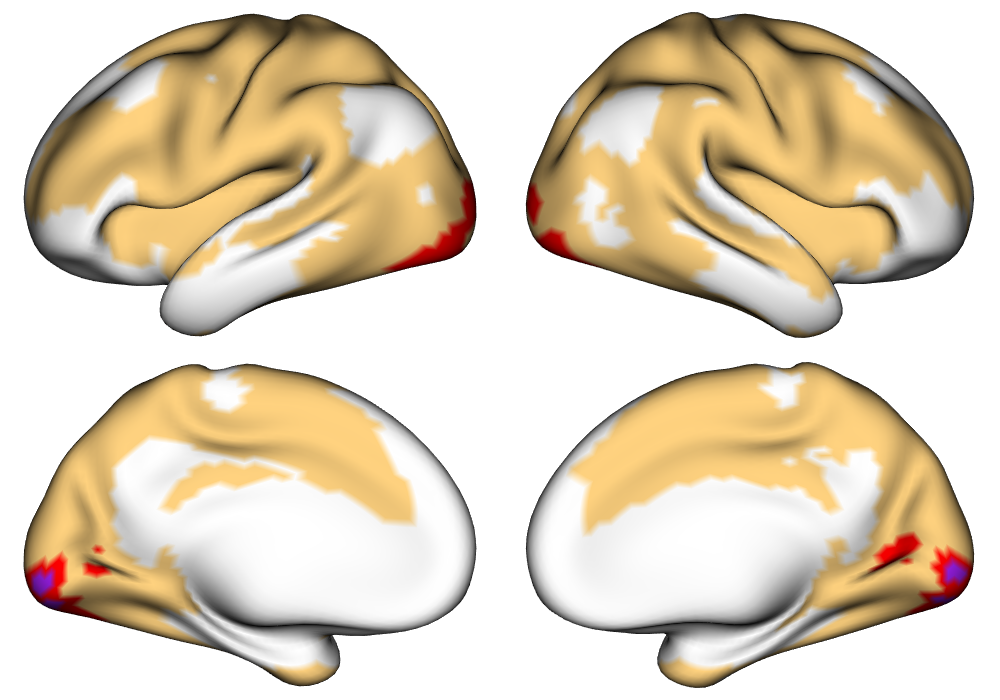}} \\ \cline{2-5}
		\rotatebox[origin=l]{90}{\qquad \qquad \quad \textbf{Tongue}} &
		\multicolumn{2}{c|}{\Includegraphics[height=4.45cm]{607_group_classical_tongue_activations.png}} &
		\multicolumn{2}{c|}{\Includegraphics[height=4.45cm]{607_group_bayes_tongue_activations.png}} \\ \cline{2-5}
		\rotatebox[origin=c]{90}{\textbf{Foot}} & 
		\Includegraphics[height=4.45cm]{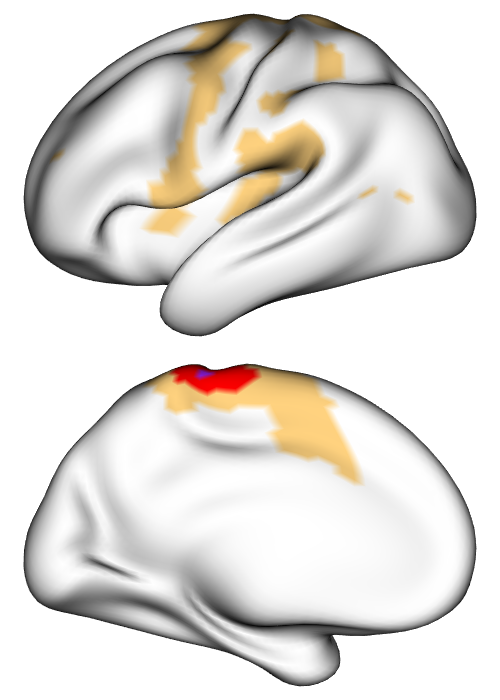} &
		\Includegraphics[height=4.45cm]{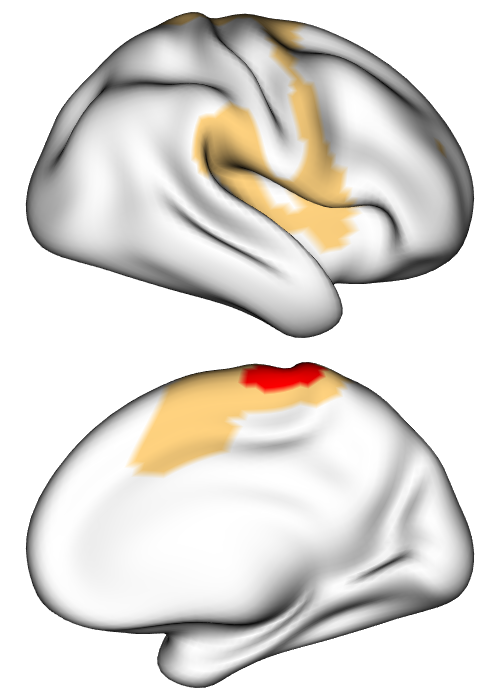} &
		\Includegraphics[height=4.45cm]{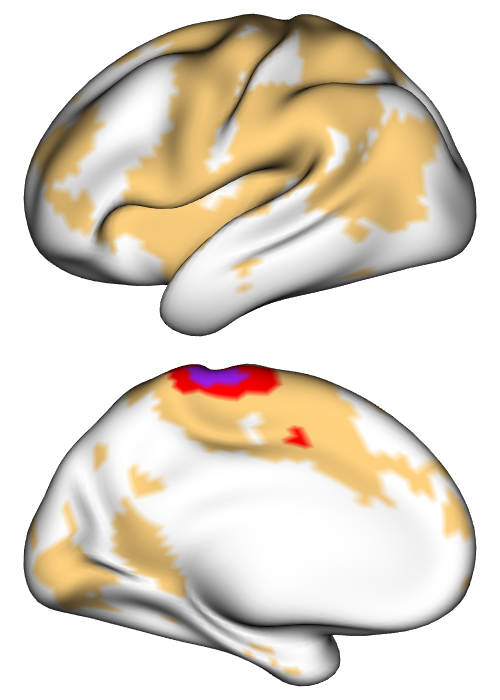} &
		\Includegraphics[height=4.45cm]{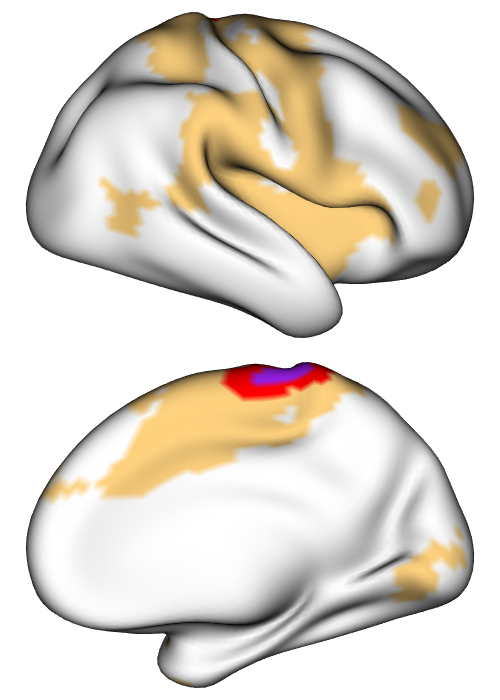} \\ 
		\cline{2-5} 
		\rotatebox[origin=c]{90}{\textbf{Hand}} & 
		\Includegraphics[height=4.45cm]{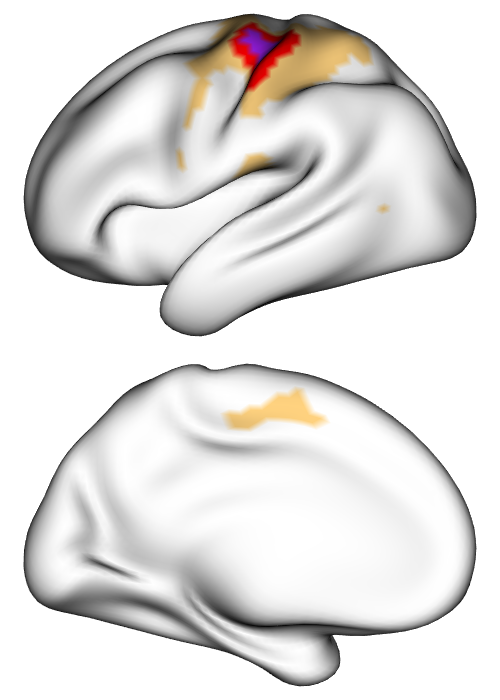} &
		\Includegraphics[height=4.45cm]{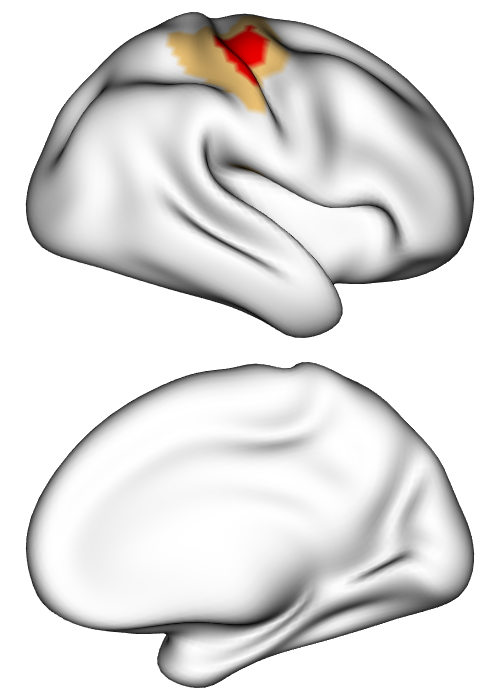} &
		\Includegraphics[height=4.45cm]{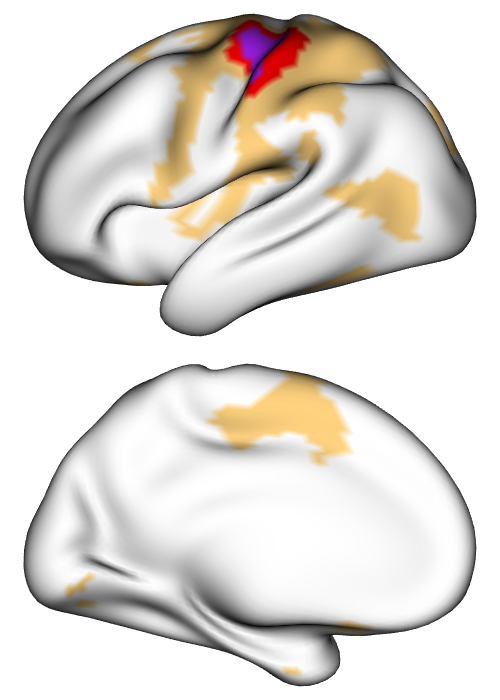} &
		\Includegraphics[height=4.45cm]{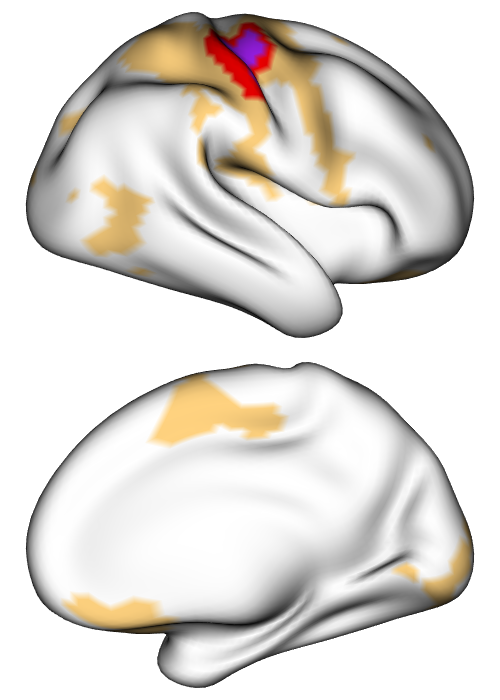} \\ 
		\cline{2-5}
		\multicolumn{1}{c}{} & \multicolumn{4}{c}{$\gamma =$ \textcolor[HTML]{FFD27F}{$\blacksquare$} 0\% 
           \textcolor[HTML]{FF0000}{$\blacksquare$} 0.5\% 
           \textcolor[HTML]{A020F0}{$\blacksquare$} 1\%}
	\end{tabularx}
	\caption{Group-level activations found for each motor task and the visual cue for three different thresholds in percent signal change, based on the average across all subjects using the test data. For lateral tasks, only the contralateral hemisphere is displayed.}
	\label{fig:alltasks_group_act}
\end{figure}

\begin{figure}[H]
    \centering
    \includegraphics[width=0.7\textwidth]{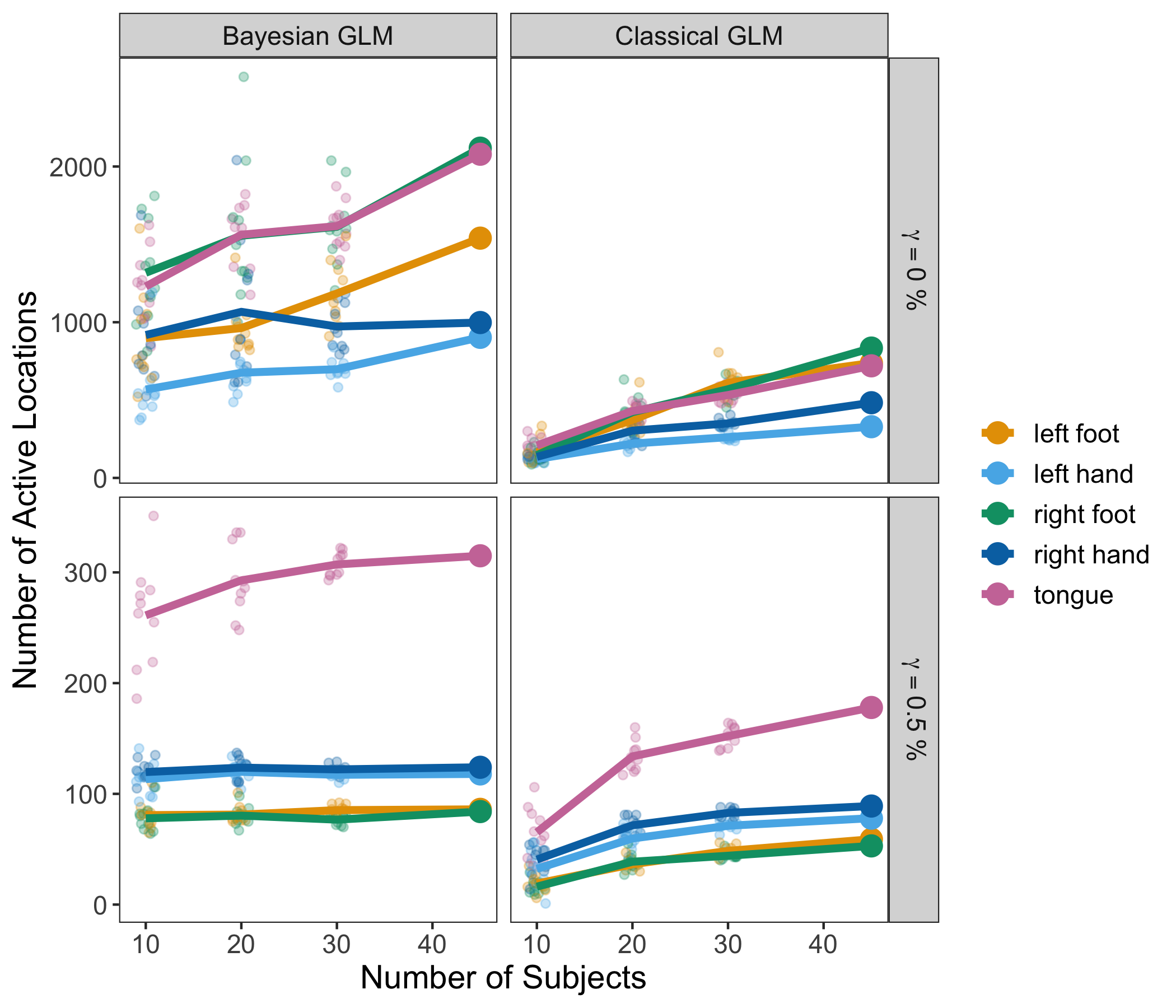}
    \caption{\textbf{Size of group-level areas of activation above 0\% and 0.5\% signal change.} Jittered dots represent random sub-samples of 10, 20 and 30 subjects; lines connect the averages within each sample size. Note that the total number of data locations is approximately 10,000.}
    \label{fig:group_power}
\end{figure}

\end{document}